\documentclass[a4paper, 10pt]{article}
\pdfoutput=1
\usepackage[utf8x]{inputenc}
\usepackage[english]{babel}
\usepackage{amsfonts, amssymb, amsmath, amsthm} 
\usepackage{stmaryrd} 
\usepackage{amsbsy} 
\usepackage{mathrsfs} 
\usepackage{genyoungtabtikz} 
\usepackage{cite} 
\usepackage{pifont} 
\usepackage{indentfirst} 
\usepackage{dsfont} 
\usepackage{relsize} 
\usepackage{geometry} 
\usepackage{graphicx}
\usepackage{tabularx}
\usepackage{fancyhdr}
\usepackage{mdframed}
\usepackage[strict]{changepage}
\usepackage{color, fancybox, empheq}
\definecolor{linkcolor}{rgb}{0,0,0.6}
\usepackage[	pdftex,colorlinks=true,	
			pdfstartview=FitV,
			linkcolor= linkcolor,
			citecolor= linkcolor,
			urlcolor= linkcolor,
			hyperindex=true,
			hyperfigures=false]
			{hyperref}

\linethickness{1.2pt}

\usepackage{lastpage}
\date{}
\newcommand{\Pd}[1]{\mathcal{P}^{(#1)}}
\newcommand{\Pf}[1]{\mathscr{P}^{(#1)}}
\renewcommand{\th}{\text{th}}
\newcommand{\pos}[2]{\underset{#2}{\underset{\uparrow}{#1}}}
\newcommand{\1}{\mathbf{1}}
\newcommand{\branching}{\underset{\so(d) \,\downarrow\,
    \so(d-1)}{\longrightarrow}}
\newcommand{\branchingeven}{\underset{\so(2r) \,\downarrow\,
    \so(2r-1)}{\longrightarrow}}
\newcommand{\id}{\mathlarger{\mathds{1}}}
\newcommand{\spa}[1]{\mathrm{span} \left\{ #1 \right\}}
\newcommand{\flimit}{\underset{\lambda \rightarrow 0}{\longrightarrow}}

\newcommand{\Sp}{\mathsf{\Sigma}}
\newcommand{\e}{\mathsf{e}}

\renewcommand{\a}{\mathfrak{a}}
\renewcommand{\b}{\mathfrak{b}}
\newcommand{\g}{\mathfrak{g}}
\newcommand{\h}{\mathfrak{h}}
\renewcommand{\k}{\mathfrak{K}}
\renewcommand{\l}{\mathfrak{l}}
\newcommand{\m}{\mathfrak{m}}
\newcommand{\n}{\mathfrak{n}}
\newcommand{\p}{\mathfrak{p}}
\renewcommand{\t}{\mathfrak{t}}
\newcommand{\so}{\mathfrak{so}}
\newcommand{\iso}{\mathfrak{iso}}

\newcommand{\blambda}{\boldsymbol{\lambda}}

\newcommand{\C}{\mathbb{C}}
\newcommand{\N}{\mathbb{N}}
\newcommand{\R}{\mathbb{R}}
\newcommand{\V}{\mathbb{V}}
\newcommand{\Y}{\mathbb{Y}}
\newcommand{\Z}{\mathbb{Z}}

\newcommand{\cA}{\mathcal{A}}
\newcommand{\cB}{\mathcal{B}}
\newcommand{\cC}{\mathcal{C}}
\newcommand{\D}{\mathcal{D}}
\newcommand{\cO}{\mathcal{O}}
\newcommand{\cS}{\mathcal{S}}
\newcommand{\U}{\mathcal{U}}
\newcommand{\cV}{\mathcal{V}}
\newcommand{\cW}{\mathcal{W}}
\newcommand{\cY}{\mathcal{Y}}

\newcommand{\dd}{\mathrm{d}}
\newcommand{\ISO}{\mathrm{ISO}}
\newcommand{\SO}{\mathrm{SO}}
\newcommand{\dS}{\mathrm{dS}}
\newcommand{\AdS}{\mathrm{AdS}}
\newcommand{\pAdS}{\mathrm{(A)dS}}
\newcommand{\diag}{\mathrm{diag}}
\newcommand{\Poinc}{\mathrm{Poinc.}}
\newcommand{\M}{\mathscr{M}}

\newtheorem{theorem}{Theorem}[section] \newtheorem*{theorem*}{Theorem}
\newtheorem{lemma}[theorem]{Lemma} \newtheorem*{lemma*}{Lemma}
\theoremstyle{definition}
\newtheorem{definition}{Definition}[section]
\newtheorem*{definition*}{Definition}

\definecolor{theRed}{rgb}{0.56,0,0}
\definecolor{theBlue}{rgb}{0,0,0.8}

\mdfdefinestyle{rem}{topline=false, rightline=false, bottomline=false,
  linewidth=2pt}

\newenvironment{remark}
{ \begin{changemargin}{1.2cm}{0.5cm} \begin{mdframed}[style=rem]
      \underline{\it \bfseries Remark.} }
{ \end{mdframed} \end{changemargin} }

\newenvironment{example}
{ \begin{center}\rule{\textwidth}{1.5pt}\end{center} 
    \underline{\sc Example:} \it }
{ \begin{center}\rule{\textwidth}{1.5pt}\end{center} }

\Yboxdim{9pt} 
\Ylinethick{1pt}
\newcommand{\Yt}[1]{\check{\mathbb{Y}}_p^{(#1)}}
\newcommand{\Yp}[1]{\hat{\mathbb{Y}}_{n,p_I}^{(#1)}}

\usepackage{tikz, tikz-cd, animate}
\usetikzlibrary{arrows, matrix}

\begin{document}

\begin{center}
  \rule{.55\textwidth}{.9pt} \bf \\
  \vspace{.3cm}
  \Large Mixed-symmetry fields in de Sitter space:\\ 
  a group theoretical glance  \\
  \rule{.55\textwidth}{.9pt} 
\end{center}

\begin{center}
Thomas Basile$^{a,b,}$\footnote{E-mail address:
  \href{mailto:thomas.basile@umons.ac.be}{\tt{thomas.basile@umons.ac.be}}},
{Xavier Bekaert$^{a,c,}$\footnote{E-mail address:
    \href{mailto:xavier.bekaert@lmpt.univ-tours.fr}{\tt{xavier.bekaert@lmpt.univ-tours.fr}}}
  and Nicolas Boulanger$^{b,}$\footnote{Research Associate of the Fund
    for Scientific Research$\,$-FNRS (Belgium);
    \href{mailto:nicolas.boulanger@umons.ac.be}{\tt{nicolas.boulanger@umons.ac.be}}}
}
\end{center}

\vspace*{.5cm}

\begin{footnotesize} 
\begin{center}
$^a$
Laboratoire de Math\'ematiques et Physique Th\'eorique\\
Unit\'e Mixte de Recherche $7350$ du CNRS\\
F\'ed\'eration de Recherche $2964$ Denis Poisson\\
Universit\'e Fran\c{c}ois Rabelais, Parc de Grandmont\\
37200 Tours, France \\
\vspace{2mm}{\tt \footnotesize }
$^{b}$
Groupe de M\'ecanique et Gravitation\\
Service de Physique Th\'eorique et Math\'ematique\\
Universit\'e de Mons -- UMONS\\
20 Place du Parc\\
7000 Mons, Belgique\\
\vspace{2mm}{\tt \footnotesize }
$^c$ B.W. Lee Center for Fields, Gravity and Strings\\ 
Institute for Basic Science\\ 
Daejeon, South Korea\\
\vspace*{.3cm}
\end{center}
\end{footnotesize}
\vspace*{.4cm}

\begin{abstract}
  We rederive the characters of all unitary irreducible representations
  of the $(d+1)$-dimensional de Sitter spacetime isometry algebra
  $\so(1,d+1)$, and propose a dictionary between those representations
  and massive or (partially) massless fields on de Sitter spacetime.
  We propose a way of taking the flat limit of representations in
  (anti-) de Sitter spaces in terms of these characters, and
  conjecture the spectrum resulting from taking the flat limit of
  mixed-symmetry fields in de Sitter spacetime. We 
  comment on a possible equivalent of the scalar singleton for the de Sitter (dS) spacetime.
\end{abstract}


\vspace{7mm}

\tableofcontents

\setcounter{footnote}{0}

\section{Introduction}
The importance of mixed-symmetry fields (i.e., fields whose physical
components carry representations of the little group described by
Young diagrams of height greater than one) is no longer to be
emphasised: whether motivated by string theory -- where they make up
most of the spectrum -- or, more generically, by quantum field theory
in arbitrary dimensions -- where they are the central objects of
interest in the sense that they are the most general fields one may
consider.\footnote{They anyway appear upon electric-magnetic duality
  transformation of fields of spin two (or higher) in spacetime
  dimensions greater than four.} At the free level, equations of
motions for \it massless \rm mixed-symmetry fields in flat spacetime
were spelled out by Labastida
\cite{Labastida:1987kw,Labastida:1986zb}\,\footnote{See also
  \cite{Siegel:1986zi} for an earlier, non minimal formulation
  (starting from the light-cone gauge).} (see
\cite{Bekaert:2002dt,Bekaert:2006ix} for a proof that his equations
and trace constraints describe the right propagating degrees of
freedom and \cite{Campoleoni:2009gs} for the fermionic case), and
later given in the unfolded form \cite{Skvortsov:2008vs}.  In anti-de
Sitter (AdS$_{d+1}$) spacetime, the study of massless mixed-symmetry
fields was mostly driven by Metsaev \cite{Metsaev:1995re,
  Metsaev:1997nj, Metsaev:1998xg} who gave both the group theoretical
description of the corresponding $\so(2,d)$ module and (partially)
gauged fixed equations, similar to Fronsdal's equation for totally
symmetric fields \cite{Fronsdal:1978vb} in the De Donder gauge,
i.e., the action of the wave operator on the field is equal to a
critical mass square, together with divergencelessness and
tracelessness conditions, and completed by similar equations on the
gauge parameters. Again, those equations were later revisited using
the unfolded approach in \cite{Boulanger:2008up, Boulanger:2008kw} for
the unitary cases and in \cite{Skvortsov:2009zu, Skvortsov:2009nv} for
the non-unitary cases, thereby generalising the Lopatin-Vasiliev
equations \cite{Lopatin:1987hz} that describe the propagation of free
massless, totally-symmetric fields around (A)dS spacetime. The
generalised Bargmann-Wigner equations for arbitrary mixed-symmetry
(partially)-massless gauge fields, both unitary and non-unitary, were
given in \cite{Boulanger:2008up, Boulanger:2008kw} in a framework
where both spacetime signatures are treated on the same
footing. Notice that the presentation of the equations of motion for
massless fields in the form of a Fierz-Pauli system was given by
Metsaev \cite{Metsaev:1995re, Metsaev:1997nj, Metsaev:1998xg} for the
AdS signature. This being said, the only difference with the dS
signature, as far as the form of the Fierz-Pauli-like equations is
concerned, resides in the sign of the eigenvalue of the
Laplace-Beltrami operator in the wave equation for the mixed-symmetry
gauge potential. However, a nontrivial difference between the positive
and negative cosmological constant cases is the question of unitarity
of the fields and their corresponding (irreducible) representations,
which is one of the main issues investigated in the present paper.

In deriving all these equations, the constraints imposed by gauge
symmetry were crucial. At the group theoretical level, the presence of
this symmetry in AdS$_{d+1}$ translates into the fact that the
representation corresponding to the gauge field is constructed as a
quotient: the gauge parameter module forms a submodule to be modded
out from the gauge field module in order to obtain an irreducible
representation (irrep) of the isometry algebra $\so(2,d)$. Unitary and
irreducible representations (UIRs) of $\so(2,d)$ are well known by
now, and the correspondence with fields in AdS$_{d+1}$ is also well
established, in the physically important cases of bounded
energy. However, a similar dictionary between $(d+1)$-dimensional de
Sitter spacetime (dS$_{d+1}$) fields and the UIRs of its isometry
algebra, $\so(1,d+1)$, is still missing in full generality. A first
step in this direction was made in \cite{Joung:2006gj, Joung:2007je}
where the authors studied UIRs of the de Sitter group corresponding to
massive and massless scalars. Arbitrary spin, and especially
mixed-symmetry massless fields remain elusive in this respect. In the
present paper, we fill this gap and relate arbitrary mixed-symmetry
fields in de Sitter spacetime to UIRs of $\so(1,d+1)$ given in the
mathematical \cite{Hirai1962, Schwarz1971} and Euclidean Conformal
Field Theory (CFT) literature \cite{Dobrev:1977qv, Todorov:1978rf}.

On top of that, mixed-symmetry gauge fields in AdS$_{d+1}$ were shown
to have quite an interesting flat limit \cite{Brink:2000ag}: starting
from a gauge field in AdS$_{d+1}$ with symmetry encoded by an
arbitrary $\so(d)$ Young diagram $\Y$ and sending the cosmological
constant $\Lambda$ to zero yields a spectrum of massless fields in
flat spacetime composed of all possible fields labelled by $\so(d-1)$
Young diagrams obtained from $\Y$ by removing boxes in each of the
last rows of each block (until it reaches the length of the row just
below), leaving the first (upper) block untouched. For proofs of this
spectrum, see \cite{Boulanger:2008up, Boulanger:2008kw,
  Alkalaev:2009vm}. This property can be reformulated as the group
theoretical statement that a massless, mixed-symmetry, irrep of
$\so(2,d)$ contracts to a direct sum of massless Poincar\'e irreps,
the spectrum of massless fields on Minkowski spacetime being given by
a truncation of the branching of $\Y$ with respect to $\so(d-1)
\subset \so(d)$. We show that a similar situation occurs in dS$_{d+1}$,
the difference being that the spectrum is given by a truncated
branching of $\Y$ where the last block (i.e., the lowest one) is left
untouched in the unitary case. In light of the recent revival of
interest for higher-spin theories formulated around flat spacetime
\cite{Campoleoni:2012th, Barnich:2013axa, Campoleoni:2015qrh,
  Sleight:2016xqq, Ponomarev:2016lrm}, such a mechanism relating
massless fields of arbitrary spin in either AdS$_{d+1}$ or dS$_{d+1}$
(which are more natural backgrounds for higher-spin gravity) to their
flat spacetime counter parts can be of great help in understanding the
subtleties of these flat spacetime formulations as limits of theories
in curved spacetime.
\noindent
This paper is organised as follows:
\begin{itemize}
\item In \hyperref[sec:2]{Section 2} we expose the classification of
  the UIRs of $\so(1,d+1)$ that can be found in the literature,
\item In \hyperref[sec:3]{Section 3} we use the previously derived
  character formulae to investigate the flat limit of (massive and)
  massless field/representations of $\so(1,d+1)$,
\item We conclude in \hyperref[sec:4]{Section 4} with some
  considerations on the possibility of a singleton type representation
  for $\so(1,d+1)$ and a corresponding Flato-Fronsdal theorem,
\item Finally, we include a few technicalities in several appendices.
\end{itemize}

\section{$\so(1,d+1)$ unitary irreducible representations}\label{sec:2}

We begin this section by reviewing the classification of the UIRs of
$\so(1,d+1)$ and spelling out their characters (derived in
\hyperref[app:char]{Appendix \ref{app:char}}). With the latter at
hand, we try to establish a dictionary between these UIRs and massive
or massless fields in de Sitter space. 

The Lie algebra $\so(1,d+1)$ is spanned by antisymmetric and Hermitian
generators $M_{AB} = -M_{BA}\, , \, (M_{AB})^{\dagger} = M_{AB}\, , \,
(A,B = 0, 1, \dots, d, d+1)$ subject to the commutation relations:
\begin{equation}
  [M_{AB}, M_{CD}] = i \left( \eta_{BC} M_{AD} + \eta_{AD} M_{BC} -
  \eta_{AC} M_{BD} - \eta_{BD} M_{AC} \right)\,,
\end{equation}
with $\eta_{AB} = \diag(-1, +1, \dots, +1)$. One can perform the
following redefinitions:
\begin{equation}
  D:= -i\,M_{0\,d+1}\, , \, P_i := M_{0i} + M_{d+1\,i}\, , \,
  K_i:=M_{0i} - M_{d+1\,i}\, , \quad(i=1,2,\cdots, d)\, ,
\end{equation}
thereby leading to the commutation relations for the conformal algebra
of the $d$-dimensional Euclidean space\,\footnote{The above generators
  of $\so(1,d+1)$ are related to those of \cite{Dobrev:1977qv} by
  $X_{ij} = i\,M_{ij}\, , \, C_i = K_i\, , \, T_j = P_j$.}:
\begin{equation}
  \left\{
  \begin{aligned}[]
    [M_{ij}, M_{kl}] = i\,\delta_{jk} M_{il} + \dots\,\, & , & [K_i,
      P_j] = 2 \left( iM_{ij} + \delta_{ij} D \right), \\ [M_{ij},
      P_k] = 2\,i\, \delta_{k[j} P_{i]}, \qquad \quad & & [M_{jk}, K_i] =
    2\,i\, \delta_{i[j} K_{k]}, \qquad \\ [D, P_i] = P_i, \qquad \qquad
    \qquad \, \, & & [D, K_i] = -K_i. \qquad \qquad \, \, \,
  \end{aligned}
  \right.
\end{equation}
In this interpretation, the subalgebra $\so(d) = \spa{M_{ij}}$
corresponds to infinitesimal rotations of the Euclidean space
${\mathbb R}^d$. Let $r := \left[ \tfrac{d}2 \right]$ denote the rank
of $\so(d)$ (with $[x]$ denoting the integer part of $x$).  The
remaining generators $D, P_i$ and $K_j$ correspond respectively to
infinitesimal dilations, translations and special conformal
transformations of the Euclidean space ${\mathbb R}^d$.

\subsection{Classification}
Let us start by recalling the classification of the generalised
Lorentz (or de Sitter) group UIRs, established in
\cite{Dobrev:1977qv, Todorov:1978rf} (see also \cite{Hirai1962,
  Schwarz1971, Thieleker1974, Gavrilik:1975ae}).  As was originally
shown by Harish-Chandra, for non-compact semisimple Lie groups these
representations can be classified in different series called
``principal'', ``complementary'' and ``discrete'' (see
\cite{Knapp2001, Knapp2002} for more details). As in the more familiar
case $\so(2,d)$ (see \hyperref[app:ads]{appendix \ref{app:ads}} for a
summary of the relevant irreps of $\so(2,d)$\,), each highest-weight
irrep of $\so(1,d+1)$ is labelled by an $\so(d)$ highest-weight $\vec
s = (s_1, \dots, s_r)$ corresponding to the spin (where the entries
$s_i \in \tfrac{1}2 \N$\,,\footnote{Strictly speaking, for $d=2r$ the
  last entry $s_r$ can be negative as well, and the irreps where the
  last two entries only differ by a sign are related by a discrete
  transformation. For this reason, this subtlety will be ignored in
  this subsection but taken into account later on.}  are such that
$s_1\geqslant s_2\geqslant \cdots\geqslant s_r$ and\,\footnote{In
  other words, the components of the $\so(d)$ highest-weight are
  either all integer or all half-integer.} $2 s_1 = \dots = 2 s_r \,
\text{mod}\, 2\,$) and an additional $\so(1,1)$ weight $\Delta_c \in
\C$ corresponding to the ``conformal weight'' of the
representation.\footnote{Notice that the conformal weight $\Delta$ is
  always a real number in the case of UIRs of $\so(2,d)$, as the
  corresponding Hermitian generator (the energy) spans $\so(2)$.} The
Young diagram $\Y$ has rows of lengths $[\,s_i\,]$ (with
$i=1,2,\cdots,r$). For tensorial representations, the entries of $\vec
s$ are integers, thus $[\,s_i\,]=s_i$ for bosonic fields. In order to
have a simpler uniform treatment (including the fermionic case), with
a slight abuse of notation the lenghts of the rows of the Young
diagram corresponding to a (tensor)-spinor representation of $\so(d)$
with half-integer entries will nevertheless be denoted $s_i$, as in
the bosonic case (although strictly speaking they are equal to
$[\,s_i\,]=s_i-\frac12$). The list of UIRs of $\so(1,d+1)$ is as
follows:

\begin{itemize}
\item {\bf Principal series:} $\Delta_c= \tfrac{d}2 + i\rho$\,, with
 $ \rho \in \R$ and $\vec s$ arbitrary.
\item {\bf Complementary series:} $s_i = 0$ for $p+1 \leqslant i
  \leqslant r$\,, where $p\in\{0, 1,2,\cdots, [\tfrac{d-1}2]\}$ is
  the number of nonvanishing entries in $\vec s$ (thus, when $d$ is
  even, at least one entry vanishes whereas when $d$ is odd, its
    last entry can be non-zero\footnote{In such a case (i.e., $s_{r}\neq 0\,$ and $d$ odd), 
    it appears \cite{Thieleker1974, Todorov:1978rf} that $s_i\in \mathbb{N}$ 
    $\forall i\,$, i.e., half-integer values for $s_{i}$ are excluded.}.) $\Delta_c =
  \tfrac{d}2+c$ with $c \in \R$ such that $0 < \,\rvert c \rvert <
  \tfrac{d}2 - p$\,.
\item {\bf Exceptional series:} $s_i = 0$ {for $p+1 \leqslant i
  \leqslant r$ where $p \in \{ 1,2,\cdots, r \}$ and $\Delta_c =d-p$\,
  or $\Delta_c=p$\,.}  They are essentially the boundary points of the
  complementary series.
\item {\bf Discrete series (only for $d=2r+1$):} $\Delta_c =\tfrac{d}2
  +k$ with $k \in \tfrac{1}2 \N$ and $0 < k \leqslant s_r$ (thus all
  entries in $\vec s$ are non-vanishing).
\end{itemize}

Notice that an irrep labelled by $[\Delta_c\,; \vec s\,]$ is
(partially) equivalent to the irrep labelled by $[d-\Delta_c\,; \vec
  s\,]$ \cite{Dobrev:1998md}.  In Euclidean CFT literature, the
representation for $[d-\Delta_c\,; \vec s\,]$ is usually referred to
as the ``shadow'' of the one for $[\Delta_c\,; \vec s\,]$.\\

\begin{remark}
  The existence of a whole series of UIRs, the discrete one, only in
  {even spacetime dimensions (i.e. odd $d$)} can seem a bit strange at
  first sight, but it can actually be explained by a standard result
  due to Harish-Chandra. Indeed, he proved that a real semisimple Lie
  group possesses a discrete series of UIRs if and only if it has a
  \it compact \rm Cartan subgroup. In the case of $\SO(1,d+1)$ of
  interest for us, which is of rank $r+1$, the maximal compact
  subgroup is $\SO(d+1)$, which has rank $\left[ \tfrac{d+1}2
    \right]$. In {even spacetime dimensions (i.e. $d = 2r+1$), the
    group} $\SO(1,d+1)$ has the same rank $r+1$ as its maximal compact
  subgroup $\SO(d+1)$ and therefore has a compact Cartan subgroup,
  namely the one of the subgroup $\SO(d+1)$. In odd spacetime
  dimensions (i.e. $d = 2r\,$) however, the rank of $\SO(d+1)$ is $r$
  and does not match that of $\SO(1,d+1)$, which means that there is
  no compact Cartan subgroup, hence the absence of a discrete series
  for $d = 2r$.
\end{remark}

\subsection{Structure and characters of the corresponding modules}
The above listed UIRs were constructed and classified using the method
of induced representations (see \cite{Todorov:1978rf}, Chap. IV,
Appendix B), a construction that we will briefly outline for the sake of
completeness. First of all, we need to introduce a few subalgebras of
$\mathfrak{g}=\so(1,d+1)$ (and the corresponding subgroups of
$G=\SO(1,d+1)\,$):
\begin{itemize}
\item $\k = \so(d+1)$ is its maximal compact subalgebra;
\item $\a = \so(1,1) = \spa{D}$ is the abelian subalgebra generated by
  the dilation operator;
\item $\m = \so(d) = \spa{M_{ij}}$ is the centraliser of $\a$ in $\k\,$,
  generated by the $d$-dimensional rotations;
\item $\n = {\mathbb R}^d = \spa{K_i}\,$ is the nilpotent (and
  abelian) subalgebra generated by the special conformal
  transformations.
\end{itemize}
Starting with the Iwasawa decomposition (i.e. the decomposition of a
semisimple Lie algebra into its maximal compact subalgebra, an abelian
and a nilpotent subalgebra) of $\g$, one can introduce the
corresponding Iwasawa decomposition at the group level $G=KAN$, with
$K$, $A$ and $N$ the Lie subgroups of which $\k$, $\a$ and $\n$ are
respectively the Lie algebras.  One can further introduce the
centraliser $M \equiv \SO(d)$ of $A$ in $K$, and define the parabolic
subgroup $P=MAN$ of $\SO(1,d+1)$ in terms of its Langlands
decomposition (the product of semisimple, abelian, and nilpotent
subgroups).  This might be better understood at the algebra level,
where a parabolic subalgebra $\p$ of some semisimple Lie algebra $\g$
is defined as any subalgebra containing the Borel subalgebra $\b$ of
$\g$, the latter being the subalgebra made out of the Cartan
subalgebra together with the subalgebra generated by the raising (or
lowering) operators (or equivalently the subalgebra dual to the space
of positive, or negative, roots). In our case, the Cartan subalgebra
of $\g=\so(1,d+1)$ that we will consider is composed of the Cartan
subalgebra of $\m=\so(d)$ and $\a=\so(1,1)$. The parabolic subalgebra we
are interested in here is $\p = \so(1,1) \inplus \iso(d) 
:=\spa{M_{jk}, K_i, D}\,$, with $\iso(d) = \so(d) \inplus
\n:=\spa{M_{jk}, K_i}$\,. 

Secondly, consider a finite-dimensional UIR $(\V_{\blambda},
\rho_{\blambda})$ of the corresponding parabolic subgroup $P$. It is
labelled by the weight $\blambda = [\Delta_c\,; \vec s\,]$\,, since a
standard lemma (c.f. Lemma 1 in Chapter 19 of \cite{Barut:1986dd})
ensures that the nilpotent subgroup $N$ acts trivially in such a
case. This irrep induces a representation
$(\mathscr{C}^\infty(G,\V_{\blambda}),{\cal R}_{\blambda})$ of $G$ on the
space $\mathscr{C}^\infty(G, \V_{\blambda})$ of functions on the group
$G$ with value in $\V_{\blambda}$ and subject to the covariance
condition:
\begin{equation}
  f(gx) = \rho_{\blambda}(x^{-1}) f(g)\, ,\quad \forall f \in
  \mathscr{C}^\infty(G,\V_{\blambda}), g \in G, x \in P\, ,
\end{equation}
via:
\begin{equation}
  \big({\cal R}_{\blambda}(g) f\big)(g') = f(g^{-1}g')\, , \quad \forall f
  \in \mathscr{C}^\infty(G,\V_{\blambda}),\, g,g' \in G\, .
\end{equation}
Following \cite{Dobrev:1977qv}, these induced representations, where
one uses the action of the group on itself, will be called the
\textit{elementary representations}. The ``subrepresentation theorem''
(see e.g. \cite{Dobrev:1977qv}, p.47) supports this terminology: every
UIR of $\SO(1,d+1)$ is (infinitesimally) equivalent to an irreducible
component of an elementary representation.

In order to classify the UIRs of $\SO(1,d+1)$, one thus has to
decompose the elementary representation into its irreducible and
unitary components, which gives rise to the above mentioned series of
representations\,\footnote{Note that the construction sketched here
  has no claim at providing an exhaustive picture of the theory of
  induced representations, nor at complete mathematical rigor. Our
  only purpose is to give an intuitive picture of the way the
  $\SO(1,d+1)$ UIRs discussed in this paper were classified and their
  relation with the corresponding algebra representations.}. The
principal series corresponds to a continuum of UIRs of $G$ that are
induced by a UIRs of $P$ in which the nilpotent part $N$ is
represented trivially, and are already irreducible as constructed
above. The discrete series corresponds to, as their name suggests, a
discrete set of UIRs induced by $P$ and appearing in the decomposition
of the elementary representation. As mentioned in the previous
subsection, the exceptional series, singled out in the classification
of $\so(1,d+1)$ UIRs actually consists of irreps with a conformal
weight $\Delta_c$ at the unitarity bound of the complementary
series. 

At the algebra level, this construction corresponds to generalised
Verma modules, reviewed in more details in \hyperref[app:bgg]{Appendix
  \ref{app:bgg}}. At the level of Lie algebras, induced
representations are constructed as follows:
\begin{itemize}
\item Given a Lie subalgebra $\h \subset \g$ and a finite-dimensional
  $\h$-module $\V$, the module $\U(\g) \otimes_{\U(\h)}
  \V$ (where $\U(\g)$ is the universal enveloping algebra of $\g$)
  makes up a representation of $\g$;
\item To see that, recall that an element $x$ of $\U(\g)
  \otimes_{\U(\h)} \V$ reads $(y_1 \dots y_k) \otimes v\,$ with $y_1,
  \dots, y_k \in \g\,$ and $v \in \V$, hence there exists a natural
  action of $\g$ on this module, namely $\rho(z)x := (z y_1 \dots y_k)
  \otimes v\,$ for $z \in \g$, induced by the associative product in
  the universal enveloping algebra;
\item Finally, the subscript $\U(\h)$ on the tensor product symbol
  simply means that $\forall x \in \U(\h)\, ,\quad \, \rho(x) (\id)
  \otimes v = (x) \otimes v = (\id) \otimes (\tilde \rho(x)v)$ where
  $\tilde \rho$ is the representation of $\U(\h)$ on $\V$ (arising
  from the one of $\h$ on $\V$).
\end{itemize}
Here we are interested in $\h = \p = \so(1,1) \inplus \iso(d)$ and
$\mathfrak{g}=\so(1,d+1)$. We will consider generalised Verma modules
based on this algebra: $\cV_{\blambda} := \U(\g) \otimes_{\U(\p)}
\V_{\blambda}$, where as previously $\blambda = [\Delta_c\,; \vec s]$ is an
$\so(1,1) \oplus \so(d)$ highest-weight and $\V_{\blambda}$ the
corresponding $\so(1,1) \oplus \so(d)$ highest-weight module. Using
the Poincar\'e-Birkhoff-Witt theorem, the generalised Verma module
$\cV_{\blambda}$ can be equivalently defined as: $\cV_{\blambda} = \U(\t)
\otimes \V_{\blambda}$, as $\t = \spa{P_i}$ is the complement of $\p$ in
$\so(1,d+1)$. In other words, one can think of a generalised Verma
module as the module obtained by acting with all the lowering
operators of the algebra that do not belong to the chosen parabolic
subalgebra (in our case, the translation generators) on a
finite-dimensional highest-weight \textit{space} $\V_{\blambda}$ of the
parabolic subalgebra $\p$ instead of a highest-weight \textit{vector},
as would be the case in the (more standard) context of Verma
modules.

The character of a generalised Verma module $\cV_{[\Delta_c\,; \vec
    s\,]}$ reads
\begin{equation}
  \chi^{\dS}_{[\Delta_c; \vec s\,]}(q, \vec x) = q^{\Delta_c}
  \chi^{\so(d)}_{\vec s}(\vec x)\, \Pd d (q, \vec x)\, ,
    \label{char_GVM}
\end{equation}
where the function $\Pd d (q, \vec x)$ is the character of the
elementary representation of trivial weight $[0\,; \vec 0\,]$ (i.e. an
$\so(2,d)$ scalar function) and is given by
\begin{equation}
  \Pd d (q, \vec x) = \prod_{i=1}^r \frac{1}{(1-qx_i)(1-qx_i^{-1})}
  \times \left\{
  \begin{aligned}
    1 \hspace{15pt} & \text{ if } d = 2r \\ \frac{1}{1-q} & \text{ if
    } d = 2r+1
    \end{aligned}
  \right.
  \label{vacuum_char}
\end{equation} 

\subsubsection{Principal series}
The representations of the principal series are induced from irreps of
$\p$ with complex $\so(1,1)$ weight $\Delta_c = \tfrac{d}2 + i\rho\,$,
where $\rho \in \R$ and arbitrary $\so(d)$ highest-weight
(i.e. arbitrary spin). The corresponding generalised Verma modules are
irreducible as $\so(1,d+1)$-modules.  The following character was
derived originally in \cite{Hirai1965}, where the author computed it
working at the $\SO(1,d+1)$ group level:
\begin{equation}
  \chi^{\dS}_{[\Delta_c; \vec s\,]}(q, \vec x) = \left(q^{\tfrac{d}2+i\rho}
  \chi^{\so(d)}_{\vec s_+}(\vec x) + q^{\tfrac{d}2-i\rho}
  \chi^{\so(d)}_{\vec s_-}(\vec x) \right) \, \Pd d (q, \vec x)\, ,
    \label{char_principal}
\end{equation}
where $\vec s_\pm$ denotes ``chiral'' pairs of $\so(d)$ highest-weights
(when the distinction is relevant), i.e. $$\vec s_\pm = (s_1, \dots,
s_{r-1}, \pm s_r\,)\quad\mbox{for}\,\, d=2r,$$ and $$\vec s_+=\vec s_-
= \vec s \quad\mbox{for}\,\, d=2r+1\,.$$ Notice that, as a consequence
of working at the group level, both chiralities (accompanied with a
conjugation of the $\so(1,1)$ weight) appear in the above
expression. The principal series of representations is known to
describe massive fields in dS$_{d+1}$ (see for instance
\cite{Joung:2006gj, Mickelsson:1972fh}) which complies with the fact
that their definition does not involve any quotient of elementary
representations and therefore do not exhibit any gauge invariance.

\subsubsection{Complementary series}
The elementary representations of the complementary series are also
irreducible from the start, and as such their $\SO(1,d+1)$ characters
\cite{Hirai1965} read:
\begin{equation}
  \chi^{\dS}_{[\Delta_c; \vec s\,]}(q, \vec x) = \left(q^{\tfrac{d}2+c}
  \chi^{\so(d)}_{\vec s_+}(\vec x) + q^{\tfrac{d}2-c}
  \chi^{\so(d)}_{\vec s_-}(\vec x) \right) \, \Pd d (q, \vec x)\, .
    \label{char_complementary}
\end{equation}
For the same reason as in the principal series, these UIRs should
correspond to massive fields. One may phrase the difference between
those two series of massive fields as follows: those in the principal
series describe ``very massive'' fields whereas those in the
complementary series correspond to ``not-so-massive'' fields. Let us
expand a little bit: when writing down a wave equation for a field in
(A)dS, one would refer to the eigenvalue $m^2$ of the Laplace-Beltrami
operator as the mass squared of this field, for lack of a
group-theoretical invariant concept as in Minkowski space where it is
exactly the quadratic Casimir operator of the Poincar\'e
group. However, this mass term is also related to the value of the
quadratic Casimir operator for $\so(1,d+1)$ or $\so(2,d)$ and thereby
it can be expressed in term of the conformal weight
$\Delta_c$\,. Having at hand the relation between $\Delta_c$, $m$ and
the spin of this field (encoded in the $\so(d)$ part of the Casimir
operator), principal series fields have a higher corresponding mass
squared $m^2$. This distinction is illustrated in the simple example
of a massive scalar field, detailed in Appendix \ref{app:example} and
sketched in Figure \ref{mass_scalar}.

\hspace{10pt}
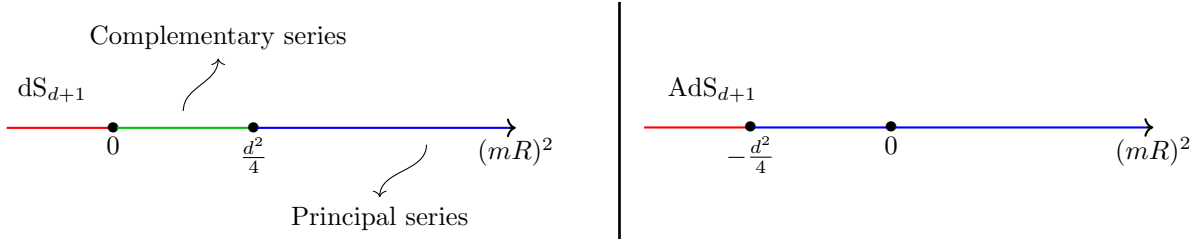
\begin{figure}[!ht]
  \center
  \begin{minipage}[c]{.45\textwidth}
    \begin{tikzpicture}
      \draw (-0.8,0.5) node {dS$_{d+1}$};
      \draw[thick, color=red] (-1.4, 0) -- (0,0);
      \draw (0,0) node {$\bullet$}; \draw (0,0) node[below]{$0$};
      \draw[thick, color=green!75!black] (0.05,0) -- (1.8,0);
      \draw (1.85,0) node {$\bullet$}; 
      \draw (1.85,0) node[below] {$\tfrac{d^2}4$};
      \draw[thick, color=blue] (1.9,0) -- (5.2,0);
      \draw[thick, ->] (5.2,0) -- (5.3,0) node[below] {$(mR)^2$};

      \node[anchor=east] at (4.8,-1.2) (text) {Principal series};
      \node[anchor=west] at (4,-0.1) (description) {};
      \draw (description) edge[out=270,in=90,->] (text);

      \node[anchor=east] at (3.2,1.2) (text) {Complementary series};
      \node[anchor=west] at (0.8,0.1) (description) {};
      \draw (description) edge[out=90,in=-90,->] (text);
    \end{tikzpicture}
  \end{minipage}
  \hspace{5pt}
  \vrule width 1pt height 50pt
  \hspace{5pt}
  \begin{minipage}[c]{.45\textwidth}
    \begin{tikzpicture}
      \draw (-0.5,0.5) node {AdS$_{d+1}$};
      \draw[thick, color=red] (-1.4, 0) -- (0,0);
      \draw (0,0) node {$\bullet$}; \draw (0,0) node[below]{$-\tfrac{d^2}4$};
      \draw[thick, color=blue] (0.05,0) -- (1.8,0);
      \draw (1.85,0) node {$\bullet$}; 
      \draw (1.85,0) node[below] {$0$};
      \draw[thick, color=blue] (1.9,0) -- (5.2,0);
      \draw[thick, ->] (5.2,0) -- (5.3,0) node[below] {$(mR)^2$};
    \end{tikzpicture}
    \vspace{2pt}
  \end{minipage}
  \caption{Unitary (blue and green), and non-unitary (red) regions for
    the squared mass of a scalar field in de Sitter (left) and anti-de
    Sitter space (right).}\label{mass_scalar}
\end{figure}
\vspace{-10pt}

\subsubsection{Exceptional series}
The representations of the exceptional series are those irreps induced
by UIRs of $\p$ with conformal weight at the unitary bound of the
complementary series, i.e.  $\Delta_c = d-p$ or $\Delta_c=p$, with $p$
the height of the Young diagram $\Y$ labeling the $\so(d)$ part of the
irrep. As a consequence, the corresponding generalised Verma module
contains null vectors, i.e. these elementary representations are
reducible. One therefore has to find all submodules contained in the
generalised Verma module constructed from the $\p$-irrep $[d-p; \vec
  s\,]$. This survey was done in \cite{Gavrilik:1975ae, Klimyk:1976ac}
at the group level. At the algebra level, one can rely on the
Bernstein-Gel'fand-Gel'fand (BGG) resolutions to perform the
same analysis. The idea is the following: given a generalised Verma
module $\cV_{\blambda}$, where ${\blambda}=[\Delta_c; \vec s\,]$ represents the
highest-weight characterising the irrep of $\p$ from which it is
built, the BGG theorem gives a criterion for an element of the Weyl
group of $\so(1,d+1)$ to yield a highest-weight defining a submodule,
when applied to $\blambda$. It furthermore provides a resolution of the
irreducible module $\D_{\blambda}$ in the form of an exact sequence
involving $\cV_{\blambda}$ and its submodules. This analysis and the BGG
resolution is known in the case of the complex algebra $\so_\C(d+2)$
and was used in \cite{Shaynkman:2004vu} to classify the possible
systems of unfolded equations invariant under the conformal algebra
$\so(2,d)$. Using these resolutions, one can derive the character of
an irreducible representation in the exceptional series in terms of
characters of $\p$, as detailed in \hyperref[app:char]{Appendix
  \ref{app:char}} (see also the appendix F of \cite{Beccaria:2014jxa}
for an earlier derivation of such a dictionary for characters).

In order to be able to write the characters in a more compact way, we
will use the following notation:
\begin{itemize}
\item $\Y_p$ will represent a Young diagram of height $p$ (with $p
  \leqslant r$), i.e.
  \begin{equation}
    \Y_p := (s_1, \dots, \pos{s_p}{p\th}, 0, \dots, 0) = \vec s\, ,
  \end{equation}
  with $s_p>0$\,;
\item $(\Y_p,\1^m)$ will represent a Young diagram of height $p+m$
  obtained by adding $m$ rows of length one below $\Y_p$, i.e.
  \begin{equation}
    (\Y_p,\1^m) := (s_1, \dots, \pos{s_p}{p\th},\pos{1,\dots,1,0,
      \dots}{(p+m)\th} , 0) \, ;
  \end{equation}
\item $\Yt i$ will represent the diagram obtained from $\Y_p$ after
  having (i) removed its $i$th row and (ii) removed one box in each of
  the row below the previously removed one, i.e.
  \begin{equation}\label{Yti}
    \Yt i := (s_1, \dots, s_{i-1}, \pos{s_{i+1}-1}{i\th}, \dots,
    \pos{s_p-1}{(p-1)\th}, 0, \dots, 0)\, .
  \end{equation}
\end{itemize}

Depending on the parity of $d$, the structure of the irreducible
module obtained from $\cV_{[d-p; \Y_p]}$ is slightly different, which
is why we need to treat both cases separately.

\begin{itemize}
\item Even spacetime dimension ($d=2r+1$):
\begin{eqnarray}
  \chi^\dS_{[d-p; \Y_p]}(q, \vec x) & = & \sum_{m=0}^{r-p} (-1)^{m}
  (q^{p+m} - q^{d-p-m})\, \chi_{(\Y_p,\1^m)}^{\so(d)}(\vec x) \, \Pd d
  (q,\vec x) \label{char_exc_odd} \\ && \hspace{120pt} -
  \sum_{\ell=1}^p (-1)^{p+1+\ell} q^{s_{\ell}+d-\ell}\,
  \chi_{\Yt{\ell}}^{\so(d)}(\vec x)\, \Pd d (q,\vec x) \nonumber \, ;
\end{eqnarray}
\item Odd spacetime dimension ($d=2r$):
  \begin{eqnarray}
    \chi^\dS_{[d-p; \Y_p]}(q, \vec x) & = & 2 \sum_{\ell=1}^p
    (-)^{p+\ell+1} q^{s_{\ell}+d-\ell} \chi^{\so(d)}_{\Yt{\ell}}(\vec
    x)\, \Pd d (q, \vec x) \nonumber \\ && \qquad \qquad +
    \sum_{n=0}^{r-p-1} (-)^{n} (q^{d-p-n}+q^{p+n}) \chi^{\so(d)}_{(\Y_p,
      \1^{n})}(\vec x)\, \Pd d (q, \vec x) \label{char_exc_even} \\ &&
    \qquad \qquad \qquad \qquad \nonumber + (-)^{r-p}\, q^{d/2} \Big(
    \chi^{\so(d)}_{(\Y_p, \1^{r-p}_+)}(\vec x) + \chi^{\so(d)}_{(\Y_p,
      \1^{r-p}_-)}(\vec x) \Big)\, \Pd d (q, \vec x)\, .
  \end{eqnarray}
\end{itemize}

\begin{remark}In even spacetime dimensions, the character
  \eqref{char_exc_odd} exactly reproduces the formula for the
  $\SO(1,d+1)$ character derived in \cite{Hirai1965}, upon rewriting
  it in way that makes the $\so(d)$ part of the character
  explicit. However, in odd spacetime dimensions,
  \eqref{char_exc_even} differs from the formula given in
  \cite{Hirai1965}, namely the first line of our formula is not
  recovered from the expression of \cite{Hirai1965}. Nevertheless, we
  want to stress that we have derived the expression
  \eqref{char_exc_even} as well as all the characters of the Lie \it
  algebra \rm $\so(1,d+1)$ presented in this paper using the BGG
  resolutions recalled in \hyperref[app:char]{Appendix
    \ref{app:char}}.
\end{remark}

Knowing the structure of the corresponding modules, we can now propose
a field theoretical interpretation. First of all, the presence of
submodules in the generalised Verma module $\cV_{[\Delta_c\,;\vec
    s\,]}$ suggests the presence of gauge invariance for the
corresponding fields, i.e. the exceptional series UIRs should
correspond to massless fields in dS$_{d+1}$. However, the simplest
massless fields that one could think of, which are the totally
symmetric, spin-$s$ gauge fields, seem to be either absent of this
series of irreps or do not have the expected conformal weight: being
labelled by the single row Young diagram $\Y=(s, 0, \dots, 0)$, the
associated conformal weight in this series would be $\Delta_c = d-1$
and not the usual $s+d-2$. Our interpretation of this apparent
contradiction is that the conformal weight and Young diagram
characterising a UIR in the exceptional series actually corresponds to
that of the \it curvature \rm (see our definition below) of the
massless field that it describes.

In order to discuss gauge field and curvatures for arbitrary Young
diagrams, we will use the following notation:
\begin{itemize}
\item A Young diagram will be generically seen as composed of $B$
  blocks, each of the them being of individual length $\ell_I$ and
  height $h_I$ ($1 \leqslant I \leqslant B$);
\item We will write the cumulated height of the first $I$ blocks $p_I
  := \sum_{J=1}^I h_J$ {(thus $p_1=h_1$)}, and hence the total height
  of the Young diagram is $p_B$, that we will write $p$ hereafter;
\item Therefore, the Young diagram will be written as
  \begin{equation}
    \vec s = (\underbrace{\ell_1, \dots, \ell_1}_{h_1},
    \underbrace{\ell_2, \dots, \ell_2}_{h_2}, \dots,
    \underbrace{\ell_B, \dots, \ell_B}_{h_B}, 0, \dots, 0) \equiv
    (\ell_1^{h_1}, \ell_2^{h_2}, \dots, \ell_B^{h_B})\, ,
  \end{equation}
  In the case of $B=3$ blocks:
  
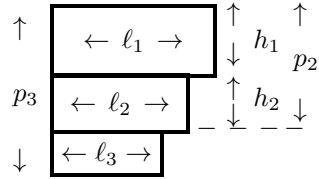
\begin{center}
    \begin{figure}[!ht]
      \begin{tikzpicture}
        \put(200,-20){\framebox(60,25){$\leftarrow\, \ell_1\, \rightarrow$}}
        \put(200,-41){\framebox(50,20){$\leftarrow\, \ell_2\, \rightarrow$}}
        \put(200,-57.5){\framebox(40,15){$\leftarrow \ell_3 \rightarrow$}}
        \put(185,-5){$\uparrow$}
        \put(185,-30){$p_3$}
        \put(185,-55){$\downarrow$}
        \put(265,0){$\uparrow$}
        \put(275,-10){$h_1$}
        \put(265,-15){$\downarrow$}
        \put(265,-28){$\uparrow$}
        \put(275,-32){$h_2$}
        \put(265,-38){$\downarrow$}
        \put(290,0){$\uparrow$}
        \put(290,-36){$\downarrow$}
        \put(290,-17){$p_2$}
        \put(252,-43){$\,-\,-\,-\,-$}
      \end{tikzpicture}
    \end{figure}
  \end{center}
  \vspace{30pt}
\end{itemize}

Recall that a massless gauge field $\varphi_\Y$ of mixed
symmetry\,\footnote{Recall that such a field corresponds to a Lorentz
  tensor whose spacetime indices have the symmetry properties of the
  $\so(1,d)$ Young diagram $\Y$ (i.e. it is completely traceless), and
  subject to a divergencelessness condition which ensures that the
  field only propagates the degrees of freedom corresponding to the
  little group representation, i.e. the $\so(d)$ Young diagram $\Y$.}
described by the Young diagram $\Y$ is subject to gauge
transformations of the form:
\begin{equation}
  \delta_\epsilon^{(I)} \varphi_\Y = \nabla_{(I)} \epsilon_{\Y'_I} +
  \text{traces},
\end{equation}
where $\Y'_I$ is the Young diagram obtained from $\Y$ by removing one
box in the last row of its $I$th block, and $\nabla_{(I)}$ means that
the derivative acting on $\epsilon$ is projected, in the sense that
the resulting object has the symmetry of $\Y$. In this paper, what we
call the \emph{curvature} corresponds to the ``primary Weyl
  tensor'' defined in \cite{Boulanger:2008up, Boulanger:2008kw}, and is
  obtained by acting with $\ell_I-\ell_{I+1}$ derivatives on the
  massless field $\varphi_\Y$ and by projecting the resulting object
  on the symmetries of the $\so(1,d)$ Young diagram\,\footnote{Let
      us stress that, here as in most of the paper, we compare
      irreducible Lorentz tensor with Young diagrams of $\so(d)$ (of
      the same shape) labelling the modules of $\so(1,d+1)$, having in
      mind that the former are usually used to represent the latter.}
  obtained by adding $\ell_I-\ell_{I+1}$ boxes to $\Y$ in the
  $(p_I+1)$th row. In other words, the row below the activated $I$th
  block is completed with derivatives until its length reaches
  $\ell_I$. The terminology is justified by the fact that the primary
  Weyl tensor is the gauge-invariant quantity of lowest order in
  derivatives (and that does not vanish on the solutions of the field
  equations).

To take into account partial masslessness\,\footnote{Recall that
    partially massless fields inherit their name from the fact that
    they propagate an intermediate number of degrees of freedom
    between those of a \it bona fide \rm massless field and a massive
    one (the canonical example being that of a spin-$s$ partially
    massless field of \it depth \rm $t$ in 4 dimensions, which
    propagates $2\,t$ helicities, namely $\pm s, \pm (s-1), \dots, \pm
    (s-t+1)$, with $t \in \{ 1,2,\cdots, s \}$).}, initially
  introduced in \cite{Higuchi:1986wu, Deser:1983mm, Deser:1983tm,
    Deser:2001us} for totally symmetric gauge fields and generalised
  to mixed-symmetry fields in \cite{Boulanger:2008up,
    Boulanger:2008kw} (see also \cite{Joung:2015jza, Gwak:2016sma,
    Gunaydin:2016amv, Brust:2016zns, Brust:2016xif} for recent works),
  the previous discussion should be refined a little bit. Indeed,
  partially massless fields are subject to higher derivative gauge
  transformations of the form
\begin{equation}
  \delta^{(I)}_\epsilon \varphi_\Y = \underbrace{\nabla \dots
    \nabla}_{t} \epsilon_{\Y_I^{'(t)}} + \text{traces}\, ,
\end{equation}  
where $\Y_I^{'(t)}$ is the Young diagram obtained from $\Y$ after
having removed $t$ boxes in the last row of the $I$th block, and
where, as in the previous case, all derivatives should be projected so
as to reconstruct an object with the symmetry of $\Y$. As a
consequence, the corresponding curvature necessitates less derivatives
than in the massless case for the same $\varphi_\Y\,$. 
More precisely, the curvature is built by
acting upon the gauge field with $\ell_I-\ell_{I+1}-t+1$ derivatives and
projecting it on the symmetry of the Young diagram obtained from $\Y$
by adding $\ell_I-\ell_{I+1}-t+1$ boxes at the end of the $(p_I+1)$th line
(i.e. the row under the activated, $I$th, block). It follows from the
above definitions that the depth of the partially massless field, that
is, the number of derivatives involved in its gauge transformation, is
now bounded by $\ell_I - \ell_{I+1}$, i.e. the difference between the length
of the $I$th block and that of the block below (if any). Notice that
the depth of a partially massless field can be read off either from
its conformal weight, which is $\Delta_c = \ell_I + d -p_I - t$, or from
the difference between the length of the $I$th block and the next one,
in its curvature Young diagram $\Y_{p_I+1}$. \\

The above described objects are illustrated in the figure
\ref{FigYdiagcurvpot}.
\begin{center}
  \begin{figure}[!ht]
    \begin{tikzpicture}
      \put(95,5){$\footnotesize \mathsf{Potential}$}
      \put(70,-45){\framebox(95,40){$\Y_u$}}
      \put(70,-86){\framebox(85,40)}
      \put(70,-117){\framebox(30,30){$\Y_d$}}

      \put(10,-65){$\footnotesize \mathit{activated,}\, \, \rightarrow$}
      \put(10,-75){$\footnotesize \mathit{ I{\rm th}\,\, block}$}
      
      \put(250,5){$\footnotesize \mathsf{Gauge}$}
      \put(220,-45){\framebox(95,40){$\Y_u$}}
      \put(220,-86){\framebox(85,40)}
      \put(281,-86){\framebox(24,6){\tiny$\times\,\dots\,\times$}}
      \put(280,-93){\scriptsize $\leftarrow \, t \, \rightarrow$}
      \put(281,-86){\framebox(5,6)}
      \put(300,-86){\framebox(5,6)}
      \put(220,-117){\framebox(30,30){$\Y_d$}}
      
      \put(395,5){$\footnotesize \mathsf{Curvature}$}
      \put(370,-45){\framebox(95,40){$\Y_u$}}
      \put(370,-86){\framebox(85,40)}
      \put(429,-84){\scriptsize $\boldsymbol{\rvert}\hspace{-3pt}\leftarrow t \rightarrow$}
      \put(401,-93){\framebox(35,6){\tiny $\nabla\,\,\,\,\,\dots\,\,\,\,\,\nabla$}}
      \put(431,-93){\framebox(5,6)}
      \put(401,-93){\framebox(5,6)}
      \put(402,-95){$\underbrace{\hspace{35pt}}_{\ell_I-\ell_{I+1}-t+1\hspace{-15pt}}$}
      \put(370,-117){\framebox(30,30){$\Y_d$}}
    \end{tikzpicture}
    \vspace{120pt}
    \caption{Young diagrams corresponding, from left to right, to a
      mixed-symmetry partially massless field of depth-$t$, its gauge parameter acting in the
      isolated middle block (with crosses indicating removed
      cells), and finally its curvature built by acting with derivatives
      in the same block. The symbols $\Y_u$ and $\Y_d$ represent arbitrary Young
      diagrams that can be glued respectively above (up) and below
      (down) the middle block.}\label{FigYdiagcurvpot}
  \end{figure}
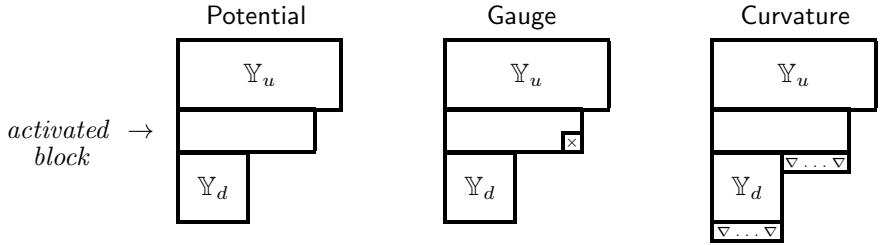
\end{center}

\vspace{-15pt} \noindent
Now with this picture in mind, the Young diagrams appearing in
\eqref{char_exc_odd} and \eqref{char_exc_even} can be interpreted as
follows:
\begin{itemize}
\item $\Y_p$ corresponds to the curvature of the (unitary) gauge field, the
  latter having the $\so(d)$ symmetry $\Y=\Yt{p}$, i.e., described by the
  Young diagram obtained after removing the last row of $\Y_p$.
\item The last block is activated for this gauge field, i.e., it is
  subject to gauge transformations generated by a gauge parameter with
  the symmetry of the Young diagram obtained after removing $t$ boxes,
  for a depth $t$ partially massless field (keeping in mind that $t=1$
  corresponds to the massless case), from the last row of $\Yt{p}$. It
  happens to be exactly the shape of the next diagram appearing in
  \eqref{char_exc_odd} and \eqref{char_exc_even}, namely $\Yt{p-1}$,
  which therefore corresponds to the gauge parameter of our gauge
  field.
\item Using the same rationale, one can convince oneself that the
  remaining Young diagram of the type $\Yt{\ell}\, , \, \ell=1, \dots,
  p-2$ describe the higher order reducibilities of the gauge parameter
  with shape $\Yt{p-1}$.
\item The last class of diagrams appearing in
  \eqref{char_exc_odd}-\eqref{char_exc_even} are of the form $(\Y_p,
  \1^m)\, ,\, m=1, \dots, r-p$, and describe a chain of Bianchi
  identities: they are obtained from $\Y_p$, the curvature of the
  gauge field, by adding a box under the last row repeatedly. The
  vanishing of such tensors would be obtained by acting repeatedly
  with $\nabla$ on the curvature.
\end{itemize}

Notice that, according to this dictionary, only fields where the last
block is ``activated'' are described by UIRs from the exceptional
series. This fact complies with the expectation that, in opposition
with the anti-de Sitter case where only mixed-symmetry fields whose
first block is activated are unitary, in de Sitter unitary
mixed-symmetry fields are those whose \it last \rm block is touched by
gauge transformations. This observation is also supported by the fact
that, if one forget about unitarity, then irreps in the exceptional
series seem to describe mixed-symmetry fields whose activated block is
not necessarily the last one (see \hyperref[app:nonunitary]{Appendix
  \ref{app:nonunitary}}). Finally, it appears from the previous
discussion that totally symmetric, partially massless, fields are
unitary in de Sitter spacetime, in any dimension, as was expected
\cite{Deser:2003gw}. Notice that the depth-$t$ partially massless
fields are those whose curvature Young diagram $\Y_p$ last row is
shorter than the preceding one, i.e., $s_p < s_{p-1}$.

\subsubsection{Discrete series}
Finally, in even spacetime dimensions (i.e., when $d=2r+1$), UIRs in
the discrete series arise from reducible generalised Verma module
induced by an irrep of $\p$ of highest-weight $[k+\frac{d}2;\vec s\,]$
where $k$ is an half-integer that we will rewrite as
$k=k'-\tfrac{1}2$, with $k'$ a positive integer for bosonic
fields\,\footnote{For fermionic fields, $k'$ should be a half-integer,
  as it is eventually related to the conformal weight of a partially
  massless field which depends on its spin.}, setting a lower bound on
the last component of $\vec s\,$: $0 < k' \leqslant s_r$, i.e., $\vec
s$ describes a maximal height Young diagram, as they also can be found
in the BGG resolutions detailed in \cite{Shaynkman:2004vu} (see
\hyperref[app:char]{Appendix \ref{app:char}}). Their character
\cite{Hirai1965} reads:
\begin{equation}
  \chi^\dS_{[k+\tfrac{d}2;\vec s\,]}(q, \vec x) = q^{k'+r}
  \chi^{\so(d)}_{\vec s}(\vec x) \Pd d (q, \vec x) + \sum_{i=1}^r
  (-1)^{r+1+i}\, q^{s_{i}+d-i}\, \chi^{\so(d)}_{\check{\Y}_{\vec s,
      k'}^{(i)}}(\vec x) \, \Pd d (q, \vec x)
  \label{char_discrete}
\end{equation}
where $\check{\Y}_{\vec s, k'}^{(i)}$ is the Young diagram obtained
from $\vec s$ after (i) having removed the $i$th row as well as one
box in all rows below the $i$th one and (ii) filling the last row with
$k'-1$ boxes, i.e., $\check{\Y}_{\vec s, k'}^{(i)} = (s_1, \dots,
s_{i-1}, s_{i+1}-1,\dots, s_r-1,k'-1)$. Writing $k'=s_r-t+1\,$ with $1
\leqslant t \leqslant s_r$, we can recognise the conformal weight of a
partially massless mixed-symmetry field with a maximal-height Young
diagram (i.e., $p=r$) and whose last block is activated, in the
exponent of the variable $q$ in the first term: $\Delta_c = k'+r = s_r
+ d - r - t$. Then, looking at the sum backward (i.e., starting with the last term
with $i=r$) we recognise as a second term the conformal weight and
Young diagram associated with the gauge parameter of the
maximal-height partially massless field: $\Delta_c = s_r+d-r$ and
$\check{\Y}^{(r)}_{\vec s, k'} = (s_1, \dots, s_{r-1}, s_r - t)$. As
usual, the removal of $t$ boxes in the last row together with the
increase in the conformal weight by $t$ units represents the gauge
symmetry enjoyed by the depth-$t$ partially massless field. With this
picture in mind, the $r-1$ remaining terms in the above expression are
naturally interpreted as the reducibilities of the gauge parameter.

It may seem surprising that, contrarily to the exceptional series,
irreps of the discrete series correspond to a description of a
massless field only in terms of the potential and its gauge symmetry,
and does not involve its curvature. This can be understood a
posteriori by the fact that those irreps are labelled by a maximal
height $\so(d)$ Young diagram, and therefore the curvature is described
by a Young diagram with $r+1$ rows, which vanishes identically as an
$\so(d)$ representation and is thus absent in \eqref{char_discrete}.

\begin{remark}
  The 4-dimensional case appears to be somewhat degenerate, in the
  sense that it can only accommodate totally symmetric fields (the
  isometry algebra is $\so(1,4)$ and therefore the relevant rotation
  subalgebra is $\so(3)$ which has rank 1), hence all massless fields
  fall in the discrete series (as they are described by maximal height
  Young diagram). As a consequence, their character only contain the
  potential part (and not the curvature part) of the gauge field, and
  therefore are similar to those of massless fields in AdS$_4$.
\end{remark}

\subsection{Masslessness: AdS vs dS}
In curved spacetime for fields with spin one or more, the definition
of mass (and, therefore, of masslessness) is ambiguous.  A standard
modern criterion for ``masslessness''\footnote{This criterion has the
  advantage to incorporate in a natural way the partially massless
  fields.} of fields on de Sitter or anti-de Sitter spacetimes is that
the corresponding irrep is not a generalised Verma module (or
elementary representation) but arises as a quotient of such modules.

In $(d+1)$-dimensional anti-de Sitter spacetime, (positive-energy unitary) massive and
massless fields are organised quite simply with respect to their
conformal weight: given an $\so(d)$ highest-weight $\Y$ of height $p-1$
corresponding to the potential, whose first row is of length
$s:=\ell_1$ and first block of height $h_1$, massive fields are those
irreps with $\Delta>s+d-h_1-1$ and massless fields lie at the boundary
of this spectrum, being characterised by
$\Delta_{s,h_1}:=s+d-h_1-1$. The reason for this repartition is the
following: for a large conformal weight, $\Delta > \Delta_{s,h_1}$, no
negative norm vector are present in the module. Then, lowering
$\Delta$, some null vectors will appear when reaching the critical
value $\Delta_{s,h_1}$, that one should get rid of by modding out the
submodule they define. Finally, negative norm state start appearing
for $\Delta < \Delta_{s,h_1}$ so these irreps are non-unitary.

In $(d+1)$-dimensional de Sitter spacetime, there seems to be a
similar distribution of (unitary) massive and massless fields as a
function of their conformal weight in de Sitter spacetime, with the
important difference that both types of field are further split into
two subcategories at the group theoretical level.  Given the same
$\so(d)$ highest-weight $\Y$ as considered previously, there is a first
continuum of massive fields -- the principal series -- labelled by a
purely complex conformal weight $\Delta_c = \tfrac{d}2+i\rho\,$, ($\rho
\in \R\,$), together with a second marginal continuum of massive fields
-- the complementary series -- with real conformal weight $p <
\Delta_c < d-p$ (taking into account the partial equivalence between
representations with $\Delta_c$ and $d-\Delta_c$). Then at the
boundary of the complementary series, $\Delta_c = d-p$ and $\Delta_c =
p$, (partially) massless fields appear as UIRs from the exceptional
series. Finally, for $d=2r+1$, another class of gauge fields is
possible, belonging to the discrete series of UIRs. They correspond to
(partially) massless fields labelled by Young diagrams of
maximal-height.\\

\noindent This repartition is illustrated in the following figure.

\hspace{10pt}
\begin{figure}[!ht]
  \center
  \begin{minipage}[c]{.45\textwidth}
    \begin{tikzpicture}
      \draw (-1.2,2.8) node {dS$_{d+1}$};
      \draw[thick] (-2,0) -- (0,0) node {$\bullet$};

      \draw[thick] (0,0) node[below] {$0$} -- (1,0)
      node[color=blue]{$\bullet$};
      \draw (1.05,0) node{$]$};
      \draw (1,-0.1) node[below]{\small $p$};
      \draw[thick, color=green!75!black] (1.05,0) -- (2,0);
      \draw (1.8,0) node[below] {$\tfrac{d}2$};
      \draw[thick, ->] (2,0) node{$\bullet$} -- (5.2,0)
      node[below]{$\Re(\Delta_c)$} -- (5.4,0);
      
      \draw[thick, ->] (0,0) -- (0,3) node[right] {$\Im(\Delta_c)$};
      \draw[thick, color=red] (2,0.08) -- (2,2);
      \draw[thick, color=red] (2,-0.08) -- (2,-0.8);
      \draw[dashed] (2,2) -- (2,3);
      \draw[dashed] (2,-0.8) -- (2,-1.4);
      
      \draw[thick, color=green!75!black] (2.08,0) -- (3.45,0)
      node{\color{black} $[$};
        \draw (3.5,-0.1) node[below] {\small $d-p$};
        \draw[color=blue] (3.5,0) node {$\bullet$};
    \end{tikzpicture}
  \end{minipage}
  \hspace{5pt}
  \vrule width 1pt height 60pt
  \hspace{5pt}
  \begin{minipage}[c]{.45\textwidth}
    \begin{tikzpicture}
      \draw (-1.2,2.65) node {AdS$_{d+1}$};
      \draw[thick] (-2,0) -- (2,0) node[color=blue] {$\bullet$};
      \draw (0,0) node {$\bullet$};
      \draw (0,0) node[below] {$0$};
      \draw (2,-0.25) node {\small $\Delta_{s,h_1}$};
      \draw[thick] (2.08,0) -- (4.9,0);
      \draw[thick, color=green!75!black] (2.08,0) -- (4.9,0);
      \draw[thick, ->] (4.9,0) -- (5,0) node[below] {$\Delta$};
    \end{tikzpicture}
    \vspace{15pt}
  \end{minipage}
  \caption{Repartition of massive and massless fields in dS$_{d+1}$
    (left) and AdS$_{d+1}$ (right) as a function of the conformal
    weight $\Delta_c / \Delta$, for a fixed diagram $\Y$ of total
    height $p-1$, and first block of height $h_1$ and length $s$. On
    the left / de Sitter side, massive field in the principal and
    complementary series are depicted respectively by a red and a
    green line, the massless field is represented by a blue dot. On
    the right / anti-de Sitter side, massive fields correspond to the
    green line and the massless fields are the blue dots. Massless fields in the 
    discrete series appear as discrete points on the green line.}
\end{figure}
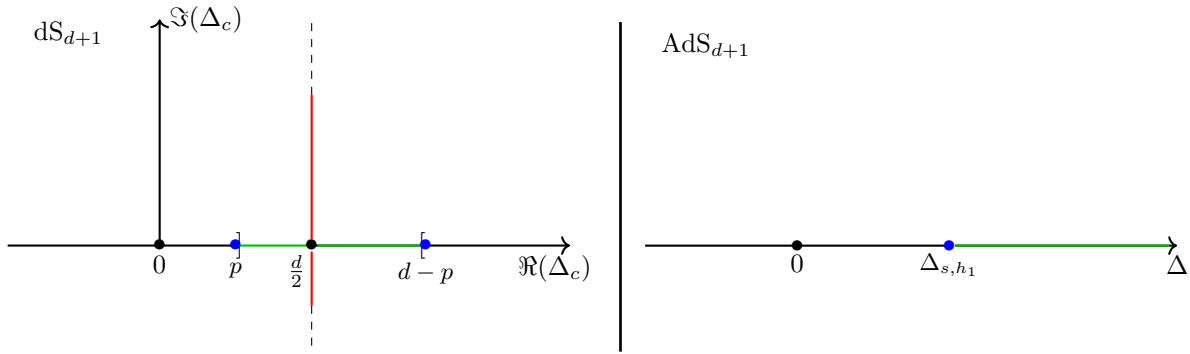
\vspace{-10pt} \\

\begin{remark}
  Notice that, because the conformal weight of a field in the
  exceptional series does not depend on the length of the first row of
  its Young diagram but only on $p\,$, the height of its first column, the
  difference between massless and partially massless fields with the
  same spin is no longer encoded in the conformal weight of the two
  corresponding representations, like in AdS$_{d+1}$, but into the
  Young diagram labeling the irreps. As this diagram corresponds to
  that of the curvature of the field, a massless and a partially
  massless field are labelled by a Young diagram whose last row are of
  different length, hence both are unitary in
  dS$_{d+1}$. In contradistinction, in AdS$_{d+1}$ the conformal
  weight of a partially massless field is lower than that of the
  corresponding massless field and therefore falls below the unitarity
  bound.
\end{remark}

\section{Flat limit}\label{sec:3}

One can recover from the (anti-)de Sitter spacetime (A)dS$_{d+1}$ the
flat Minkowski spacetime $\M_{d+1}$ by just sending its curvature
radius to infinity, $R \rightarrow \infty$, or equivalently by sending
to zero the reduced cosmological constant $\lambda^2 := -\sigma
\,\tfrac{2\Lambda}{d(d-1)}$ where $\sigma=-\Lambda/|\Lambda|$, making
this quantity always positive for both sign of the cosmological
constant $\Lambda$ (since $\sigma=-1$ corresponds to dS$_{d+1}$, and
$\sigma=+1$ to AdS$_{d+1}$). The flat limit $\lambda=1/R \rightarrow
0$ corresponds to a contraction of the (A)dS$_{d+1}$ isometry algebras
to that of Minkowski spacetime, i.e., the Poincar\'e algebra $\iso(1,d)
= \so(1,d) \inplus\R^{d+1}$. Indeed, exhibiting the Lorentz
subalgebra, common to these three isometry algebras, they can be
presented as:
\begin{equation}
  [M_{ab}, M_{cd}] = i\,\eta_{bc} M_{ad} + \dots\, , \quad [M_{ab},
    P_c] = 2\,i\, \eta_{c[b} P_{a]}\, , \quad [P_a, P_b] = i\, \sigma
  \lambda^2 M_{ab}\, .
\end{equation}
It is clear from this presentation that sending the inverse radius
$\lambda$ of (A)dS to zero, the transvection generators $P_a$ become
the usual flat spacetime translation generators, which span the
abelian ideal $\R^{d+1}$ of the Poincar\'e algebra. On the (A)dS side,
one of these generators belongs to the Cartan subalgebra, and the
``energy'' label ($\Delta_c$ or $\Delta$ for respectively dS$_{d+1}$
or AdS$_{d+1}$) is the eigenvalue of this particular
generator. However, on the flat side, there is no longer a Cartan
subalgebra, since the Poincar\'e algebra is not semisimple.

A heuristic way to translate this feature on the characters is to
rescale the variable carrying the weight coming from the corresponding
transvection generator:
\begin{equation}
  q \rightarrow e^{-\iota \beta \lambda}
\end{equation}
with $\iota := \sqrt{\sigma}$ and $\beta\in\mathbb R$ some constant
(that we could sometimes interpret as the inverse temperature in the
case of negative cosmological constant) with dimension of length; and
then to send $\lambda \rightarrow 0\,$:
\begin{equation}
  \chi^{\text{(A)dS}}_{[\Delta_{(c)}; \vec s\,]}(e^{-\iota \beta
    \lambda}, \vec x) \,\underset{\lambda \rightarrow 0}{\leadsto}\,
  \sum_{[m;\vec \sigma] \in \Sp([\Delta_{(c)}; \vec s\,])}
  \chi^{\Poinc}_{[m; \vec \sigma]}(\beta, \vec x)\, ,
  \label{limit_char}
\end{equation}
where $[m\,; \vec \sigma\,]$ denotes a UIR of the Poincar\'e group
labelled by its mass $m$ and a little group (i.e., $\SO(d)$ for massive
irreps, $\SO({d-1})$ for massless helicity ones) highest-weight $\vec
\sigma$ and $\Sp([\Delta_{(c)}, \vec s\,])$ denotes the set of
Poincar\'e irreps resulting from the flat limit (or contraction) of
the (A)dS representation labelled by $[\Delta_{(c)}, \vec
  s\,]$. Before going into more details on the de Sitter case, which
is the main purpose of the present paper, we will start by revisiting
the by-now well understood case of mixed-symmetry fields in anti-de
Sitter spacetime whose flat limit is controlled by the
Brink-Metsaev-Vasiliev (BMV) mechanism.

\subsection{Anti-de Sitter case: the Brink-Metsaev-Vasiliev mechanism}
Brink, Metsaev and Vasiliev conjectured in \cite{Brink:2000ag} that a
single massless mixed-symmetry fields in anti-de Sitter spacetime is
mapped to a set of mixed-symmetry massless fields in flat space;
conjecture later proven in \cite{Boulanger:2008up, Boulanger:2008kw,
  Alkalaev:2009vm}.

Consider a unitary mixed-symmetry gauge field on AdS$_{d+1}$ of
symmetry characterised by the conformal weight $\Delta_{\ell_1, h_1}
:= \ell_1 + d - h_1 - 1$ and the Young diagram $\Y= (\ell_1^{h_1},
\ell_2^{h_2}, \dots, \ell_B^{h_B})$ where $\ell_I^{h_I}$ represents
the $I$th block of length $\ell_I$ and height $h_I$, and $B$ stands
for the number of blocks of the diagram. The total height of $\Y$ is
$p=\sum_{I=1}^B h_I$.  The flat limit of this single massless
mixed-symmetry field on AdS$_{d+1}$ is the following:
\begin{equation}
  \Y = (\ell_1^{h_1}, \ell_2^{h_2}, \dots, \ell_B^{h_B}) \flimit \{
  (\ell_1^{h_1}, \ell_2^{h_2-1}, \ell_2 - n_2, \dots, \ell_B^{h_B-1},
  \ell_B - n_B)\}\,,
  \label{BMV_spectrum}
\end{equation}
where the set of massless fields on $\M_{d+1}$ is determined by the
numbers $n_I$ of boxes removed from the $\ell_I$th column with $n_1=0$
and
\begin{equation}
  0 \leqslant n_I \leqslant \ell_I - \ell_{I+1}\, , \quad \forall I
  \in \{ 2,3,\cdots, B \}\,.
\end{equation}
This limit is essentially\footnote{The important distinction with a
  genuine branching rule of the orthogonal group is that the first
  block is not touched here.} a branching rule of the orthogonal
group: on the AdS$_{d+1}$ side, the spin is given by the
highest-weight of the $\so(d)$ subalgebra of $\so(2,d)$ whereas in flat
spacetime, the spin is given by the highest-weight of the orthogonal
(sub)algebra of the little algebra, that is $\so(d-1)$ for massless
fields in $(d+1)$-dimensional Minkowski space. When performing the
flat limit from AdS$_{d+1}$ to $\M_{d+1}$, one basically trades the
energy/conformal weight for the mass which is obviously zero for
massless fields, meaning that they are entirely characterised by their
spin. As a consequence, one needs to relate the $\so(d)$ part of the
representation of $\so(2,d)$ leftover after having sent $\lambda$ to
zero by branching them onto $\so(d-1)$ in order to have a proper
interpretation in terms of flat massless fields. To understand the
structure of the massless mixed-symmetry representations of
$\so(2,d)\,$ in more details, it is quite convenient to have a look at
their characters (recalled in \hyperref[app:ads]{Appendix
  \ref{app:ads}}):
\begin{equation}
  \chi^{\AdS}_{[\Delta_{s,h_1}; \vec s\,]}(q, \vec x) =
  q^{\Delta_{s,h_1}} \, \left( \chi^{\so(d)}_{\vec s}(\vec x) +
  \sum_{k=1}^{h_1} (-q)^{k} \chi^{\so(d)}_{\vec s_k}(\vec x) \right) \,
  \Pd d (q, \vec x)\, .
\end{equation}
The above formula should be read as follows: the module corresponding
to a massless mixed-symmetry field in AdS$_{d+1}$ described by the
$\so(d)$ highest-weight $\vec s$ is obtained by a succession of
quotients of generalised Verma modules with $\so(d)$ highest-weight
$\vec s_k$ ($k=1, \dots, h_1$) obtained from the Young diagram $\vec
s$ by removing from it the last box on the last $k$  rows in the first
block (of height $h_1$) and increasing the conformal weight by one
unit each time a box is removed. This structure is the
group-theoretical description underlying the gauge symmetry available
for unitary mixed-symmetry fields: they have gauge (for gauge)
parameters with the symmetry of each of the $(h_1-1)$ diagrams in the
chain obtained from removing a box from the previous
diagram. Schematically, this can be depicted as:
\pagebreak
\begin{center}
  \begin{figure}[!h]
    \begin{tikzpicture}
      \put(20,-20){$q^{\Delta_{s,h_1}}$}
      \put(48,-35){\framebox(45,35)} 
      \put(48,-61.5){\framebox(25,25){$\Y_d$}}
      
      \put(100,-20){$-q^{\Delta_{s,h_1}+1}$}
      \put(190,-35){\framebox(5,5){\mbox{\tiny $\times$}}}
      \put(150,-35){\framebox(45,35)} 
      \put(150,-61.5){\framebox(25,25){$\Y_d$}}
      
      \put(200,-20){$+q^{\Delta_{s,h_1}+2}$}
      \put(290,-29){\framebox(5,5){\mbox{\tiny $\times$}}}
      \put(290,-35){\framebox(5,5){\mbox{\tiny $\times$}}}
      \put(250,-35){\framebox(45,35)}
      \put(250,-61.5){\framebox(25,25){$\Y_d$}}
      
      \put(300,-20){$-\dots +(-1)^{h_1} q^{\Delta_{s,h_1}+h_1}$}
      \put(445,-5){\framebox(5,5){\mbox{\tiny $\times$}}}
      \put(445,-28){\framebox(5,21.5){$\vdots$}}
      \put(446,-20.5){$\rvert$}
      \put(445,-35){\framebox(5,5){\mbox{\tiny $\times$}}}
      \put(405,-35){\framebox(45,35)}
      \put(405,-61.5){\framebox(25,25){$\Y_d$}}

      \put(465,-20){$(*)$}
    \end{tikzpicture}
    \label{BMV_idea}
  \end{figure}
\end{center}
\vspace{40pt} 

Now, as mentioned above, when taking the flat limit this becomes an
alternated sum of $\so(d)$ characters. Branching each one of these $h_1$
diagrams will produce a number of $\so(d-1)$ Young diagrams but this
precise sequence is such that only those obtained by deleting boxes in
the last rows (until reaching the length of the row just below) in
each one of the blocks except the first one. Indeed, branching the
first diagram will yield:\\

\vspace{-10pt}
\begin{center}
  \begin{figure}[!ht]
    \begin{tikzpicture}
      \put(125,5){$\leftarrow\,\,\, \ell_1 \,\,\, \rightarrow$}
      \put(125,-35){\framebox(45,35)}
      \put(160,-10){\footnotesize $\uparrow$}
      \put(160,-20){\footnotesize $h_1$}
      \put(160,-30){\footnotesize $\downarrow$}
      \put(132,-30){\footnotesize $\ell_2$}
      \put(125,-34){\footnotesize $\leftarrow \,\,\,\,\, \rightarrow$}
      \put(125,-61.5){\framebox(25,25){$\Y_d$}}

      \put(190,-25){$\branching$}
      
      \put(250,-25){$\bigoplus\limits_{n=\ell_2}^{\ell_1}\,\,
        \bigoplus\limits_{\Y_{d\downarrow}\in\cB(\Y_d)}$}
     
      \put(362,-8){\footnotesize $\uparrow$}
      \put(350,-18){\footnotesize $h_1-1$}
      \put(362,-27){\footnotesize $\downarrow$}

      \put(330,-30){\framebox(45,30)}
      \put(330,-37.5){\framebox(40,6){\mbox{\footnotesize $n$}}}
      \put(330,-64){\framebox(25,25){$\Y_{d\downarrow}$}}
    \end{tikzpicture}
    \vspace{40pt}
  \end{figure}
\end{center}
\noindent
where $\ell_1$ is the length of the first block, $\ell_2$ is the
length of the first row of the second block. The diagram below the
first block is represented by $\Y_d$, while
  $\cB(\Y_d)$ represents all diagrams obtained from branching $\Y_d$
. Branching the second Young diagram in \hyperref[BMV_idea]{$(*)$}
will produce exactly the same sum of diagrams, but with $n$ running
now from $\ell_2$ to $\ell_1-1$ instead of $\ell_1$. As a consequence,
only the diagrams where the first block is left intact and the second
is branched onto $\so(d-1)$ will survive, which is exactly what the
BMV limit tells us. At this stage, one has to notice that the
branching of the second diagram will also produce another sum of
diagrams, similar to the previous one with $n=\ell_2, \dots, \ell_1-1$
but where one extra box is removed in the first block, at the
$(h_1-1)$th row. It turns out that those diagrams will be suppressed
when branching the third diagrams, and this mechanism of cancellation
will repeat itself until the last (the $h_1$th) diagram, so that in
the end one is left only with diagrams produced by the branching rule
of $\so(d)$ onto $\so(d-1)$ except that the first block is intact.

\begin{example}
  Let us consider the example of a mixed-symmetry field in AdS$_{d+1}$
  (in dimension greater than or equals to 8) with $\vec s =
  (s,s,2,1,0,\dots,0)$, and look at its flat limit. From the above
  discussion, it appears that it contracts to the following sequence
  of $\so(d)$ representation:
  \begin{equation}\small
    \D(s+d-3; \vec s) \flimit \gyoung(_5s,_5s,;;,;) -
    \gyoung(_5s,_4{s-1};\times,;;,;) +
    \gyoung(_4{s-1};\times,_4{s-1};\times,;;,;)
  \end{equation}
  where the boxes containing a $\times$ symbol should be considered as
  absent (this notation is intended to remind us of the fact that
  these quotients signify the presence of gauge symmetry). When
  branching the $\so(d)$ Young diagrams onto $\so(d-1)$ ones, one
  produces all Young diagrams obtained by deleting boxes in the last
  row of each block, until reaching the length of the next rows. For
  instance, the first diagram branches as:
  \begin{equation}\small
    \gyoung(_5s,_5s,;;,;) \branching \bigoplus_{n=2}^{s} \Bigg(
    \ \gyoung(_5s,_4n,;;,;) \oplus \gyoung(_5s,_4n,;,;) \oplus
    \gyoung(_5s,_4n,;;) \oplus \gyoung(_5s,_4n,;)\, \Bigg) ,
    \label{first_diagram}
  \end{equation}
  whereas the second and third diagrams yield:
  \begin{eqnarray}\small
    \gyoung(_5s,_4{s-1};\times,;;,;) \branching &
    \bigoplus\limits_{n=2}^{s-1} & \Bigg( \ \gyoung(_5s,_3n,;;,;)
    \oplus \gyoung(_5s,_3n,;,;) \oplus \gyoung(_5s,_3n,;;) \oplus
    \gyoung(_5s,_3n,;) \nonumber \\ && \ \oplus
    \ \gyoung(_4{s-1},_3n,;;,;) \oplus \gyoung(_4{s-1},_3n,;,;) \oplus
    \gyoung(_4{s-1},_3n,;;) \oplus \gyoung(_4{s-1},_3n,;)\, \Bigg) ,
    \label{second_diagram}
  \end{eqnarray}
  and
  \begin{equation}\small
    \gyoung(_4{s-1};\times,_4{s-1};\times,;;,;) \branching
    \bigoplus_{n=2}^{s-1} \Bigg( \ \gyoung(_4{s-1},_3n,;;,;) \oplus
    \gyoung(_4{s-1},_3n,;,;) \oplus \gyoung(_4{s-1},_3n,;;) \oplus
    \gyoung(_4{s-1},_3n,;)\, \Bigg) .
    \label{third_diagram}
  \end{equation}
  The first line of diagrams in \eqref{second_diagram} obtained after
  branching the second diagram in the original sequence cancels 
  those appearing in \eqref{first_diagram} (the branching of the first
  diagram of the sequence) with less than $s$ boxes in the second
  line, i.e., those where the first block is left untouched. The second line
  of diagrams in \eqref{second_diagram} is identical to those
  appearing in \eqref{third_diagram}, the branching of the third
  diagram of the sequence, thereby leaving as expected all Young
  diagrams obtained from branching the original $\so(d)$ Young diagram
  $(s,s,2,1,0,\dots,0)$ onto $\so(d-1)$ leaving the first block
  (composed of the two first rows, in this case) intact.
\end{example}

\subsection{Principal and complementary series}
It was shown in \cite{Mickelsson:1972fh} that the principal series of
representations of the Lorentz group $\SO(1,d+1)$ contracts to the
direct sum of two massive representations of the Poincar\'e group
$\ISO(1,d)$ of left and right chirality (when it exists, i.e., for
$d=2r$), where the mass is given by $\rho\,$. In practice, we consider
the limit process \eqref{limit_char} in Formula \eqref{char_principal}, 
keeping finite the product
$\lambda \rho=m$ (in accordance with \cite{Mickelsson:1972fh})\,:
\begin{eqnarray}
  \chi^{\dS}_{[\tfrac{d}2+i\rho; \vec s\,]}(q, \vec x) & = & q^{d/2} \Pd
  d (q, \vec x) \, \left(q^{i\rho} \chi^{\so(d)}_{\vec s_+}(\vec x) +
  q^{-i\rho} \chi^{\so(d)}_{\vec s_-}(\vec x) \right)\, \\ & \flimit &
  \chi^{\Poinc}_{[m; \vec s\,]}(\beta, \vec x) = \left(e^{-\beta m}
  \chi^{\so(d)}_{\vec s_+}(\vec x) + e^{\beta m} \chi^{\so(d)}_{\vec
    s_-}(\vec x) \right) \Pf d (\vec x)
  \label{char_flat_princ}
\end{eqnarray}
and
\begin{equation}
  \Pf d (\vec x) := \prod_{i=1}^r \frac{1}{(1-x_i)(1-x_i^{-1})}
  \left\{
  \begin{aligned}
    1 \hspace{30pt} & \text{ if } d=2r\,
    ,\\ \left.\frac{1}{1-\alpha}\right|_{\alpha \rightarrow 1} &
    \text{ if } d=2r+1\, ,
  \end{aligned} \right.
  \label{flat_pfunc}
\end{equation}

The resulting expression coincides with the Poincar\'e characters
computed in any dimensions in \cite{Campoleoni:2015qrh}, reviewed in
\hyperref[app:poincare]{Appendix \ref{app:poincare}}. The situation is
similar for the complementary series of representations, 
see Formula \eqref{char_complementary}, 
where
$\Delta_c=\tfrac{d}2 + c$ ($0 < \rvert c \rvert < \tfrac{d}2-p$)
except that the product $\lambda c \flimit 0$ vanishes in the flat
limit, so one should set $m=0$ in \eqref{char_flat_princ} and branch
the $\so(d)$ characters appearing onto $\so(d-1)$ (using the branching
rules for the orthogonal algebra recalled in
\hyperref[app:branching]{Appendix \ref{app:branching}}).

\subsection{Exceptional series}\label{proof_exc_odd}
The flat limit of UIRs in the exceptional series is a bit more subtle,
but at the same time richer. It is to be excepted, if our
identification of this series of irreps with massless fields in de
Sitter spacetime is correct: having the BMV mechanism in mind, one
would anticipate that the spectrum of massless fields in flat space
resulting from the flat limit of a mixed-symmetry field in de Sitter
spacetime to be composed of a plethora of fields falling into irreps
of $\so(d-1)$ related to those appearing in the branching rule of the
$\so(d)$ Young diagram of the original field. In order to see if these
expectations are met, we will perform the flat limit of the characters
slightly differently than before: after having set $q=1$, or
equivalently sent $\lambda \rightarrow 0$, we will branch all $\so(d)$
irreps onto $\so(d-1)$, as it characterises entirely the massless
Poincar\'e irreps of helicity type.

\subsubsection{Even spacetime dimensions}
For $d=2r+1$, the flat limit of \eqref{char_exc_odd} yields
\begin{eqnarray}
  \chi^\dS_{[d-p;\Y_p]}(q,\vec x) \flimit \sum_{\sigma_1 = s_2}^{s_1}
  \sum_{\sigma_2=s_3}^{s_2} \dots \sum_{\sigma_{p-1}=s_{p}}^{s_{p-1}}
  \chi^{\so(2r)}_{(\sigma_1,\dots,\sigma_{p-1})}(\vec x) \Pf d (\vec
  x)\,.
\end{eqnarray}

\begin{proof}
  After having set $q=1$ in \eqref{char_exc_odd}, only the following
  alternating sum of $\so(d)$ characters is left:
  \begin{equation}
    \chi^\dS_{[d-p;\Y_p]}(q,\vec x) \flimit \sum_{\ell=0}^{p-1}
    (-1)^\ell \chi^{\so(2r+1)}_{\Yt{p-\ell}}(\vec x) \Pf{d} (\vec x)
    \label{flat_sequence}
  \end{equation}
  where $\Yt i$ was defined in \eqref{Yti}.  Notice that we
  deliberately used $\Yt{p-\ell}$ instead of $\Yt{\ell}$ in the above
  sum so that the first diagram is the one where the last row was
  deleted, and consequently the last diagram is the one where the
  first row was removed. As mentioned previously, in order to figure
  out the actual field content in flat spacetime, one should branch
  these diagrams onto $\so(2r)$\,, the massless little
  algebra. Because the branching rules for $\so(2r+1)$, given in
  \hyperref[app:branching]{Appendix \ref{app:branching}}, do not
  involve any additional factors on top of the characters of the
  irreps appearing in the branching (contrarily to the $\so(2r)$
  case), we can trade the characters for the corresponding Young
  diagrams without loss of information. We will look at the Young
  diagrams appearing in \eqref{flat_sequence} in three groups: we will
  start by treating the first two diagrams together, then we will look
  at the last two diagrams, and finally an arbitrary triplet of
  consecutive diagrams appearing in the above alternate sum.\\

  Let the last three rows (the $(p-2)$th, $(p-1)$th and $p$th) of
  $\Y_p$ be respectively of length $s$, $t$ and $v$. Starting with the
  first two diagrams in \eqref{flat_sequence}, i.e., $\Yt{p}$ and
  $\Yt{p-1}$ and branching them onto $\so(d-1)$, we obtain on the one
  hand for $\Yt{p}$:\\
  
  \begin{center} \vspace{-10pt}
    \begin{figure}[!ht]
      \begin{tikzpicture}
        \put(70,-30){\framebox(35,25){$\Y'$}}
        \put(70,-37.5){\framebox(30,6){\mbox{ \footnotesize $s$}}}
        \put(70,-45){\framebox(25,6){\mbox{ \footnotesize $t$}}}
        
        \put(115,-30){$\branching$}
        
        \put(180,-30){$\bigoplus\limits_{\sigma=t}^s\,
          \bigoplus\limits_{\Y'_\downarrow \in \cB(\Y')} \Bigg($}
        
        \put(245,-30){$\bigoplus\limits_{\nu=v}^t$}
        
        \put(280,-30){\framebox(35,25){$\Y_\downarrow'$}}
        \put(280,-37.5){\framebox(30,6){\mbox{\scriptsize $\sigma$}}}
        \put(280,-45){\framebox(25,6){\mbox{\scriptsize $\nu$}}}
        
        \put(330,-30){$\oplus$}
        \put(340,-30){$\bigoplus\limits_{\nu=0}^{v-1}$}
        
        \put(370,-30){\framebox(35,25){$\Y_\downarrow'$}}
        \put(370,-37.5){\framebox(30,6){\mbox{\scriptsize $\sigma$}}}
        \put(370,-45){\framebox(25,6){\mbox{\scriptsize $\nu$}}}
        \put(410,-30){$\Bigg)$}

        \put(480,-30){$(\mathsf{A})$}
      \end{tikzpicture}
      \label{fig:a}
    \end{figure}
  \end{center}
  \vspace{20pt}

  where $\Y'$ designates the first $p-3$ rows from the Young diagram
  of total height $p$ that we are considering and $\Y_\downarrow'$ all
  the Young diagrams onto which it branches; and on the other hand
  $\Yt{p-1}$:
  \vspace{10pt}
  \begin{center} \vspace{-20pt}
    \begin{figure}[!ht]
      \begin{tikzpicture}
        \put(70,-30){\framebox(35,25){$\Y'$}}
        \put(70,-37.5){\framebox(30,6){\mbox{ \footnotesize $s$}}}
        \put(70,-45){\framebox(25,6){\mbox{ \footnotesize $v-1$}}}
        
        \put(115,-30){$\branching$}
        
        \put(180,-30){$\bigoplus\limits_{\nu=0}^{v-1}\,
          \bigoplus\limits_{\Y'_\downarrow \in \cB(\Y')} \Bigg($}
        
        \put(245,-30){$\bigoplus\limits_{\sigma=v-1}^{t-1}$}
        
        \put(280,-30){\framebox(35,25){$\Y_\downarrow'$}}
        \put(280,-37.5){\framebox(30,6){\mbox{\scriptsize $\sigma$}}}
        \put(280,-45){\framebox(25,6){\mbox{\scriptsize $\nu$}}}
        
        \put(325,-30){$\oplus$}
        \put(340,-30){$\bigoplus\limits_{\sigma=t}^{s}$}
        
        \put(370,-30){\framebox(35,25){$\Y_\downarrow'$}}
        \put(370,-37.5){\framebox(30,6){\mbox{\scriptsize $\sigma$}}}
        \put(370,-45){\framebox(25,6){\mbox{\scriptsize $\nu$}}}
        \put(410,-30){$\Bigg)$}

        \put(480,-30){$(\mathsf{B})$}
      \end{tikzpicture}
      \label{fig:b}
    \end{figure}
  \end{center}
  \vspace{30pt} 

  The second parts of the above branched diagrams in Figure
  \hyperref[fig:a]{A} and \hyperref[fig:b]{B} are common to both of
  them, and therefore will disappear in the alternating sum
  \eqref{flat_sequence}. The first part of the branching from the
  first diagram describes exactly the field content left after the
  flat limit, and indeed, we will see that the other parts all cancel
  each other.\\

  Next, we can have a look at the last two diagrams in the sum
  \eqref{flat_sequence}, which both have the form:
  \begin{center}
    \begin{figure}[!ht]
      \begin{tikzpicture}
        \put(130,-10){\framebox(45,6){\mbox{ \footnotesize $m$}}}
        \put(130,-36.5){\framebox(30,25){$\Y'$}}
        \put(130,-45){$\leftarrow y \rightarrow$}
        
        \put(190,-20){$\branching$}
        
        \put(255,-20){$\bigoplus\limits_{n=y}^m\,
          \bigoplus\limits_{\Y_\downarrow'\in\cB(\Y')}$}
        
        \put(320,-10){\framebox(35,6){\mbox{ \footnotesize $n$}}}
        \put(320,-36.5){\framebox(30,25){$\Y_\downarrow'$}}
      \end{tikzpicture}
    \end{figure}
  \end{center}
  \vspace{25pt} 

  with respectively $m=s_2-1$ (for the last diagram $\Yt{1}$) and
  $m=s_1$ (for the second to last diagram $\Yt{2}$), and where now
  $\Y'$ is the Young diagram made out of the $p-2$ last rows of
  $\Yt{1}$ (or $\Yt{2}$, as they only differ by their first row), and
  $y=s_3-1$ is the length of the first row of $\Y'$ for these two
  diagrams. Because the first row of the last diagram in
  \eqref{flat_sequence} is shorter than the one of the preceding
  diagram (since $s_2-1<s_1$), all the diagrams resulting from the
  branching of the last diagram $\Yt{1}$ will also be a part of the
  branching of the preceding diagram $\Yt{2}$. Hence, all diagrams
  produced by the branching of the last one in \eqref{flat_sequence}
  are cancelled by the branching of the second to last one.\\

  Finally, let us consider a triplet of diagrams appearing in the sum
  \eqref{flat_sequence}:\\
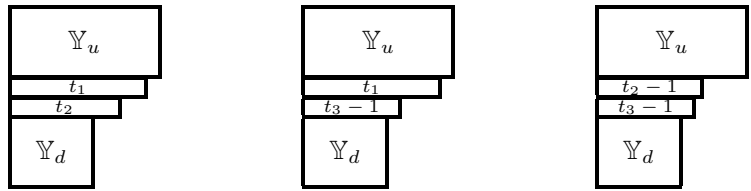
  \begin{center}
    \vspace{-15pt}
    \begin{figure}[!ht]
      \begin{tikzpicture}
        \put(100,-36.5){\framebox(55,25){$\Y_u$}}
        \put(100,-44){\framebox(50,6){\mbox{\scriptsize $t_1$}}}
        \put(100,-51.5){\framebox(40,6){\mbox{\scriptsize $t_2$}}}
        \put(100,-78){\framebox(30,25){$\Y_d$}}
        
        \put(210,-36.5){\framebox(55,25){$\Y_u$}}
        \put(210,-44){\framebox(50,6){\mbox{\scriptsize $t_1$}}}
        \put(210,-51.5){\framebox(35,6){\mbox{\scriptsize $t_3-1$}}}
        \put(210,-78){\framebox(30,25){$\Y_d$}}
        
        \put(320,-36.5){\framebox(55,25){$\Y_u$}}
        \put(320,-44){\framebox(38,6){\mbox{\scriptsize $t_2-1$}}}
        \put(320,-51.5){\framebox(35,6){\mbox{\scriptsize $t_3-1$}}}
        \put(320,-78){\framebox(30,25){$\Y_d$}}
      \end{tikzpicture}
    \end{figure}
  \end{center}
  \vspace{60pt}

  The second diagram in this triplet branches as follows:
  
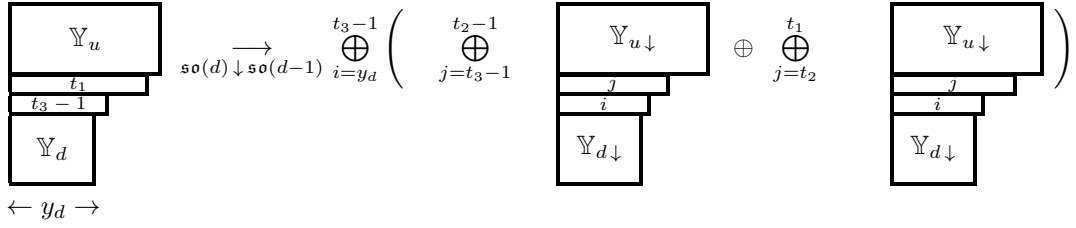
\begin{center}
    \begin{figure}[!ht]
      \begin{tikzpicture}
        
        \put(35,-36.5){\framebox(55,25){$\Y_u$}}
        \put(35,-44){\framebox(50,6){\mbox{\scriptsize $t_1$}}}
        \put(35,-51.5){\framebox(35,6){\mbox{\scriptsize $t_3-1$}}}
        \put(35,-78){\framebox(30,25){$\Y_d$}}
        \put(33,-90){$\leftarrow y_d \rightarrow$}
        
        \put(103,-30){$\branching$}
        
        \put(170,-30){$\bigoplus\limits_{i=y_d}^{t_3-1}\,
          \bigoplus\limits_{\tiny \begin{array}{c}\Y_{u\downarrow} \in \cB(\Y_u) \\
           \Y_{d\downarrow} \in \cB(\Y_d)\end{array}}\Bigg($}
        
        \put(250,-30){$\bigoplus\limits_{j=t_3-1}^{t_2-1}$}
        
        \put(285,-36.5){\framebox(55,25){$\Y_{u\, \downarrow}$}}
        \put(285,-44){\framebox(40,6){\mbox{\scriptsize $j$}}}
        \put(285,-51.5){\framebox(33,6){\mbox{\scriptsize $i$}}}
        \put(285,-78){\framebox(30,25){$\Y_{d\, \downarrow}$}}
        
        \put(355,-30){$\oplus$}
        \put(370,-30){$\bigoplus\limits_{j=t_2}^{t_1}$}
        
        \put(405,-36.5){\framebox(55,25){$\Y_{u\, \downarrow}$}}
        \put(405,-44){\framebox(45,6){\mbox{\scriptsize $j$}}}
        \put(405,-51.5){\framebox(33,6){\mbox{\scriptsize $i$}}}
        \put(405,-78){\framebox(30,25){$\Y_{d\, \downarrow}$}}      
        \put(465,-30){$\Bigg)$}
      \end{tikzpicture}
    \end{figure}
  \end{center}
  \vspace{70pt}

  where $y_d$ is the length of the first row of $\Y_d$. Now one can
  notice that the second part of this sum will be contained in the
  branching of the first diagram of the above triplet, whereas the
  first part of the sum will be contained in the branching of the
  third diagram in the triplet. As a consequence, every diagrams in
  the sequence \eqref{flat_sequence}, obtained by branching onto
  $\so(d-1)$, is cancelled by those coming from the branching of the
  preceding and following diagram, leaving only those announced above
  (coming from branching the first diagram in \eqref{flat_sequence}).
\end{proof}

As explained in \hyperref[app:poincare]{Appendix \ref{app:poincare}},
the characters of massless helicity Poincar\'e UIRs, in even
$d+1=2(r+1)$ dimensions have the form:
\begin{equation}
  \chi^\Poinc_{[0;\Y]}(\vec x) = \chi^{\so(2r)}_{\Y}(\vec x)\,
  \Pf{d}(\vec x)\, ,
\end{equation}
hence the character obtained from the flat limit of the character of
an UIR in the exceptional series of $\so(1,d+1)$ associated with an
$\so(d)$-weight $\Y_p = (s_1,\dots,s_p)$ can be rewritten as:
\begin{equation}
  \chi^\dS_{[d-p;\Y_p]}(q, \vec x) \flimit \sum_{\Y' \in \Sp(\Y_p)}
  \chi^\Poinc_{[0;\, \Y']}(\vec x)\, ,
\end{equation}
with
\begin{equation}
  \Sp(\Y_p) := \Big\{ \Y' = (\sigma_1,\dots,\sigma_{p-1}) \, \rvert \,
  s_{i+1} \leqslant \sigma_i \leqslant s_i, \, i \in \{ 1,2,\cdots, p-1
  \} \Big\}\, ,
\end{equation}
describing the spectrum of massless fields appearing in the flat
limit. Just as the BMV spectrum in AdS$_{d+1}$, this set of fields is
a truncation of the branching rule of the Young diagram of the gauge
potential, i.e., $\Yt{p}$.

\begin{itemize}
\item \textbf{Massless fields:} In the particular case of
  \textit{massless} fields, one block is left untouched as in the BMV
  case, but this time it is the last block instead of the first
  one. Indeed, the Young diagram $\Y_p$ describes the shape of the
  curvature of the gauge field with symmetry $\Yt{p}$, thus the last
  row has the same length as the previous, i.e., $s_p=s_{p-1}$ for
  massless fields. As a consequence, no box can be removed in the last
  row of $\Yt{p}$, and its last block is ``protected''.\\

  In order to emphasise the analogy with the BMV mechanism, the
  spectrum $\Sp(\Y_p)$ of massless fields in flat spacetime can be
  rewritten in a way closer to \eqref{BMV_spectrum} so as to make
  explicit the blocks of the Young diagram of the massless fields: let
  $\Yt{p} = (\ell_1^{h_1}, \dots, \ell_{B}^{h_{B}-1})$, then
  \begin{equation}
    \Sp(\Y_p) = \Big\{ \Y' = (\ell_1^{h_1-1}, n_1, \dots,
    \ell_{B-1}^{h_{B-1}-1}, n_{B-1}, \ell_{B}^{h_{B}-1})\, , \,
    \ell_{I+1} \leqslant n_I \leqslant \ell_I\, , \, I \in \{
    1,2,\cdots, B-1\} \Big\}\, .
    \label{BMV_spectrum_odd}
  \end{equation}
\item \textbf{Partially massless fields:} For depths $t$ strictly
  higher than one, the situation is similar, up to a minor
  modification: additional fields can contribute to the above flat
  spacetime spectrum, namely massless fields with Young diagrams of
  the same shape as those contained in $\Sp(\Y_p)$ and in which up to
  $t-1$ boxes were removed on the last line. More precisely, the
  spectrum of fields is given by the set:
  \begin{eqnarray}
    \Sp(\Y_p;t) & = & \Big\{ \Y' = (\ell_1^{h_1-1}, n_1, \dots,
    \ell_{B-1}^{h_{B-1}-1}, n_{B-1}, \ell_{B}^{h_{B}}-k)\, \,|
    \nonumber \\ && \qquad \qquad \qquad \qquad \quad \ell_{I+1}
    \leqslant n_I \leqslant \ell_I\, , \, I \in \{ 1,2,\cdots, B-1
    \}\, , \, k=0,1,\dots,t-1 \Big\}\, .
  \end{eqnarray}
  In particular, this proves what was conjectured in eq. (3.78) of
  \cite{Boulanger:2008kw}.
\end{itemize}

\subsubsection{Odd spacetime dimensions}
Unfortunately, for even $d$ the situation is not as neat as the
previous one. It seems that taking the flat limit at the character
level in the same fashion as was done for odd dimensions previously
does not produce a natural spectrum of fields in flat space. Indeed,
in odd spacetime dimension, the flat limit of \eqref{char_exc_even}
yields:
\begin{equation}
  \chi^\dS_{[d-p;\Y_p]}(q, \vec x)\, \flimit (-)^{r-p}\,\sum_{k=1}^r
  \cA^{(r)}_k (\vec x)\,\, \big(2-\xi_k(1)\big)
  \sum_{\sigma_1=s_2}^{s_1} \dots \sum_{\sigma_{p-1}=s_p}^{s_{p-1}}
  \sum_{\sigma_p=1}^{s_p}
  \chi^{\so(2r-1)}_{(\sigma_1,\dots,\sigma_p,\1^{r-1-p})}(\hat{\vec
    x}_k) \Pf{2r}(\vec x)
  \label{even_flat}
\end{equation}
where $\xi_k$ and $\hat{\vec x}_k$ are defined in
\hyperref[app:branching]{Appendix \ref{app:branching}}.\\

Let us first show how it is obtained, before discussing its
significance (or present lack thereof).
\begin{proof}
  After setting $q=1$ in \eqref{char_exc_even}, what is left is the
  following sum of $\so(2r)$ characters:
  \begin{eqnarray}
    \chi^\dS_{[d-p;\Y_p]}(q, \vec x) & \flimit & 2\,\sum_{n=0}^{r-p-1}
    (-)^{n} \chi^{\so(2r)}_{(\Y_p, \1^{n})}(\vec x)\, \Pf{2r} (\vec x)
    + (-)^{r-p}\, \Big( \chi^{\so(2r)}_{(\Y_p, \1^{r-p}_+)}(\vec x) +
    \chi^{\so(2r)}_{(\Y_p, \1^{r-p}_-)}(\vec x) \Big)\, \Pf{2r} (\vec
    x) \nonumber \\ && \qquad \qquad -2 \sum_{\ell=0}^{p-1} (-1)^\ell
    \chi^{\so(2r)}_{\Yt{p-\ell}}(\vec x) \Pf{2r} (\vec x)
    \label{sum_flat}
  \end{eqnarray}
  The second line in the above equation will produce the same
  expression as in the odd-dimensional case (though with a
  multiplicity two) once all irreps of $\so(2r)$ are branched onto
  $\so(2r-1)$. Therefore, what we need to look is the sequence of
  $\so(2r)$ Young diagrams appearing in the first line.\\

  Let us start with the first two diagrams and their branching, for
  $\Y_p$:
  \begin{center} \vspace{-15pt}
    \begin{figure}[!ht]
      \begin{tikzpicture}
        \put(125,-36.5){\framebox(30,25){$\Y_u$}}
        \put(125,-43.5){\framebox(25,6){\footnotesize $s$}}

        \put(165,-30){$\branching$}

        \put(225,-30){$\bigoplus\limits_{n=0}^s\,
          \bigoplus\limits_{\Y_{u\downarrow}\in\cB(\Y_u)}$}

        \put(295,-36.5){\framebox(30,25){$\Y_{u\, \downarrow}$}}
        \put(295,-43.5){\framebox(25,6){\footnotesize $n$}}
      \end{tikzpicture}
    \end{figure}
  \end{center}
  \vspace{20pt}

  where $\Y_u$ represents the Young diagram made out of the $p-1$
  first rows of $\Y_p$, and for $(\Y_p,\1)$

  \begin{center}\vspace{-20pt}
    \begin{figure}[!ht]
      \begin{tikzpicture}
        \put(110,-36.5){\framebox(30,25){$\Y_u$}}
        \put(110,-43.5){\framebox(25,6){\footnotesize $s$}}
        \put(110,-51){\framebox(6,6)}

        \put(145,-30){$\branching$}

        \put(205,-30){$\bigoplus\limits_{n=1}^s\,
          \bigoplus\limits_{\Y_{u\downarrow}\in\cB(\Y_u)} \Big($}

        \put(280,-36.5){\framebox(30,25){$\Y_{u\, \downarrow}$}}
        \put(280,-43.5){\framebox(25,6){\footnotesize $n$}}

        \put(320,-30){$\oplus$}

        \put(340,-36.5){\framebox(30,25){$\Y_{u\, \downarrow}$}}
        \put(340,-43.5){\framebox(25,6){\footnotesize $n$}}
        \put(340,-51){\framebox(6,6)}
        \put(380,-30){$\Big)$}
      \end{tikzpicture}
    \end{figure}
  \end{center}
  \vspace{25pt}

  From the branching of the first diagram, only the part with $n=0$ of
  the sum survives in the reduction of \eqref{sum_flat}, that is all
  diagrams obtained from branching $\Y_p$ and removing its last
  row. This makes up all of the diagrams appearing in the flat limit
  for $d=2r+1$. They will therefore cancel exactly those coming from
  branching the second line of \eqref{sum_flat}. Just as in the
  previous odd-dimensional case, the rest of the sequence is such that
  all other diagrams, when branched, produce a set of diagrams that
  will, for the most part, be cancelled. Indeed, in general a diagram
  of the form $(\Y_p, \1^m)$ branches as:
    
  \begin{center} \vspace{-15pt}
    \begin{figure}[!ht]
      \begin{tikzpicture}
        \put(100,-36.5){\framebox(30,25){$\Y_u$}}
        \put(100,-43.5){\framebox(25,6){\footnotesize $s$}}
        \put(100,-60){\framebox(6,15)}
        \put(100,-67){\framebox(6,6)}

        \put(138,-30){$\branching$}

        \put(205,-30){$\bigoplus\limits_{n=1}^s\,
          \bigoplus\limits_{\Y_{u\downarrow}\in\cB(\Y_u)} \Big($}

        \put(280,-36.5){\framebox(30,25){$\Y_{u\, \downarrow}$}}
        \put(280,-43.5){\framebox(25,6){\footnotesize $n$}}
        \put(280,-60){\framebox(6,15)}

        \put(315,-30){$\oplus$}

        \put(335,-36.5){\framebox(30,25){$\Y_{u\, \downarrow}$}}
        \put(335,-43.5){\framebox(25,6){\footnotesize $n$}}
        \put(335,-60){\framebox(6,15)}
        \put(335,-67){\framebox(6,6)}
        \put(375,-30){$\Big)$}
      \end{tikzpicture}
    \end{figure}
  \end{center}
  \vspace{45pt}
  
  The first sum of diagrams in the right hand side is exactly the
  second sum of diagrams that appear when branching the Young diagram
  $(\Y_p, \1^{m-1})$, therefore it will be cancelled. The last thing
  to check is that the last diagram, the one with maximal height $r$
  will not bring diagrams that cannot be cancelled by previous
  terms. As the branching process here is to go from $\so(2r)$ to
  $\so(2r-1)$, a Young diagram with maximal height will branch only
  onto diagrams where the last row was removed (so that it has the
  correct height for an $\so(2r-1)$ Young diagram). Concretely:
  \begin{center} \vspace{-20pt}
    \begin{figure}[!ht]
      \begin{tikzpicture}
        \put(140,-36.5){\framebox(30,25){$\Y_u$}}
        \put(140,-43.5){\framebox(25,6){\footnotesize $s$}}
        \put(140,-60){\framebox(6,15)}
        \put(125, -22){\Large $\uparrow$}
        \put(125, -42){$r$}
        \put(125, -65){\Large $\downarrow$}
        \put(140,-67){\framebox(6,6)}

        \put(115,-45){(0,-0.4) -- (0,-2.3)};
        
        \put(180,-30){$\branchingeven$}

        \put(250,-30){$\bigoplus\limits_{n=1}^s\,
          \bigoplus\limits_{\Y_{u\downarrow}\in\cB(\Y_u)}$}

        \put(315,-36.5){\framebox(30,25){$\Y_{u\, \downarrow}$}}
        \put(315,-43.5){\framebox(25,6){\footnotesize $n$}}
        \put(315,-60){\framebox(6,15)}
      \end{tikzpicture}
    \end{figure}
  \end{center}    
  \vspace{50pt}

  However, taking into account the fact that the branching rule for an
  $\so(2r)$ character of a Young diagram of maximal height involves
  an additional factor compared to the non-maximal Young diagrams, see
  \eqref{even_branching}, the $\so(2r-1)$ diagrams produced by
  branching the $\so(2r)$ diagram $(\Y_p,\1^{r-p})$ are not
  cancelled, rather they come with a factor $2-\xi(1)$.
\end{proof}

In order to illustrate the mechanism explained above, let us detail a
concrete case in the example below.

\begin{example}
  Let us consider a massless, totally symmetric, spin-$s$ field in
  dS$_{d+1}\,$ when $d=2r\,$, for the sake of simplicity. Its
  character reads:
  \begin{eqnarray}
    \chi^\dS_{[d-2; s,s]}(q, \vec x) & = & \sum_{m=0}^{r-3} (-)^m
    (q^{d-2-m} + q^{2+m}) \chi^{\so(2r)}_{(s,s,\1^{m})} \Pd d (q, \vec
    x) - 2\, q^{s+d-2} \left( \chi^{\so(2r)}_{(s)}(\vec x) - q\,
    \chi^{\so(2r)}_{(s-1)}(\vec x) \right) \Pd d (q, \vec x) \nonumber
    \\ & & \qquad \qquad +\, (-)^r\,q^r \left(
    \chi^{\so(2r)}_{(s,s,\1^{r-2}_+)}(\vec x) +
    \chi^{\so(2r)}_{(s,s,\1^{r-2}_-)}(\vec x) \right) \Pd d (q, \vec
    x)
  \end{eqnarray}
  In the flat limit considered so far in this paper, $q \rightarrow
  1$, it becomes:
  \begin{eqnarray}
    \chi^\dS_{[d-2; s,s]}(q, \vec x) & \flimit & \Bigg(
    2\,\sum_{m=0}^{r-3} (-)^m \chi^{\so(2r)}_{(s,s,\1^{m})} - 2\,
    \left[ \chi^{\so(2r)}_{(s)}(\vec x) - \chi^{\so(2r)}_{(s-1)}(\vec
      x) \right] \\ && \qquad \qquad \qquad \qquad \qquad \quad +
    (-)^r \left[ \chi^{\so(2r)}_{(s,s,\1^{r-2}_+)}(\vec x) +
      \chi^{\so(2r)}_{(s,s,\1^{r-2}_-)}(\vec x) \right] \Bigg) \Pf d
    (\vec x) \nonumber
  \end{eqnarray}
  Now using the branching rules derived in
  \hyperref[app:branching]{Appendix \ref{app:branching}}:
  \begin{eqnarray}
    \chi^{\so(2r)}_{(s)}(\vec x) - \chi^{\so(2r)}_{(s-1)}(\vec x) & =
    & \sum_{k=1}^r \cA_k^{(r)}(\vec x) \left( \sum_{\sigma=0}^s
    \chi_{(\sigma)}^{\so(2r-1)}(\hat{\vec x}_k) -
    \sum_{\sigma=0}^{s-1} \chi_{(\sigma)}^{\so(2r-1)}(\hat{\vec x}_k)
    \right) \\ & = & \sum_{k=1}^r \cA_k^{(r)}(\vec x)
    \chi_{(s)}^{\so(2r-1)}(\hat{\vec x}_k)\, ,
  \end{eqnarray}
  as was observed in, for instance, \cite{Campoleoni:2015qrh}. Now
  turning to the curvature and Bianchi identities contributions:
  \begin{equation}
    \chi^{\so(2r)}_{(s,s)}(\vec x) = \sum_{k=1}^r \cA_k^{(r)}(\vec x)
    \sum_{\sigma=0}^{s} \chi^{\so(2r-1)}_{(s,\sigma)}(\hat{\vec x}_k)
  \end{equation}
  \begin{equation}
    \chi^{\so(2r)}_{(s,s,\1^m)}(\vec x) = \sum_{k=1}^r \cA_k^{(r)}(\vec
    x) \sum_{\sigma=1}^{s} \left(
    \chi^{\so(2r-1)}_{(s,\sigma,\1^m)}(\hat{\vec x}_k) +
    \chi^{\so(2r-1)}_{(s,\sigma,\1^{m-1})}(\hat{\vec x}_k) \right) \,
    , \qquad (m=1,\dots,r-3)
  \end{equation}
  \begin{equation}
    \chi^{\so(2r)}_{(s,s,\1^{r-2}_+)}(\vec x) +
    \chi^{\so(2r)}_{(s,s,\1^{r-2}_-)}(\vec x) = \sum_{k=1}^r
    \cA_k^{(r)}(\vec x)\, \xi_k(1)\, \sum_{\sigma=1}^{s}
    \chi^{\so(2r-1)}_{(s,\sigma,\1^{r-3})}(\hat{\vec x}_k)
  \end{equation}
  Making use of these 3 equations, one ends up with:
  \begin{eqnarray}
    2\,\sum_{m=0}^{r-3} (-)^m \chi^{\so(2r)}_{(s,s,\1^{m})} & + &
    (-)^r \left[ \chi^{\so(2r)}_{(s,s,\1^{r-2}_+)}(\vec x) +
      \chi^{\so(2r)}_{(s,s,\1^{r-2}_-)}(\vec x) \right] \\ & = &
    \sum_{k=1}^r \cA_k^{(r)}(\vec x)\, \left( 2\,
    \chi_{(s)}^{\so(2r-1)}(\hat{\vec x}_k) + (-)^r
    \big(2-\xi_k(1)\big) \sum_{\sigma=1}^s
    \chi_{(s,\sigma,\1^{r-3})}^{\so(2r-1)}(\hat{\vec x}_k) \right)
  \end{eqnarray}
  Putting all the pieces together, the flat limit now reads:
  \begin{equation}
    \chi^\dS_{[d-2; s,s]}(q, \vec x) \flimit (-)^r \sum_{k=1}^r
    \cA_k^{(r)}(\vec x)\, \big(2-\xi_k(1)\big) \sum_{\sigma=1}^s
    \chi_{(s,\sigma,\1^{r-3})}^{\so(2r-1)}(\hat{\vec x}_k)\, ,
  \end{equation}
  which is, to say the least, confusing, having nothing to do with the
  quite coherent and expected spectrum produced by the flat limit of
  $\so(1,d+1)$ characters when $d=2r+1$.
\end{example}

As announced at the beginning of this subsection, this flat limit is a
bit puzzling, as it cannot be interpreted naturally as a BMV-type
spectrum in de Sitter. Although the characters of exceptional series
representations have the same structure (that is, as explained in
\hyperref[sec:2]{Section 2}, it contains information about the gauge
fields, its gauge parameter and their reducibility, as well as the
curvature and its Bianchi identities), there are two problems arising
when considering their flat limit as we proposed:
\begin{itemize}
\item[(i)] The two sequences of $\so(d)$ Young diagrams appearing in the
  character and describing on one side the gauge field and its gauge
  parameter, and on the other side the curvature and its Bianchi
  identities, both produce the expected spectrum $\Sp(\Y_p)$ when
  branched onto $\so(d-1)$ but come with a relative minus sign, hence
  they cancel each other.
\item[(ii)] In the ``Bianchi'' sequence, the presence of maximal
  height $\so(2r)$ Young diagram of the form $(\Y_p, \1_\pm^{r-p})$
  whose characters, when branched onto $\so(2r-1)$ involve an
  additional factor $\xi(1)$ with respect to non-maximal Young
  diagrams (see \hyperref[app:branching]{Appendix
    \ref{app:branching}}). As a consequence, we are left with some
  maximal height diagrams of $\so(2r-1)$ which have, to our
  knowledge, no interpretation as massless fields resulting from a
  flat limit of the original gauge field.
\end{itemize}

A possible resolution of these difficulties could be brought by the
following argument: as mentioned in \hyperref[sec:2]{Section 2}, the
classification of UIRs -- classification that we related to a field
theoretic classification of massive and massless fields in de Sitter
-- was obtained at the Lie group $\SO(1,d+1)$ level.  To obtain such a
dictionary, we looked for the corresponding UIRs at the Lie algebra
$\so(1,d+1)$ level, UIRs for which we could write down the
corresponding characters.  In turn, these characters gave us some
insight into the structure of each of these representations. This
being said, it is a well known fact that not all Lie algebra
representations extend to group representations. Therefore one could
speculate that the reason why we did not find the irrep in the known
classification of $\SO(1,d+1)$ UIRs that would correspond to the
potential module only (i.e., the module made out of the potential,
quotiented by its gauge parameter and its higher-order reducibilities)
is precisely because this representation of the Lie algebra does not
extend to a \textit{unitary} representation of the Lie
group.

If this happens to be correct, then in both odd ($d=2r$) and even
($d=2r+1$) spacetime dimension, for a (partially) massless field of
given $\so(d)$ type, we prescribe to consider only the gauge potential
part of the module for which the character in $\dS$ is exactly the
same as the character for $\so(2,d)$ irreps corresponding to the same
$\so(d)$ type.  Therefore, the flat limit of the purely potential
$\so(1,d+1)$ module will produce the sum of $\iso(1,d)$ characters
corresponding to the BMV spectrum $\Sp$ we found in
\eqref{BMV_spectrum_odd} for the unitary case in dS. For the
non-unitary cases in both dS and AdS, see
\hyperref[app:nonunitary]{Appendix \ref{app:nonunitary}}. The only
difference is the protected block for unitary fields (the first one in
AdS$_{d+1}$, the last one in dS$_{d+1}$).
 
Our above interpretation is supported by the fact that the technique
of the proof presented in \cite{Boulanger:2008up, Boulanger:2008kw}
holds for both $\AdS$ and $\dS\,$, irrespectively of the parity of the
dimension. Actually, in \cite{Boulanger:2008up, Boulanger:2008kw} the
whole procedure was presented for both signs of the cosmological
constant. Only the computations of the critical masses were performed
for the $\AdS$ signature, though there is nothing that prevents 
one from computing the critical masses for the other signature.

\subsection{Discrete series}
The flat limit of UIRs in the discrete series is quite similar to that
of exceptional series representations in odd spacetime
dimensions. Indeed, the ``Bianchi-identity part'' (i.e., containing the
Young diagrams $(\Y_p, \1^m)\, , \, m=0,\dots,r-p$\,) of the character
of exceptional series irreps vanishes (due to the factor $q^{\Delta_c}
- q^{d-\Delta_c}\stackrel{q\to 1}{\to}0$ coming in front of it),
therefore we are only left with the usual\,\footnote{Usual in the
  sense that it is the only one AdS$_{d+1}$ characters appearing and
  is therefore the part that dS$_{d+1}$ characters have in common with
  the former.} set of Young diagrams corresponding to the field, its
gauge parameter and its reducibility. As a consequence, one has to
branch the same type of sequence of $\so(d)$ characters as for
exceptional series characters in odd spacetime dimensions, with the
only difference that here the first Young diagram (corresponding to
the massless field itself) is of maximal height. This last specifity
does not change the argument presented in the previous section for the
flat limit of UIRs in the exceptional series. Thence, we obtain the
following flat limit of UIRs in the discrete series or, equivalently,
(partially) massless fields with maximal-height Young diagrams:
\begin{equation}
  \chi^\dS_{[s_r+d-r-t; \Y_{\vec s, t}]}(q, \vec x) \flimit
  \sum_{\sigma_1=s_2}^{s_1} \dots \sum_{\sigma_r=s_r-t+1}^{s_r}
  \chi^{\so(2r)}_{(\sigma_1, \dots, \sigma_r)}(\vec x)\,
  \Pf{2r+1}(\vec x) = \sum_{\Y' \in \Sp(\Y_{\vec s, t})}
  \chi^\Poinc_{[0;\, \Y']}(\vec x)\, .
\end{equation}
with, upon rewriting $\Y_{\vec s, t}$ as $(\ell_1^{h_1}, \dots,
\ell_B^{h_B})$ to exhibit its various blocks (with $\sum_{I=1}^B h_I =
r$ and $\ell_B=s_r$), the flat space spectrum:
\begin{eqnarray}
  \Sp(\Y_{\vec s,t}) := \Big\{ \Y' & = & (\ell_1^{h_1-1},\ell_1-n_1,
  \dots, \ell_{B-1}^{h_{B-1}-1},\ell_{B-1}-n_{B-1},\ell_B^{h_B-1},
  \ell_B-m)\, , \\ && \qquad \qquad \qquad 0 \leqslant n_I \leqslant
  \ell_I - \ell_{I+1} \, , \, I=1,\dots,B-1\, , \, m=0, \dots, t-1
  \Big\}\, . \nonumber
\end{eqnarray}
As previously, it appears that for massless fields (i.e., $t=1$) the
last block is protected whereas for partially massless fields, the
corresponding flat spacetime spectrum can also contain fields where up
to $t-1$ boxes are removed from the last block.

\section{Conclusions}\label{sec:4}

In this paper, we investigated the representation theory of
$\so(1,d+1)$ and tried to give a field theoretic interpretation of the
list of UIRs known for this algebra. We proposed a dictionary between
(partially) massless mixed-symmetry fields of \it arbitrary \rm shape
and representations in the exceptional and discrete series, thereby
extending and completing the work of \cite{Joung:2006gj, Joung:2007je}
concerning scalar fields. A byproduct of this identification is to
confirm the anticipated unitarity of partially massless fields in de
Sitter. More precisely, we found for gauge fields of arbitrary shape,
that unitary fields in dS$_{d+1}$ are those whose gauge symmetry
involves the lowest block of their Young diagram. This generalises the
analysis of unitarity of mixed-symmetry massless fields on
de Sitter spacetime from the case of two-column Young diagrams
\cite{deMedeiros:2002qpr} (see \cite{Joung:2016naf} for a study of 
two-column partially massless fields) to the generic case; see also
\cite{Zinoviev:2003dd} where some types of massive mixed-symmetry
fields in (A)dS and various massless limits were analysed starting
from Lagrangian formulations.

In the process of studying $\so(1,d+1)$ irreps, we were able to derive
their character, which gives us some insight into the structure of the
corresponding modules. Inspired by the BMV mechanism in anti-de Sitter
spacetime \cite{Brink:2000ag, Boulanger:2008up, Boulanger:2008kw,
  Alkalaev:2009vm}, we proposed a way of taking the flat limit of
those characters and read off the resulting flat spacetime spectrum by
recognising characters of the Poincar\'e group. Although this
procedure fails for UIRs in the exceptional series when $d$ is even,
this method yields a fairly coherent picture of the flat limit of
massless fields in de Sitter. In AdS$_{d+1}$, the BMV spectrum of
unitary massless mixed-symmetry fields in flat spacetime is given by
the $\so(d)$ branching rules of the field's Young diagram, where the
first block, activated by the gauge transformations, is left
untouched. A similar situation occurs in de Sitter spacetime, but
instead of the upper block being protected, it is now the lowest one
that is left untouched when branching the field's Young diagram onto
$\so(d-1)$. As argued in \hyperref[app:nonunitary]{Appendix
  \ref{app:nonunitary}}, this BMV-type spectrum should hold for
generic massless fields, even non unitary ones: mixed-symmetry fields
whose block affected by gauge transformations is \textit{not} the
first one in AdS$_{d+1}$ or the last one in dS$_{d+1}$, should produce
in the flat limit all massless field labelled by a Young diagram
contained in the branching rule of the original (A)dS$_{d+1}$ field's
diagram, leaving the block activated by the gauge symmetry
untouched.

Our proposition should, however, be considered with caution. Despite
the quite coherent landscape of fields described by UIRs of
$\so(1,d+1)$ according to our identifications and the consistent
BMV-type spectrum obtained in even spacetime dimensions, the failure
to obtain a similar one in odd spacetime dimensions is puzzling, and
definitely calls for further investigation. As we explained in
\hyperref[sec:3]{Section 3}, we think that the resolution of this
puzzle (namely the fact that from the field theory point of view the
parity of the spacetime dimension does not bring any difference in the
treatment or behaviour of massless fields, whereas we observe a
drastic distinction at the group theoretical level) is the distinction
between group and algebra irreps. In order to make contact with the
known classification of UIRs of $\SO(1,d+1)$, we had to look at the
group irreps which seem to be formulated in terms of the curvature of
the massless fields, but at the level of the algebra it may be
possible to consider irreps describing only the gauge field (as we are
used to in AdS$_{d+1}$). Nevertheless, having at hand this proposed
dictionary of $\so(1,d+1)$ irreps and the corresponding characters
opens several possibilities, such as the construction of a
Flato-Fronsdal theorem \cite{Flato:1978qz} for de Sitter. The
decomposition of the product of two ``shortest representations'' (that
are the Dirac scalar and spinor singletons) into irreducible
representations, as a tower of massless spin-$s$ fields in
AdS$_{d+1}$, is at the heart of the higher-spin AdS/CFT correspondence
\cite{Sezgin:2002rt, Klebanov:2002ja}. A similar theorem in de Sitter
spacetime would provide a similar kinematical evidence in favour of
the proposal \cite{Anninos:2011ui} of a higher-spin dS/CFT
correspondence. Even though we did not find an obvious unitary
singleton-type representation in the list of known $\so(1,d+1)$
irreps, one would expect that such UIRs exist because of their r\^ole
in the definition of the higher-spin algebra in (A)dS$_{d+1}$ (see for
instance \cite{Iazeolla:2008ix, Sezgin:2012ag, Joung:2014qya} for nice
overviews), which is insensitive to the signature. In fact, we
identified a natural candidate for the singleton representation and
their higher-order generalisations \cite{Limic1966}. Another potential
evidence in this direction is the fact that in dS$_4$, massless
totally symmetric fields have the same character as their AdS$_4$
counter part, and therefore summing the characters of massless
spin-$s$ fields on all spins will yield the square of the scalar
singleton character. It is therefore natural to expect this type of
decomposition to remain true in any dimensions.

\section*{Acknowledgments}

It is a pleasure to thank Andrea Campoleoni for discussions on the
characters of the Poincar\'e group. We also thank Philippe Spindel 
and Per Sundell for discussions. T.B. is supported by a joint grant ``50/50'' Universit\'e
Fran\c{c}ois Rabelais Tours -- R\'egion Centre / UMONS. The research
of X.B. was supported by the Russian Science Foundation grant
14-42-00047 in association with the Lebedev Physical Institute.

\appendix

\section{Massive scalar field on (anti) de Sitter spacetime}\label{app:example}
Let us consider a massive scalar field in (anti-) de Sitter spacetime,
subject to Klein--Gordon's equation
\begin{equation}
  \left(\nabla^2_{(\text{A})\dS_{d+1}} - m^2\right) \phi = 0\, ,
\end{equation}
where $\nabla^2 := g^{\mu \nu} \nabla_\mu \nabla_\nu$ is the
Laplace--Beltrami operator in $(d+1)$-dimensional (anti-) de Sitter
spacetime. In Poincar\'e coordinates, the metric looks like:
\begin{equation}
  \dd s_{\pAdS_{d+1}}^2\, = \frac{R^2}{z^2} \Big( \sigma \dd z^2 +
  \delta^{(\sigma)}_{\mu \nu} \dd x^\mu \dd x^\nu \Big)\, ,
\end{equation}
where $R$ is the curvature radius and $\delta^{(\sigma)} =
\diag(-\sigma, 1, \dots, 1)$. A more convenient patch for the purpose
of holographic reconstruction \cite{Skenderis:2002wp} can be obtained
by redefining the $z$ coordinate:
\begin{equation}
  \sqrt{R \rho} = z \Rightarrow \frac{\dd z}{z} = \frac{\dd
    \rho}{2\rho}\, ,
\end{equation}
after which the metric takes the form:
\begin{equation}
  \dd s_{\pAdS_{d+1}}^2 = R^2\left(\sigma \frac{\dd \rho^2}{4\rho^2} +
  \frac{1}{R\, \rho} \delta^{(\sigma)}_{\mu \nu} \dd x^\mu \dd x^\nu
  \right)\, .
\end{equation}
In these coordinates, the Klein-Gordon equation reads:
\begin{equation}
  \nabla^2_{\pAdS_{d+1}} \phi(\rho, \vec x) = \frac{1}{R^2}
  \left[4\sigma\rho^2 \partial_\rho^2 - 4\sigma\big(\tfrac{d}2 -
    1\big) \rho \partial_\rho + R\,\rho\, \Box_{(\sigma)} \right]
  \phi(\rho, \vec x) = m^2 \phi(\rho, \vec x)
  \label{box}
\end{equation}
with $\Box_{(\sigma)} = \delta_{(\sigma)}^{\mu \nu}
\frac{\partial}{\partial x^\mu} \frac{\partial}{\partial x^\nu}$, the
wave operator on the (conformal) boundary of (A)dS$_{d+1}$, i.e., the
Laplacian $\Delta_{\R^d}$ on the $d$-dimensional Euclidean space for
the de Sitter case, or the d'Alembert operator $\Box$ on
$d$-dimensional Minkowski space for the anti-de Sitter space.  Using
the ansatz:
\begin{equation}
  \phi(\rho, \vec x) = \rho^{\Delta/2} \varphi(\rho, \vec x)\, ,
  \label{reconstruction_ansatz}
\end{equation}
with $\varphi(\rho, \vec x)$ a scalar field, well-behaved at the
conformal boundary of (A)dS$_{d+1}$ (located at $\rho \rightarrow 0$),
plugging this into \eqref{box}, and evaluating the resulting
expression at the boundary yields:
\begin{equation}
  (mR)^2 = \sigma\Delta(\Delta-d) \quad\Rightarrow\quad \Delta_{\pm} =
  \frac{d}2 \pm \sqrt{\frac{d^2}4 + \sigma (mR)^2}\, .
  \label{mass_delta}
\end{equation}
At this point, an important difference appears between AdS$_{d+1}$ and
dS$_{d+1}$. In the former case, the exponent $\Delta$ is always real
(as the term under the square root will always be positive), however
in the latter case $\Delta$ can be complex. When the scalar field is
``very massive'', i.e., has $mR > d/2$, in de Sitter space the exponent
$\Delta$ reads:
\begin{equation}
  \Delta = \tfrac d2 + i\rho\, , \quad \text{with} \quad \rho = \pm
  \sqrt{(mR)^2 - \tfrac{d^2}4} \Leftrightarrow (mR)^2 = \tfrac{d^2}4 +
  \rho^2\, .
\end{equation}
As we will see later on, this corresponds to scalar fields in the
``principal series'' of UIRs of the de Sitter isometry group
$\SO(1,d+1)$, whereas ``not-so-massive'' fields, i.e., those with a
mass such that $0 \leqslant mR < d/2 \Leftrightarrow \tfrac{d}2 <
\Delta_c < d$ will be identified with scalar fields of the
``complementary series'', where we introduced $\Delta_c =
\left. \Delta \right|_{\dS}$. Plugging \eqref{mass_delta} in
\eqref{reconstruction_ansatz} transforms it into a differential
equation on $\varphi$:
\begin{equation}
  \Big( 4\sigma \rho^2 \partial^2_\rho + 2\sigma (2\Delta_\pm-d+2)
  \rho \partial_\rho - R\rho \Box_{(\sigma)} \Big) \varphi(\rho, \vec
  x) = 0\, .
  \label{eq_dev}
\end{equation}
From now on, we will restrict ourselves to $0\leqslant mR < d/2$, so
that the exponent $\Delta$ is real in both dS and AdS. Following the
analysis of e.g. \cite{Skenderis:2002wp, Bekaert:2013zya}, the next
step is then to expand $\varphi$ in powers of $\rho$:
\begin{equation}
  \varphi(\rho, \vec x) = \sum_{n=0}^\infty \rho^n \varphi_n(\vec x)\,
  ,
\end{equation}
where $\varphi_n(\vec x)$ are also well-behaved fields. By plugging
this expansion in \eqref{eq_dev}, we obtain the following recursion
relation among the modes of $\varphi$:
\begin{equation}
  2n \sigma \big( 2\Delta_\pm - d + 2n \big) \varphi_n(\vec x) = R\,
  \Box_{(\sigma)} \varphi_{n-1}(\vec x)\, .
\end{equation}
Assuming $2\Delta-d+2n \neq 0\, , \, \forall n \in \N$, the $n$th mode
of $\varphi$ can be expressed in terms of the boundary value
$\varphi_0$ as:
\begin{equation}
  \varphi_n(\vec x) = \frac{R^n}{4^n\, n! (\Delta - \tfrac d2 +1)_n}
  \Box_{(\sigma)}^n \varphi_0(\vec x)\, ,
\end{equation}
where $(a)_n := a(a+1) \dots (a+n-1)$ denotes the (increasing)
Pochhammer symbol.\\

\begin{figure}[!ht]
  \center
  \begin{tikzpicture}
    \draw (-1.2,2.8) node {dS$_{d+1}$};
    \draw[thick] (-2,0) -- (0,0) node {$\bullet$};
    \draw[thick, color=green!75!black] (0.05,0) -- (2,0);
    \draw (1.8,0) node[below] {$\tfrac{d}2$};
    \draw[thick, ->] (4,0) node{$\bullet$} -- (5.2,0)
    node[below]{$\Re(\Delta_c)$} -- (5.4,0);
    
    \draw[thick, ->] (0,0) -- (0,3) node[right] {$\Im(\Delta_c)$};
    \draw[thick, color=red] (2,0.08) -- (2,2);
    \draw[thick, color=red] (2,-0.08) -- (2,-0.8);
    \draw[dashed] (2,2) -- (2,3);
    \draw[dashed] (2,-0.8) -- (2,-1.4);
    
    \draw[color=blue] (0.5,0) node{$\boldsymbol \times$};
    \draw[color=blue] (1,0) node{$\boldsymbol \times$};
    \draw[color=blue] (1.5,0) node{$\boldsymbol \times$};
    \draw[color=blue] (2.5,0) node{$\boldsymbol \times$};
    \draw[color=blue] (3,0) node{$\boldsymbol \times$};
    \draw[color=blue] (3.5,0) node{$\boldsymbol \times$};
    
    \draw (2,0) node{$\bullet$};
    \draw[thick, color=green!75!black] (2.06,0) -- (3.94,0);
    \draw (4,-0.1) node[below]{\small $d$};
    \draw (-0.1,-0.1) node[below]{\small $0$};
  \end{tikzpicture}
  \caption{Repartition of the massive scalar fields in dS$_{d+1}$
    discussed above, as a function of the conformal weight $\Delta_c$.
    Massive field in the principal and complementary series are
    depicted respectively by a red and a green line, whereas the blue
    dots indicate a discrete collection of representations in the
    complementary series with $\Delta_c = \tfrac{d}2 \pm \ell\,$ that
    would correspond to higher order singletons and their shadows in
    AdS$_{d+1}$.}\label{figure}
\end{figure}
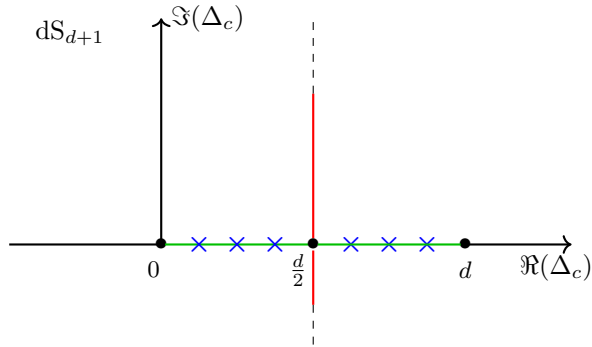

However, both in AdS$_{d+1}$ and in dS$_{d+1}$, this recursion
relation can break down if $\Delta = \frac{d}2 - \ell$ for some
integer $\ell \geqslant 1$, which is possible for $\Delta_-$ if
$\sqrt{\frac{d^2}4 + \sigma (mR)^2} = \ell \in \N_0$. In this case, a
possible solution is \cite{Bekaert:2013zya} to impose the polywave
equation as a constraint $\Box_{(\sigma)}^\ell \varphi_0(\vec x) = 0$
on the lowest order term. The power series expansion in this case
reads:
\begin{equation}
  \phi(\rho, \vec x) = \rho^{\Delta_-/2} \Big( \varphi_0(\vec x) +
  \rho\, \varphi_1(\vec x) + \dots \Big) + \rho^{\Delta_+/2} \Big(
  \tilde \varphi_0(\vec x) + \rho\, \tilde\varphi_1(\vec x) + \dots
  \Big)\, ,
\end{equation}
where $\tilde \varphi(\vec x)$ is left unconstrained. Notice that, as
in the AdS$_{d+1}$ case treated in \cite{Bekaert:2013zya}, it is the
branch $\varphi$ that is the leading one when $\rho \rightarrow 0$
(i.e., the branch that goes more slowly to zero as $\rho \rightarrow
0$).  The fact that this branch is the leading one toward the
boundary, added to the fact that this part of the series expansion of
the field is always a solution to \eqref{box} leads to the conclusion
that, for the space of solution of \eqref{box} to be an irreducible
$\so(2,d)$ or $\so(1,d+1)$ module, one has to quotient by the subspace
of subleading solutions $\tilde \varphi$. Effectively, only leading
solutions remain in the module, which thereby defines a (higher order)
singleton: a scalar field propagating no local degrees of freedom in
the bulk (i.e., ``confined'' at the conformal boundary) and defining a
boundary conformal scalar obeying the (poly)wave equation
$\Box_{(\sigma)}^\ell \varphi_0 = 0$. It complies with the fact that
one would expect a would-be de Sitter singleton to be the fundamental
field of the conformal field theory dual to the higher-spin theory in
dS$_{d+1}$. According to the proposed duality in
\cite{Anninos:2011ui}, the field should fall in a non-unitary irrep of
$\so(1,d+1)$. In our case, the field reconstructed previously should
belong to the unitary component of complementary series of
$\so(1,d+1)$ UIRs. This UIR was actually studied originally in
\cite{Limic1966}. The generic landscape of the scalar field is
summarised in Figure \ref{figure}.
	
\hspace{10pt}
\vspace{-10pt}

\section{Classification of $\so(2,d)$ unitary irreducible representations \& their characters}\label{app:ads}

We will focus on highest-weight representations, which are the
physically most relevant ones. Indeed, their energy spectrum is, by
construction, bounded from above or below.\footnote{However, remember
  that there are no such representations for the de Sitter case since
  there is no global timelike Killing vector field on de Sitter
  spacetime.} They are characterised by a highest-weight: $\blambda =
(\Delta, \vec s)$ with $\Delta$ the conformal weight, or minimal
energy in AdS$_{d+1}$, and $\vec s = (s_1, s_2, \dots, s_r)$ a
$\so(d)$ highest-weight labeling the spin of the representation. The
highest-weight UIRs of $\so(2,d)$ can be described as fields on
AdS$_{d+1}$ and classified as follows:
\begin{itemize}
\item \textbf{Massive representations:} $\Delta > s_1 + d - h_1 - 1$
  with $s_1 = \dots = s_{h_1} > s_{h_1+1}$, or $\Delta > \tfrac{d-2}2$
  for $\vec s = \vec{0}$ and $\Delta > \tfrac{d-1}2$ for $\vec s =
  (\tfrac{1}2,\cdots,\tfrac{1}2)$, whose character reads:
  \begin{equation}
    \chi_{[\Delta; \vec s\,]}^{\AdS}(q, \vec x) = q^\Delta \,
    \chi_{\vec s}^{\so(d)}(\vec x) \, \Pd d (q, \vec x)
  \end{equation}
  with $\Pd d (q, \vec x)$ given by \eqref{vacuum_char}.
\item \textbf{Massless representations:} $\Delta = \Delta_{s,h_1}: = s
  + d - h_1 -1$, with $\vec s$ such that $s_1 = \dots = s_{h_1} \equiv
  s > s_{h_1+1} \geqslant \dots \geqslant \rvert s_r \rvert$, analysed
  in \cite{Metsaev:1995re, Metsaev:1997nj} and whose character is
  given by \cite{Dolan:2005wy}:
  \begin{equation}
    \chi^{\AdS}_{[\Delta_{s,h_1}; \vec s\,]}(q, \vec x) =
    q^{\Delta_{s,h_1}} \, \left( \chi^{\so(d)}_{\vec s}(\vec x) +
    \sum_{k=1}^{h_1} (-q)^{h_1+1-k} \chi^{\so(d)}_{\vec s_k}(\vec x)
    \right) \, \Pd d (q, \vec x)
    \label{massless_mixed}
  \end{equation}
  with $\vec s_k = (s, \dots, s, \pos{s-1}{k\text{th entry}}, s - 1,
  \dots, \pos{s - 1}{h_1\text{th entry}}, s_{h_1+1}, \dots, s_r)$
\end{itemize}

\begin{example}
  Usually, totally symmetric massless fields are considered,
  i.e., massless fields with $\Delta=\Delta_s:=s+d-2$ and $\vec s = (s,
  0, \dots, 0)$ for $s \in 1+\N$ or $\vec s = (s, \tfrac{1}2, \dots,
  \tfrac{1}2)$ for $s \in \tfrac{1}2 + \N$, that we will both denote
  $(s)$. Accordingly, their character is given by the above formula in
  the special case $h_1=1$:
  \begin{equation}
    \chi^{\AdS}_{[\Delta_{s}; (s)]}(q, \vec x) = q^{\Delta_{s}} \,
    \left( \chi^{\so(d)}_{(s)}(\vec x) -q\, \chi^{\so(d)}_{(s-1)}(\vec
    x) \right) \, \Pd d (q, \vec x)
  \end{equation}
	
 Another class of physically interesting representations (although not
 unitary on AdS, contrarily to the above irreps) are the so-called
 (totally symmetric) partially-massless fields of spin-$s$ and
 depth-$t$, with $\Delta=\Delta_{s}^{(t)} := s + d - t - 1$ and $s
 \geqslant t>1$. Their character reads:
  \begin{equation}
    \chi^{\AdS}_{[\Delta_{s}^{(t)}; (s)]}(q, \vec x) =
    q^{\Delta_{s}^{(t)}} \, \left( \chi^{\so(d)}_{(s)}(\vec x) -q^t
    \chi^{\so(d)}_{(s-t)}(\vec x) \right) \, \Pd d (q, \vec x)
  \end{equation}
\end{example}

\section{Verma interlude}\label{app:bgg}

In this section, we recall some basic definitions on Verma module and
generalised Verma modules, as well as BGG resolutions in both
context. Once again, we make no attempt at full mathematical rigor,
but hope to give an intuitive picture of these concept to the
unfamiliar reader (for more details, see for instance
\cite{Leopwsky1977}).

\subsection{Verma module}
\begin{definition}[Verma module]
  Let $\g$ be a semisimple Lie algebra, with Cartan subalgebra $\h
  \subset \g$, $\Delta$ (resp. $\Delta_\pm \subset \Delta$) its root
  (resp. positive/negative root) system, $\g_\pm$ the subalgebra dual
  to the positive/negative root system $\Delta_\pm$ and $\b := \h
  \oplus \g_+$ its Borel subalgebra. Furthermore, let $\U(\g)$ denote
  the universal enveloping algebra of $\g$ and $\blambda \in \Delta$
  denote a weight. Let $v_{\blambda}$ be a one-dimensional
  $\b$-module, then the Verma module $\cV_{\blambda}$ is defined as:
  \begin{equation}
    \cV_{\blambda} := \U(\g) \otimes_{\U(\b)} v_{{\blambda}} \cong
    \U(\g_-) \otimes v_{{\blambda}}\, .
  \end{equation}
\end{definition}

In other words, a Verma module is a representation space of $\g$
constructed from a highest-weight vector, i.e., an eigenvector of the
Cartan subalgebra which is annihilated by all raising operators. In
the language of the above definition, this highest-weight vector is a
one-dimensional representation of the Borel subalgebra, which is
composed of the Cartan subalgebra and the subalgebra spanned by
raising operators, as they have a definite action on it. In turn, any
elements of $\cV_{\blambda}$ is of the form $\prod_{\alpha \in
  \Delta_-} (E_\alpha)^{n_\alpha} v_{\blambda}\, , \, n_\alpha \in \N$
where $E_\alpha$ is a lowering operator associated to the negative
root $\alpha$.\\

The BGG theorem for Verma modules gives a criterion for a Verma module
to contain a submodule, namely it gives a condition on the
highest-weight defining a submodule in a given Verma module for it to
be a proper submodule. This criterion is given in the following
theorem:

\begin{theorem}[Bernstein-Gel'fand-Gel'fand]
  Let $\g$ be a Lie algebra and $\cV_{\boldsymbol{\mu}}$ a Verma
  $\g$-module with highest-weight $\boldsymbol{\mu}$, then the two
  following statements are equivalent:
  \begin{itemize}
  \item $\cV_{\boldsymbol{\mu}} \subset \cV_{\blambda}$,
  \item $\exists\, \alpha_1, \dots, \alpha_n \in \Delta_+$ such that
    $\boldsymbol{\mu} = w_{\alpha_n} \dots w_{\alpha_1} \cdot
    {\blambda}$ and
    \begin{equation}
      \big(\alpha_k^{\vee}, w_{\alpha_{k-1}} \dots
      w_{\alpha_{1}}({\blambda}+{\bf\rho})\big) \in \N, \qquad \forall
      \ k \in \{ 1,2,\cdots, n \}\, .
      \label{bgg_condition}
    \end{equation}
  \end{itemize}
\end{theorem}

In the above definition, $w_\alpha \cdot {\blambda}$ denotes the
affine action of an element of the Weyl group of $\g$ associated to a
root $\alpha \in \Delta$, i.e., $w_\alpha \cdot {\blambda} :=
{\blambda} - \alpha(\alpha^{\vee}, {\blambda} + {\bf\rho}) =
w_\alpha({\blambda}+{\bf\rho})-{\bf\rho}\, , \, \alpha^{\vee} :=
\tfrac{2\alpha}{(\alpha, \alpha)}\,$, where $(\cdot\,, \cdot)$ denotes
the Killing form on $\g$, and $\rho$ its Weyl vector.\\

\begin{example}
  Let us consider a very simple case to illustrate this
  theorem. Taking $\mathfrak{sl}(2)\cong\so(3)$ and a Verma module
  based on the highest-weight $\lambda = s \in \tfrac{1}2 \N$, we can
  start looking for submodules using the BGG theorem. The Weyl group
  of $\so(3)$ is $\Z_2$ and its Weyl vector $\rho = \tfrac{1}2$. As a
  consequence, the only non trivial element of the Weyl group, that we
  will note $w$, is the one flipping the sign of the weight on which
  it will act and is associated to the positive root (the unit basis
  vector of the root space, as $\so(3)$ is of rank 1, that we do not
  bother writing), and therefore the only weight that can be obtained
  from an action of the Weyl group on $\lambda$ is $w \cdot \lambda =
  \lambda - 2 (\lambda + \rho) = -(\lambda+1)$. Having looked at a
  particularly simple case of rank 1, the relevance of the criterion
  \eqref{bgg_condition} cannot be fully seen, however, we were able to
  recover the standard result that $J_+(J_-)^{s+1}\,v_\lambda = 0$
  (where $v_\lambda$ is the highest-weight vector defining the Verma
  module), i.e., $(J_-)^{s+1} v_\lambda$ defines a submodule that needs
  to be modded out in order to obtain an irreducible representation of
  $\so(3)$, without having to compute explicitly the action of the
  ladder operators.
\end{example}

Recall that an integral dominant weight ${\blambda}$ is defined to be
a weight such that, for all positive root $\alpha \in \Delta_+$, it
verifies $({\blambda}, \alpha^\vee) \in \N$. An important property for
us is that for integral dominant weights, \it every \rm element of the
Weyl group verifies the condition \eqref{bgg_condition}.\\

Finally, we need to introduce the notion of length of a Weyl group
element, in order to define the BGG resolution of an irreducible Verma
module.

\begin{definition}[Length of a Weyl group element]
  Let $\g$ be a Lie algebra, and $w \in \cW$ an element of its Weyl
  group. The length of $w$, noted $\ell(w)$ is defined to be the
  minimal number of reflections $w_{\alpha^{(i)}}$ associated to
  simple roots $\alpha^{(i)}$ such that $w$ is given as a product of
  these reflections, i.e., $w = w_{\alpha^{(i_1)}} \dots
  w_{\alpha^{(i_n)}}$ and $n=\ell(w)$.
\end{definition}

\begin{theorem}[Bernstein-Gel'fand-Gel'fand resolution]
  Let ${\blambda}$ be an integral dominant highest-weight and
  $\D_{\blambda}$ the corresponding finite-dimensional and irreducible
  highest-weight module. There exists a long exact sequence:
  \begin{equation}
    0 \rightarrow \bigoplus\limits_{w \in \cW,\, \ell(w)=n} \cV_{w
      \cdot {\blambda}} \rightarrow \dots \rightarrow
    \bigoplus\limits_{w \in \cW,\, \ell(w)=1} \cV_{w \cdot {\blambda}}
    \rightarrow \cV_{\blambda} \rightarrow \D_{\blambda} \rightarrow 0
  \end{equation}
  where $n$ is the maximal length of elements of $\cW$.
\end{theorem}

\subsection{Generalised Verma module}
Now we can turn to the case of a generalised Verma module, which is
the one relevant for this paper, and we start by recalling the
definition of such modules:

\begin{definition}[Generalised Verma module]
  Let $\g$ be a finite dimensional Lie algebra, $\p$ a parabolic
  subalgebra and $\V_{{\blambda}}$ a finite dimensional representation
  space of $\p$ with highest-weight ${\blambda}$. Then the generalised
  Verma module $\cV_{{\blambda}}$ is defined as:
  \begin{equation}
    \cV_{{\blambda}} := \U(\g) \otimes_{\U(\p)} \V_{{\blambda}}
  \end{equation}
\end{definition}

\vspace{2mm}

The BGG resolution for generalised Verma modules is then quite similar
to the one previously exposed for Verma modules. The main difference
comes from a decomposition of the Weyl group induced by the choice of
a parabolic subalgebra. One way to single out a parabolic
  subalgebra of $\g$ is to consider a subset $\Sigma$ of the set of
  simple roots $\Phi_s$, and defining
\begin{equation}
  \p_\Sigma := \l_\Sigma \oplus \n_\Sigma\,,
  \label{parabolic}
\end{equation}
with
\begin{equation}
  \l_\Sigma := \h \oplus \bigoplus_{\alpha\in\Phi_\Sigma} \g_\alpha\,,
  \quad \n_\Sigma := \bigoplus_{\beta\in\Phi_+\backslash\Phi_\Sigma^+}
  \g_\beta\,,
  \label{levi_decompo}
\end{equation}
where $\Phi_\Sigma$ is the root system generated by $\Sigma$ and
$\Phi_\Sigma^+ = \Phi_+ \cap \Phi_\Sigma$ is the set of positive roots
of this root system. In this construction, the Borel subalgebra $\b_+$
is automatically part of $\p_\Sigma$ and only the root spaces
corresponding to negative roots having components in the directions
spanned by $\Sigma$ are removed. The piece $\l_\Sigma$ is a reductive
subalgebra (called the Levi subalgebra) and $\n_\Sigma$ is the
nilradical of $\p_\Sigma$, i.e., its maximal nilpotent ideal. In fact,
all parabolic subalgebras can be constructed as above. \\

With this decomposition at hand, one can define the subgroup
  $\overline \cW$ of the Weyl group of $\g$ generated by reflections
  associated to the subset of simple roots of $\Sigma$. This subgroup
  corresponds to the Weyl group of $\l_\Sigma$. Another subset $\cW'$
is the one composed of elements of $\cW$ such that any of the positive
roots of $\Phi_\Sigma$ is obtained by applying an element of $\cW'$
to a positive root in $\Phi_+$, i.e.,
\begin{equation}
  \cW' := \big\{ w \in \cW\, \rvert\, \Phi_\Sigma^+ \subset w\, \Phi_+
  \big\}\,.
\end{equation}
A property is that every element of the full Weyl group $\cW$ can be
decomposed as a product of elements of those two subgroups: $\forall w
\in \cW\, , \, \exists\, \bar w \in \overline \cW\, , \, w' \in \cW'$
such that $w = \bar w\, w'$.\\

The BGG resolution for a generalised Verma module with highest-weight
${\blambda}$ being an integral dominant weight, is defined almost as
in the case of Verma module, except for the fact the the full Weyl
group should be substituted with the subgroup $\cW'$: the long exact
sequence is \cite{Leopwsky1977}
\begin{equation}
  0 \rightarrow \bigoplus\limits_{w \in \cW',\, \ell(w)=n} \cV_{w
    \cdot {\blambda}} \rightarrow \dots \rightarrow
  \bigoplus\limits_{w \in \cW',\, \ell(w)=1} \cV_{w \cdot {\blambda}}
  \rightarrow \cV_{\blambda} \rightarrow \D_{{\blambda}} \rightarrow
  0\, .
\end{equation}

\begin{remark}
  In this paper, we are interested in the algebra $\g=\so(2+d)$ and
  its parabolic subalgebra $\p=\so(2) \inplus \iso(d)$. The positive
  root system $\Phi_+$ of the orthogonal algebras, in the orthonormal
  basis, are given by
  \begin{equation}
    \Phi_+ = \left\{
    \begin{aligned}
      \big\{ \e_i \pm \e_j\, \quad | \quad 0 \leqslant i < j \leqslant
      r \big\} \qquad \qquad \qquad \quad & \quad \text{for }
      d=2r\,,\\ \big\{ \e_k\,, \,\, \e_i \pm \e_j\, \quad | \quad 0
      \leqslant i < j \leqslant r\,, \,\, 0 \leqslant k \leqslant r
      \big\} & \quad \text{for } d=2r+1\,,
    \end{aligned}
    \right.
  \end{equation}
  corresponding to the set of simple roots
  \begin{equation}
    \Phi_s = \big\{ \e_i - \e_{i+1}\,, \quad 0 \leqslant i \leqslant
    r-1\big\} \cup \left\{
    \begin{aligned}
      \{ \e_{r-1} + \e_r\} & \quad \text{for } d=2r\,, \\ \{ \e_r \}
      \qquad \quad & \quad \text{for } d=2r+1\,.
    \end{aligned}
    \right.
  \end{equation}
  The parabolic subalgebra $\p=\so(2) \inplus \iso(d)$ is obtained
  through the construction \eqref{parabolic}--\eqref{levi_decompo} by
  choosing the subset of simple roots
  \begin{equation}
    \Sigma = \Phi_s \backslash \{ \e_0 - \e_1 \}\,.
  \end{equation}
  In this case, the root system $\Phi_\Sigma$ generated by $\Sigma$
  corresponds to that of $\so(d)$ and hence $\l_\Sigma = \so(2) \oplus
  \so(d)$, while the subgroup $\overline \cW$ is the Weyl group of
  $\so(d)$. The subgroup $\cW'$ is then generated by reflection with
  respect to the roots $\e_0 \pm \e_k$, and $\e_0$ when $d=2r+1$.
\end{remark}

\section{Characters from Bernstein-Gel'fand-Gel'fand resolutions}\label{app:char}

In \cite{Shaynkman:2004vu}, the structure of $\so_\C(2+d)$ modules
(where the subscript $\C$ is used to denote the complexified algebra) was spelled out using BGG
resolutions for generalised Verma modules (see
\cite{Gavrilik:1975ae,Klimyk:1976ac} for earlier similar results at
the group level). On the representation theory side (as recalled in
\hyperref[app:bgg]{Appendix \ref{app:bgg}}), they consist of a series
of homomorphisms between generalised Verma modules, induced by
particular elements of the Weyl group and such that the module in the
image of each of these maps is a submodule of the previous one. \\

Let us also introduce the following notations:
\begin{itemize}
\item A height-$p$ Young diagram, with $p \leqslant r$ will be
  denoted:
  \begin{equation}
    \Y_p := (s_1, s_2,\dots, \pos{s_p}{p\th}, 0, \dots, 0) = \vec s
  \end{equation}
\item An important operation on Young diagrams when dealing with the
  exceptional series is to remove one row from it and to delete one
  box in each of the following rows (i.e., situated below the one that
  was just removed). We will denote the diagram obtained from $\Y_p$
  after having performed the above modifications as:
  \begin{equation}
    \Yt i := (s_1, \dots, s_{i-1}, \pos{s_{i+1}-1}{i-\th}, \dots,
    \pos{s_p-1}{(p-1)-\th}, 0, \dots, 0)
  \end{equation}
\item A generalised Verma module based on the $\so(2) \oplus \so(d)$
  highest-weight ${\blambda}$ will generically be denoted
  $\cV_{\blambda}$, except when it is irreducible in which case we
  will write $\D_{\blambda}$. The translation rule from this $\so(2)
  \oplus \so(d)$ highest-weight to the conformal-weight/lowest-energy
  $\Delta$ and the $\so(d)$ highest-weight $\vec s$ is:
  \begin{equation}
    [\Delta\,; \vec s\,] = {\blambda} =
    (\lambda_0,\lambda_1,\cdots,\lambda_r) = (-\Delta,s_1,\cdots,s_r)
  \end{equation}
\item Finally, our elementary building block in writing characters for
  $\so_\C(2+d)$ are the characters of irreducible $\so(2) \oplus \so(d)$
  modules. We will introduce for them the notation:
  \begin{equation}
    \cY_{[\Delta\,; \vec s\,]}(q, \vec x) = q^{\Delta}
    \chi^{\so(d)}_{\vec s}(\vec x)\,.
  \end{equation} 
  The character of the generalised Verma module induced by the irrep
  $[\Delta; \vec s\,]$ of $\so(2) \oplus \so(d)$ is $$ \cY_{[\Delta;
      \vec s]}(q, \vec x) \Pd d (q, \vec x)\,,$$ with $\Pd d (q, \vec
  x)$ given by \eqref{vacuum_char}.
\end{itemize}

\vspace{6mm} For the sake of self-containedness, for each relevant
case we summarise the results of \cite{Shaynkman:2004vu} on
generalised Verma modules and then deduce the corresponding character:
\paragraph{\underline{Odd $d=2r+1$:}}  
Defining the sequence of $\so_\C(2+d)$ weights, where the first entry
is the $\so(2)$ weight and the $r$ following entries are the
components of the $\so(d)$ highest-weight:
\begin{eqnarray}
  (\lambda)_N & = & (\lambda_N - N, \lambda_0 + 1, \dots,
  \lambda_{N-1} + 1, \lambda_{N+1}, \dots, \lambda_r), \quad\qquad\qquad (N=0,
  \dots, r)\,,\\ (\lambda)_{K+r} & = & (-\lambda_{r+1-K}-K-r,
  \lambda_0 + 1, \dots, \lambda_{r-K} + 1, \lambda_{r+2-K}, \dots,
  \lambda_r), \quad (K = 1, \dots, r)\, , \\ (\lambda)_{2r+1} & = &
  (-\lambda_0-2r-1, \lambda_1, \dots, \lambda_r)\, ,
\end{eqnarray}
the following sequence is exact:
\begin{equation}
  0 \rightarrow \cV_{(\lambda)_{2r+1}} \rightarrow \cV_{(\lambda)_{2r}}
  \rightarrow \dots \rightarrow \cV_{(\lambda)_1} \rightarrow
  \cV_{(\lambda)_0} \rightarrow 0
\end{equation}
It can be shown that in odd dimensions, no subsingular module can
arise.  In other words, the above exact sequence implies the following
short exact sequences:
\begin{equation}
  0 \rightarrow \cV_{(\lambda)_{2r}} \rightarrow \D_{(\lambda)_{2r+1}}
  \rightarrow 0\, ,
\end{equation}
and
\begin{equation}
  0 \rightarrow \D_{(\lambda)_{N+1}} \rightarrow \cV_{(\lambda)_{N}}
  \rightarrow \D_{(\lambda)_N} \rightarrow 0\, , \quad\qquad (N=0,
  \dots, 2r)\, .
\end{equation}
This implies that the irreducible highest-weight module in the above
sequence with weight $(\lambda)_N$ is given by the quotient:
\begin{equation}
  \D_{(\lambda)_N} = \frac{\cV_{(\lambda)_N}}{\D_{(\lambda)_{N+1}}}\, .
\end{equation}
At the character level, this translates as:
\begin{eqnarray}
  \chi_{(\lambda)_N} (q, \vec x) & = & \cY_{(\lambda)_N}(q, \vec x)
  \Pd d (q, \vec x) - \chi_{(\lambda)_{N+1}} (q, \vec x) \\ & = &
  \sum_{k=0}^{2r+1-N} (-1)^k \cY_{(\lambda)_{N+k}} (q, \vec x) \Pd d
  (q, \vec x)
  \label{char_STV_odd}
\end{eqnarray}

\paragraph{\underline{Even $d=2r$:}}
The sequence of weights is modified in this case:
\begin{equation}
  \left\{
  \begin{aligned}
    (\lambda)_{-k} & = & (\lambda_{r-k}-r+k\,, \lambda_0+1, \dots,
    \lambda_{r-k-1}+1, \lambda_{r-k+1}, \dots, \lambda_r)\, , & \quad
    & (k=1, \dots, r) \\ (\lambda)_0 & = & (\lambda_r-r, \lambda_0+1,
    \dots, \lambda_{r-1} + 1)\, , \hspace{115pt} & & \\ (\lambda)_{0'}
    & = & (-\lambda_r-r, \lambda_0+1, \dots, -\lambda_{r-1} - 1)\,
    , \hspace{100pt} & & \\(\lambda)_{+k} & = & (-\lambda_{r-k}-r-k,
    \lambda_0, \dots, \lambda_{r-k-1}+1, \lambda_{r-k+1}, \dots,
    -\lambda_r)\, , & \quad & (k=1, \dots, r)
  \end{aligned}
  \right. 
\end{equation}
which differs from the odd-dimensional case by the presence of
non-standard (NS) homomorphisms and by a rhombus\,\footnote{The
  appearence of this rhombus is due to the fact that there exist two
  elements of the Weyl group $\cW'$ with the same length for
  $d=2r$. This can be seen in (A.16) of \cite{Shaynkman:2004vu}.} in
the middle of the sequence, yielding:
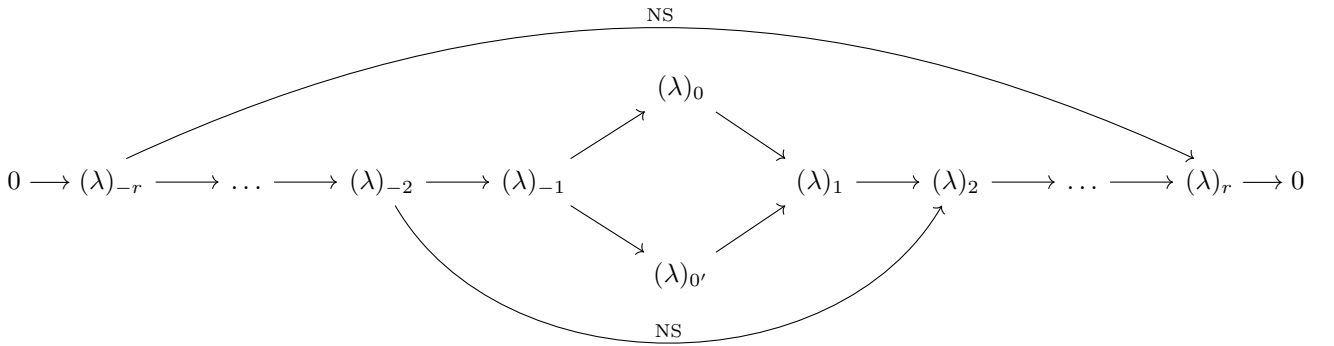
\begin{center}
  \begin{tikzcd}
    \quad & \quad & \quad & \quad & (\lambda)_0 \arrow{rd} \\ 0
    \longrightarrow (\lambda)_{-r} \arrow{r} \arrow[bend
      left=25]{rrrrrrrr}{\text{NS}} & \dots
    \arrow{r} & (\lambda)_{-2} \arrow{r} \arrow[bend
      right=60]{rrrr}{\text{NS}} & (\lambda)_{-1} \arrow{ru}
    \arrow{rd} & & (\lambda)_1 \arrow{r} & (\lambda)_2 \arrow{r} &
    \dots \arrow{r} & (\lambda)_{r} \longrightarrow 0 \\ \quad & \quad
    & \quad & \quad & (\lambda)_{0'} \arrow{ru}
  \end{tikzcd}
\end{center}

The main difference with the odd-dimensional case is the possibility
of subsingular modules, but no subsubsingular ones. The above sequence
leads to the following short exact sequences:
\begin{equation}
  \left\{
  \begin{aligned}
    0 \rightarrow \U_{(\lambda)_{N+1}} \rightarrow \cV_{(\lambda)_N}
    \rightarrow \D_{(\lambda)_N} \rightarrow 0\, , & \\ 0 \rightarrow
    \cV^*_{(\lambda)_{-N}} \rightarrow \U_{(\lambda)_{N+1}} \rightarrow
    \D_{(\lambda)_{N+1}} \rightarrow 0\, , & \\
  \end{aligned}
  \right. \hspace{15pt} (N=-1, \dots, -r)\, ,
  \label{doubling}
\end{equation}
together with
\begin{equation}
  0 \rightarrow \D_{(\lambda)_{N+1}} \rightarrow \cV_{(\lambda)_N}
  \rightarrow \D_{(\lambda)_N} \rightarrow 0\, , \qquad (N=0, \dots,
  r)\, ,
  \label{low_chain}
\end{equation}
and
\begin{equation}
  0 \rightarrow \cV^*_{(\lambda)_1} \rightarrow \U_{(\lambda)_0}
  \rightarrow \D_{(\lambda)_0} \oplus \D_{(\lambda)_{0'}} \rightarrow
  0 \, ,
  \label{rhombus}
\end{equation}
where $\cV_{(\lambda)}^*$ denotes the contragradient module. The
sequence \eqref{doubling} expresses the irreducible module
$\D_{(\lambda)_N}$ for $N=-1, \dots, -r+1$ as two different quotients:
\begin{equation}
  \D_{(\lambda)_N} = \frac{\cV_{(\lambda)_N}}{\U_{(\lambda)_{N+1}}}
	 = \frac{\U_{(\lambda)_N}}{\cV^*_{(\lambda)_{-N+1}}}\qquad (N=-1, \dots, -r+1)
\end{equation}
which can be translated into characters, yielding:
\begin{equation}
  \cY_{(\lambda)_N}(q, \vec x) \, \Pd d (q, \vec x) -
  \cC_{(\lambda)_{N+1}}(q, \vec x) = \cC_{(\lambda)_N}(q, \vec x) -
  \cY_{(\lambda)_{-N+1}}(q, \vec x) \, \Pd d (q, \vec x)
\end{equation}
where\, $\cC_{(\lambda)_N}$ is the character of the reducible module
$\U_{(\lambda)_N}$. This can be used to compute
$\cC_{(\lambda)_{-k}}$\, ($k=1, \dots, r-1$):
\begin{eqnarray}
  \cC_{(\lambda)_{-k}}(q, \vec x) & = & \big(\cY_{(\lambda)_{-k}}(q,
  \vec x) + \cY_{(\lambda)_{k+1}}(q, \vec x)\,\big) \Pd d (q, \vec x)
  - \cC_{(\lambda)_{-k+1}}(q, \vec x) \\ & = & \sum_{n=0}^{k-1} (-)^n
  \big(\cY_{(\lambda)_{-k+n}}(q, \vec x) + \cY_{(\lambda)_{k+1-n}}(q,
  \vec x)\,\big)\Pd d (q, \vec x) + (-)^k \cC_{(\lambda)_0}(q, \vec
  x)\, .
  \label{char_red_mod}
\end{eqnarray}
Using \eqref{rhombus}, we can express $\cC_{(\lambda)_0}(q, \vec x)$
as:
\begin{equation}
  \cC_{(\lambda)_0}(q, \vec x) = \chi_{(\lambda)_0}(q, \vec x) +
  \chi_{(\lambda)_{0'}}(q, \vec x) + \cY_{(\lambda)_1}(q, \vec x) \,
  \Pd d (q, \vec x)\, .
\end{equation}
Now as both modules $\D_{(\lambda)_0}$ and $\D_{(\lambda)_{0'}}$ are
resolved by the same short sequence as in the odd-dimensional case,
there character can be straightforwardly computed:
\begin{equation}
  \chi_{(\lambda)_0}(q, \vec x) = \sum_{N=0}^r (-)^N
  \cY_{(\lambda)_N}(q, \vec x) \, \Pd d (q, \vec x)\, ,
\end{equation}
idem for $\chi_{(\lambda)_{0'}}(q, \vec x)$ with the sum starting at
$N=0'$. Plugging this back into \eqref{char_red_mod}, we finally
obtain the explicit expression of $\cC_{(\lambda)_{-k}}$ in terms of
the factors $\cY_{(\lambda)_N}$ and $\Pd d$.  This formula can then be
used to express the character of the \it irreducible \rm module
$\D_{(\lambda)_{-k}}$:
\begin{eqnarray}
  \chi_{(\lambda)_{-k}}(q, \vec x) & = & \cY_{(\lambda)_{-k}}(q, \vec
  x) \, \Pd d (q, \vec x) - \cC_{(\lambda)_{-k+1}}(q, \vec
  x) \label{char_STV_even} \\ & = & \sum_{n=0}^k (-)^{k+n} \Big(
  \cY_{(\lambda)_n}(q, \vec x) + \cY_{(\lambda)_{-n}}(q, \vec x)
  \Big)\, \Pd d (q, \vec x) + 2 \sum_{n=k+1}^r (-)^{k+n}
  \cY_{(\lambda)_n}(q, \vec x) \, \Pd d (q, \vec x) \nonumber
\end{eqnarray}
where the term $\cY_{(\lambda)_{-n}}$ for $n=0$ has to be understood as
$\cY_{(\lambda)_{0'}}\,$, and the last sum is absent when $k=r$.

\paragraph{Identifying the exceptional series.}
Starting from:
\begin{equation}
  (\lambda_0, \lambda_1, \dots, \lambda_r) = (s_1-1,s_2-1, \dots,
  s_p-1, 0, \dots, 0) = (s_1-1, \Yt 1)\, ,
\end{equation}
as the weight of the long exact sequence, then the different weights
enumerated above for $d=2r+1$ take the form:
\begin{equation}
  (\lambda)_N = \left\{
  \begin{aligned}
    (s_{N+1} -(N+1), \Yt{N+1}), \quad \quad \quad & \quad & 0
    \leqslant N \leqslant p-1 \\ (-N, \Y_p, \1^{N-p}), \qquad \qquad
    \qquad \quad & \quad & p \leqslant N \leqslant r \\ (-N, \Y_p,
    \1^{d-p-N}), \qquad \qquad \qquad & \quad & r+1 \leqslant N
    \leqslant d-p \\ (-(s_{d+1-N} + N + 1), \Yt{d+1-N}), & \quad &
    d+1-p \leqslant N \leqslant d
  \end{aligned}
  \right.
\end{equation}
Now let us show that the character of the module at level $N=p$
reproduces the formula of characters for the exceptional series
in \cite{Hirai1965}:
\begin{eqnarray}
  \chi_{(\lambda)_p}(q, \vec x) & = & \sum_{k=p}^{2r+1} (-1)^{p+k}
  \cY_{(\lambda)_k}(q,\vec x) \, \Pd d (q,\vec x)\\ & = & \sum_{k=p}^r
  (-1)^{k+p} q^k \, \chi_{(\Y_p,\1^{k-p})}^{\so(d)}(\vec x)\, \Pd d
  (q,\vec x) + \sum_{k'=r+1}^{d-p} (-1)^{k'+p} q^{k'}\,
  \chi^{\so(d)}_{(\Y_p, \1^{d-p-k'})}(\vec x)\, \Pd d (q,\vec x)
  \nonumber \\ && \qquad \qquad \qquad + \sum_{k''=d+1-p}^d
  (-1)^{k''+p} q^{s_{d+1-k''}+k''-1}\, \chi_{\Yt{d+1-k''}}^{\so(d)}(\vec
  x)\, \Pd d (q,\vec x)\\ & = & \sum_{i=1}^{r-p} (-1)^{i} q^{p+i} \,
  \chi_{(\Y_p,\1^{i})}^{\so(d)}(\vec x)\, \Pd d (q,\vec x) +
  \sum_{j=1}^{r-p} \underbrace{(-1)^{d+j}}_{=-(-1)^j} q^{d-p-j}\,
  \chi^{\so(d)}_{(\Y_p, \1^{j})}(\vec x)\, \Pd d (q,\vec x) \\ && +
  \big(q^p\, \chi^{\so(d)}_{\Y_p}(\vec x) -
  q^{d-p}\,\chi^{\so(d)}_{\Y_p}(\vec x)\,\big) \Pd d (q,\vec x) +
  \sum_{\ell=1}^p \underbrace{(-1)^{d+1+h+\ell}}_{=-(-1)^{p+1+\ell}}
  q^{s_{\ell}+d-\ell}\, \chi_{\Yt{\ell}}^{\so(d)}(\vec x)\, \Pd d
  (q,\vec x) \nonumber \\ & = & (q^p - q^{d-p})\,
  \chi^{\so(d)}_{\Y_p}(\vec x)\, \Pd d (q,\vec x) - \sum_{\ell=1}^p
  (-1)^{p+1+\ell} q^{s_{\ell}+d-\ell}\, \chi_{\Yt{\ell}}^{\so(d)}(\vec
  x)\, \Pd d (q,\vec x) \nonumber \\ && \qquad \qquad +
  \sum_{m=1}^{r-p} (-1)^{m} (q^{p+m} - q^{d-p-m})
  \chi_{(\Y_p,\1^m)}^{\so(d)}(\vec x) \, \Pd d (q,\vec x)
\end{eqnarray}
where we used $i=k-p$, $j=d-p-k'$ and $\ell=d+1-k''$ when going from
the second to the third equality. \\

Turning to the $d=2r$ case, we now have the following series of
weights:
\begin{equation}
  \left\{
  \begin{aligned}
    (\lambda)_{-n} & = (s_{r-n+1}-1-(r-n), \Yt{r-n+1}), & \ n=r,
    \dots, r-p+1\\ (\lambda)_{-n} & = (-(r-n), \Y_h, \1^{r-n-p}), &
    \ n=r-p, \dots, 1\\ (\lambda)_0 & = (-r, \Y_p, \1^{r-p}_+), &
    (\lambda)_{0'} = (-r, \Y_p, \1^{r-p}_-) \\ (\lambda)_n & =
    (-(r+n), \Y_p, \1^{r-n-p}), & \ n=1, \dots, r-p \\ (\lambda)_n & =
    (-s_{r-n+1}-(r+n)+1, \Yt{r-n+1}), & \ n=r-p+1, \dots, r
  \end{aligned}
  \right.
\end{equation}
where $\1_\pm^m$ denote the $m$ last components of the $\so(2r)$
weight, these components all being egal to $1$ except for the last one
which can be $\pm 1$.\\

Using \eqref{char_STV_even}, we can write the character of the
irreducible module corresponding to the highest-weight
$(\lambda)_{-(r-p)}$ which we identified as the character of the
exceptional series in odd spacetime dimension:
\begin{eqnarray}
  \chi_{(\lambda)_{-(r-p)}}(q, \vec x) & = & \sum_{n=0}^{r-p}
  (-)^{r-p+n} \Big( \cY_{(\lambda)_n}(q, \vec x) +
  \cY_{(\lambda)_{-n}}(q, \vec x) \Big)\, \Pd d (q, \vec x) \\ &&
  \qquad \qquad \qquad + 2 \sum_{n=r-p+1}^r (-)^{r-p+n}
  \cY_{(\lambda)_n}(q, \vec x) \, \Pd d (q, \vec x) \nonumber \\ & = &
  \sum_{n=1}^{r-p} (-)^{r-p+n} (q^{r+n}+q^{r-n}) \chi^{\so(d)}_{(\Y_p,
    \1^{r-p-n})}(\vec x)\, \Pd d (q, \vec x) \nonumber \\ && \qquad +
  (-)^{r-p}\, q^r \Big( \chi^{\so(d)}_{(\Y_p, \1^{r-p}_+)}(\vec x) +
  \chi^{\so(d)}_{(\Y_p, \1^{r-p}_-)}(\vec x) \Big)\, \Pd d (q, \vec x)
  \nonumber \\ && \qquad \qquad + 2 \sum_{n=r-p+1}^r (-)^{r-p+n}
  q^{s_{r-n+1}+r+n-1} \chi^{\so(d)}_{\Yt{r-n+1}}(\vec x)\, \Pd d (q,
  \vec x) \\ & = & \sum_{n=0}^{r-p-1} (-)^{n} (q^{d-p-n}+q^{p+n})
  \chi^{\so(d)}_{(\Y_p, \1^{n})}(\vec x)\, \Pd d (q, \vec x) + 2
  \sum_{\ell=1}^h (-)^{p+\ell+1} q^{s_{\ell}+d-\ell}
  \chi^{\so(d)}_{\Yt{\ell}}(\vec x)\, \Pd d (q, \vec x) \nonumber \\ &&
  \qquad \qquad + (-)^{r-p}\, q^{d/2} \Big( \chi^{\so(d)}_{(\Y_p,
    \1^{r-p}_+)}(\vec x) + \chi^{\so(d)}_{(\Y_p, \1^{r-p}_-)}(\vec x)
  \Big)\, \Pd d (q, \vec x)
\end{eqnarray}

\paragraph{Identifying the discrete series}
Starting in $d=2r+1$ from:
\begin{equation}
  (\lambda)_0 = (\lambda_0, \lambda_1, \dots, \lambda_r) = (s_1-1,
  \Y_{r,k})\, ,
\end{equation}
with
\begin{equation}
  \Y_{r,k} = (s_2-1, s_3-1, \dots, s_r-1, k-1)\, ,
\end{equation}
leads to the following sequence of weights:
\begin{equation}
  (\lambda)_{r+1} = (-k-r, s_1, \dots, s_r)\, ,
\end{equation}
and 
\begin{equation}
  (\lambda)_{r+K} = (-s_{r+2-K}-r-K+1, s_1, \dots, s_{r+1-K},
  s_{r+3-K}-1, \dots, s_r-1, k-1)\, , \qquad K=2, \dots, r+1\,.
\end{equation}
It turns out that the character corresponding to the irreducible model
at the level $r+1$ in this sequence, computed with
\eqref{char_STV_odd}, exactly reproduces the one given by Hirai in
\cite{Hirai1965} for the direct sum of two discrete series
representations based on the highest-weight vector $\vec s = (s_1,
\dots, s_r)$ and whose conformal weight is determined by the integer
$k$:
\begin{eqnarray}
  \chi_{(\lambda)_{r+1}}(q, \vec x) & = & \sum_{j=0}^r (-1)^j
  \cY_{(\lambda)_{r+1+j}}(q, \vec x)\, \Pd d (q, \vec x) \nonumber
  \\ & = & q^{k+r} \chi^{\so(d)}_{\vec s}(\vec x) \Pd d (q, \vec x) +
  \sum_{j=1}^r (-1)^j \, q^{s_{r+1-j}+r+j}\, \chi^{\so(d)}_{(s_1, \dots,
    s_{r-j}, s_{r+2-j}-1, \dots, s_r-1, k-1)}(\vec x) \, \Pd d (q,
  \vec x) \nonumber \\ & = & q^{k+r} \chi^{\so(d)}_{\vec s}(\vec x) \Pd
  d (q, \vec x) + \sum_{i=1}^r (-1)^{r+1+i}\, q^{s_{i}+d-i}\,
  \chi^{\so(d)}_{\check{\Y}_{\vec s, k}^{(i)}}(\vec x) \, \Pd d (q, \vec
  x)
  \label{char_dis}
\end{eqnarray}
where we introduced the notation $\check{\Y}_{\vec s, k}^{(i)} = (s_1,
\dots, s_{i-1}, s_{i+1}-1,\dots, s_r-1,k-1)$ in the last line.

\section{Poincar\'e characters revisited}\label{app:poincare}

As clearly recalled in \cite{Campoleoni:2015qrh, Oblak:2016eij}, the
characters of the Poincar\'e group $\text{ISO}(1,d) = \text{SO}(1,d)
\ltimes \R^{d+1}$ follow from Frobenius formula for semi-direct
product groups:
\begin{equation}\label{Frobenius}
  \chi[(\Lambda, \alpha)] = \int_{\cO_p} d\mu(k) \, \delta_\mu(k,
  \Lambda \cdot k) \, e^{i\langle k, \alpha \rangle} \,
  \chi_{\mathcal{R}}(g^{-1}_k \Lambda g_k)
\end{equation}
where $(\Lambda,\alpha) \in \ISO(1,d)$, with $\Lambda \in \SO(1,d)$
and $\alpha \in \R^{d+1}$, is a generic element of the Poincar\'e
group.  The integral \eqref{Frobenius} is defined over the orbit of
the momentum $p\in (\R^{d+1})^*$:
\begin{equation}
  \cO_p = \left\{\Lambda \cdot p\, \rvert\, \Lambda \in \SO(1,d)
  \right\} \subset (\R^{d+1})^*
\end{equation}
In the integral \eqref{Frobenius}, the symbols $d\mu(k)$ and
$\delta_\mu(k,k')$ denote, respectively, the invariant measure on
$\cO_p$ and the associated Dirac distribution, $\chi_{\mathcal{R}}$ is
the character of an irreducible representation $\mathcal{R}$ of the
little group labeled in what follows by the highest-weight $\vec s$,
and $\langle k, \alpha \rangle:=k_\mu\alpha^\mu$. The map
\begin{equation}
  g \, :\, \cO_p \longrightarrow \SO(1,d) \,:\, q \longmapsto g_q
\end{equation}
is such that $g_q \cdot p = q\, , \, \forall q \in \cO_p$. Notice that
when integrating over the orbit $\cO_p$, because of the delta function
forcing $\Lambda$ to be an element of the little group of $p$,
$g_k^{-1} \Lambda g_k$ runs through the equivalence class of such
elements.

\subsection{Massive representations}

In this case, the orbit is $\cO_p = \left\{ k \in (\R^{d+1})^* \rvert
-m^2 = \eta_{\mu \nu} k^\mu k^\nu \right\}$. The corresponding little
group is $\SO(d)$. The mass-$m$ spin-$\vec s$ massive UIR will be
denoted $[m; \vec s\,]$.

When $d=2r$, we can take $\Lambda$ of the form:
\begin{equation}
  \Lambda =
  \begin{pmatrix}
    1 & 0 & \dots & 0 & 0 \\ 0 & R(\theta_1) & 0 & \dots & 0 \\ \vdots
    & 0 & \ddots & 0 & \vdots \\ 0 & \vdots & 0 & R(\theta_{r-1})
    & 0 \\ 0 & 0 & \dots & 0 & R(\theta_r) \\
  \end{pmatrix}
\end{equation}
where the matrices $R(\theta_i), \, i = 1, \dots, r-1$ are
usual $\SO(2)$ elements:
\begin{equation}
  R(\theta_i) = 
  \begin{pmatrix}
    \cos(\theta_i) & - \sin(\theta_i)\\
    \sin(\theta_i) & \cos(\theta_i)
  \end{pmatrix}
\end{equation}
When $d=2r+1$ however, we will consider an element $\Lambda$ of the form:
\begin{equation}
  \Lambda =
  \begin{pmatrix}
    1 & 0 & \dots & 0 & 0 \\
    0 & R(\theta_1) & 0 & \dots & 0 \\ 
    \vdots & 0 & \ddots & 0 & \vdots \\
    0 & \vdots & 0 & R(\theta_{r-2}) & 0 \\
    0 & 0 & \dots & 0 & R'(\theta_r,\varphi) \\
  \end{pmatrix}
\end{equation}
with $R'(\theta_r, \varphi)$ the $\SO(3)$ matrix:
\begin{eqnarray}
  R'(\theta_r, \varphi) & = &
	  \begin{pmatrix}
    \cos(\theta_r) & -\sin(\theta_r) & 0 \\
    \sin(\theta_r) & \cos(\theta_r) & 0 \\
    0 & 0 & 1
  \end{pmatrix}
  \begin{pmatrix}
    1 & 0 & 0 \\
    0 & \cos(\varphi) & -\sin(\varphi) \\
    0 & \sin(\varphi) & \cos(\varphi)
  \end{pmatrix}
  \nonumber
  \\ & = &
  \begin{pmatrix}
    \cos(\theta_r) & -\cos(\varphi) \sin(\theta_r) & \sin(\theta_r)
    \sin(\varphi) \\ \sin(\theta_r) & \cos(\varphi) \cos(\theta_r) & -
    \cos(\theta_r) \sin(\varphi) \\ 0 & \sin(\varphi) & \cos(\varphi)
  \end{pmatrix}
\end{eqnarray}
This differs slightly from \cite{Campoleoni:2015qrh} where $\varphi =
0$ from the beginning. We believe this provides a convenient
regularisation of the character, adapted to the flat limit. We can now
compute the character:
\begin{eqnarray}
  \chi^{\Poinc}_{[m; \vec s\,]}([f,\alpha]) & = & \int_{\cO_p} d^{d}k \,
  \delta^{(d)}(\left[\id - \Lambda \right] k) e^{i\langle k , \alpha
    \rangle} \chi^{\so(d)}_{\vec s}(g^{-1}_k \Lambda g_k) \\ & = &
  e^{-\beta m} \frac{1}{\det \left| \id - \Lambda \right|} \,
  \chi^{\so(d)}_{\vec s}(\Lambda) \\ & = & e^{-\beta m} \,
  \chi^{\so(d)}_{\vec s}(\vec \theta) \, \prod_{j=1}^r \frac{1}{\left|
    1 - e^{i\theta_j} \right|^2} \, \left\{ \begin{aligned}
    1 \hspace{50pt} , & \text{ if } d = 2r
    \\ \left.\frac{1}{1-\cos\varphi}\right|_{\varphi \rightarrow 0},
    & \text{ if } d = 2r+1
  \end{aligned}
  \right.
\end{eqnarray}
where $\beta:=i\alpha_0$ and, to derive the last equality when
$d=2r+1$, we used:
\begin{eqnarray}
  \det \left| \id - R'(\theta, \varphi) \right| & = &
  (1-\cos\varphi) \left(
  (1-\cos\theta)(1-\cos\varphi\cos\theta) + \cos\varphi
  \sin^2\theta \right) \\ & & - \sin\varphi \left(
  (1-\cos\theta)\cos\theta \sin\varphi - \sin\varphi
  \sin^2\theta \right)
\end{eqnarray}
\begin{equation}
  \Rightarrow \left. \frac{1}{\det \left| \id - R'(\theta, \varphi)
    \right|} \right|_{\varphi \rightarrow 0} =
  \frac{1}{2(1-\cos\theta)} \left. \frac{1}{1-\cos \varphi}
  \right|_{\varphi \rightarrow 0} = \frac{1}{\left| 1 - e^{i\theta}
    \right|^2} \left. \frac{1}{1-\cos \varphi} \right|_{\varphi
    \rightarrow 0}
\end{equation}


\begin{remark}
  When $\varphi=0$, $\Lambda$ is an element of the Cartan subgroup of
  $\SO(d)$. At the algebra level, the character is defined as:
  \begin{equation}
    \chi_V(\mu) = \sum_{\tau \in \Delta_V} e^{\langle \tau, \,\mu
      \rangle}
    \label{lie_alg_char}
  \end{equation}
  where $\Delta_V$ is the set of weights of the representation $V$,
  $\langle \cdot , \cdot \rangle$ is the scalar product on weight
  space and $\mu$ is an arbitrary weight. For semisimple algebras, the
  weight space has the structure of an Euclidean space, therefore one
  can write $e^{\langle \tau, \,\mu \rangle} = \prod_{j=1}^r
  x_j^{\tau_j}$ where $r$ is the dimension of the weight space
  (i.e., the rank of the algebra), $\tau_j$ the $j$th component of the
  weight $\tau$ and $x_j := e^{\mu_j}$. Hence, by definition, the
  above character encodes all the weights (i.e., eigenvalues of the
  Cartan subalgebra generators when acting on vectors in $V$)
  occurring in $V$. To compare the \it group \rm character, one has to
  evaluate the latter on an element of the Cartan subgroup. The Cartan
  subalgebra being abelian, elements of the Cartan subgroup are of the
  form $\prod_{i=1}^r \exp(\theta_i H_i)$, where $H_i$ are the Cartan
  generators. Seeing the character as (a generalisation of) the trace
  of a group element, it is clear that the character of an element of
  the Cartan subgroup will coincide with the Lie algebra character
  \eqref{lie_alg_char}, upon identifying the parameter $\theta_i$ with
  the components $\mu_i$ of the weight $\mu$ on which the latter
  character is evaluated.
\end{remark}

\subsection{Massless representations}
The massless case is a bit more subtle: in this case, the little group
is the Euclidean group $\ISO(d-1)$. However, for ``discrete'' spin (or
``helicity'') representations, the translation are represented
trivially and therefore the corresponding representation of
$\ISO(d-1)$ reduces to a representation of $\SO(d-1)$. The characters
corresponding to these massless, totally symmetric, spin-$s$
representations were also computed in \cite{Campoleoni:2015qrh},
however, when deriving them, one encounters a few difficulties in the
form of divergences to be regularised. Even tough, as could be
expected, the resulting formulae essentially contain the information
about the irrep of the little group labeling these Poincar\'e UIRs in
the form of a character of $\so(d-1)$, some regularising factors
complicate the expression obtained and make their interpretation
somewhat elusive. As the authors of \cite{Campoleoni:2015qrh} pointed
out, the characters derived by the purely group theoretical approach
do not exactly coincide with the corresponding flat spacetime
partition functions computed using heat kernel method, despite the
well known fact that the two objects are identical. It turns out that
the character part of the partition functions spelled out in
\cite{Campoleoni:2015qrh} are not plagued with as severe regularising
factors as the corresponding ones obtained with the Frobenius formula
outlined previously, and on top of that, arise naturally as flat limit
of AdS$_{d+1}$ characters. Having these facts in mind, we will assume
that for massless Poincar\'e irreps, the characters are given by the
result coming from partition function calculations, which reads:
\begin{equation}
  \chi^{\Poinc}_{[0; \vec s\,]}(\beta, \vec \theta) = \prod_{j=1}^r
  \frac{1}{\left| 1 - e^{i\theta_j} \right|^2 } \left\{
  \begin{aligned}
    \sum_{k=1}^r \cA_k^{(r)}(\vec \theta)\, \chi^{\so(d-1)}_{\vec s}
    (\hat \theta_k), & \ \text{ if } d = 2r
    \\ \left. \frac{\chi^{\so(d-1)}_{\vec s}(\vec \theta)}{1 -
      \cos(\varphi)}\right|_{\varphi \rightarrow 0}, \hspace{10pt} &
    \ \text{ if } d = 2r+1\,.
  \end{aligned}
  \right.
\end{equation}
where $\hat \theta_k$ in the first line indicates that $\theta_k$ is
removed.  Making the identification $q = e^{-\beta}$ and $x_j =
e^{i\theta_j}$, we recognise in the above formula the function $\Pf d
(\vec x)$ defined in \eqref{flat_pfunc} and appearing as the flat
limit of $\Pd d (q, \vec x)$ (the factor
$\tfrac{1}{1-\cos(\varphi)}\rvert_{\varphi \rightarrow 0}$ should only
be understood as a way of treating the divergence appearing in the
expression of the character for $d=2r+1$, and as such can be traded
for $\tfrac{1}{1-q}\rvert_{q \rightarrow 1}$ appearing in the flat
limit of (A)dS$_{d+1}$ character, as both encode the same type of
divergence to be regulated). We can therefore rewrite the Poincar\'e
characters for massless irreps as:
\begin{equation}
  \chi^{\Poinc}_{[0; \vec s\,]}(q, \vec x) = \Pf d (\vec x) \left\{
  \begin{aligned}
    \sum_{k=1}^r \cA_k^{(r)}(\vec x)\,
    \chi^{\so(d-1)}_{\vec s}(\hat{\vec x}_k), & \ \text{ if } d = 2r
    \\ \chi^{\so(d-1)}_{\vec s}(\vec x) \hspace{25pt}, & \ \text{ if
    } d = 2r+1\,,
  \end{aligned}
  \right.
\end{equation}
where $\Pf d (\vec x)$ is defined by \eqref{flat_pfunc} and $\hat{\vec
  x}_k := (x_1, \dots, x_{k-1}, x_{k+1}, \dots, x_r)$.

\section{Branching rules for $\so(d)$}\label{app:branching}

In this appendix, we derive the branching rules obeyed by $\so(d)$
characters. Before doing so, let us recall the expression of $\so(d)$
characters, written in terms of the $\xi$ and $\zeta$ variables, as
well as the Vandermonde determinant $\Delta^{(r)}(\,\vec\xi\,)$, used
for instance in \cite{Fulton1991}:
\begin{equation}
  \begin{aligned}
    \zeta_i(\alpha) := x_i^{\alpha} - x_i^{-\alpha}\, & \, , \, &
    \xi_i(\alpha) := x_i^{\alpha} + x_i^{-\alpha}\, & \, , \, &
    \Delta^{(r)}(\,\vec\xi\,) = \prod_{1 \leqslant i < j \leqslant r}
    (\xi_i - \xi_j)\, .
  \end{aligned}
\end{equation}
Then, in terms of these building blocks, the characters of a $\so(d)$
irrep labelled by the highest-weight $\vec s = (s_1, \dots, s_r)$
are
\begin{itemize}
  \item For $d=2r+1$,
    \begin{equation}
      \chi_{\vec s}^{\so(2r+1)}(x_1,\dots,x_r) =
      \frac{1}{\Delta^{(r)}(\,\vec\xi\,)\, \prod_{k=1}^r
        \zeta_k(\tfrac{1}2)} \, \det\big[ \zeta_j(s_i+r-i+\tfrac{1}2)
      \big]\,;
    \end{equation}
\item For $d=2r$,
  \begin{equation}
    \chi_{\vec s}^{\so(2r)}(x_1,\dots,x_r) =
    \frac{1}{2\,\Delta^{(r)}(\,\vec\xi\,)} \Big( \det\big(
    \xi_j(s_i+r-i) \big) + \det\big( \zeta_j(s_i+r-i) \big) \Big)\,,
    \label{char_so_even}
  \end{equation}
\end{itemize}
where $\det(A_{ij})$ denote the determinant of the matrix with entries
$A_{ij}$, and indices $i,j$ run from $1$ to $r$. The branching rules
for $\so(d)$, that we want to rederive at the character level, are at
the level of irreps:
\begin{equation}
  \D_{\vec s}^{\so(2r+1)} = \bigoplus\limits_{s_1 \geqslant \lambda_1
    \geqslant s_2 \geqslant \dots \geqslant \lambda_{r-1} \geqslant
    s_r \geqslant \rvert \lambda_r \rvert} \D_{\vec\lambda}^{\so(2r)}
\end{equation}
\begin{equation}
  \D_{\vec s}^{\so(2r)} = \bigoplus\limits_{s_1 \geqslant \lambda_1
    \geqslant s_2 \geqslant \dots \geqslant \lambda_{r-1} \geqslant
    \rvert s_r \rvert} \D_{\vec\lambda}^{\so(2r-1)}
\end{equation}
In the second formula, we denoted, with a slight abuse of notation,
the $\so(2r-1)$ weight by $\vec\lambda$ altough it actually stands for
$(\lambda_1,\ldots,\lambda_{r-1})$, i.e., a vector with $r-1$
components (since this is the rank of $\so(2r-1)$\,). We will consider
the two cases separately, starting with the odd dimensional one, and
the main things we will need are the two identities gathered hereafter
in a lemma.\\

\begin{minipage}[l]{.05\textwidth}
  \begin{flushright}
    \vrule width 2pt height 90pt
  \end{flushright}
\end{minipage}
\hspace{.2pt}
\begin{minipage}[r]{.85\textwidth}
  \begin{lemma}\label{LemmaF}
    \begin{equation}
      \sum_{\mu=\lambda}^{\lambda'} \zeta(\mu+\alpha) =
      \zeta^{-1}(\tfrac{1}2) \Big( \xi( \lambda' + \alpha + \tfrac{1}2
      )-\xi( \lambda + \alpha - \tfrac{1}2 ) \Big)
    \end{equation}
    
    \begin{equation}
      \sum_{\mu=\lambda}^{\lambda'} \xi(\mu+\alpha) =
      \zeta^{-1}(\tfrac{1}2) \Big( \zeta( \lambda' + \alpha +
      \tfrac{1}2 ) - \zeta( \lambda + \alpha - \tfrac{1}2 ) \Big)
    \end{equation}
  \end{lemma}
\end{minipage}\\

\subsection{$\so(2r+1) \downarrow \so(2r)$}
In odd dimensions, the branching rules at the character level reads:
\begin{equation}
  \chi^{\so(2r+1)}_{\vec s}(\vec x) = \sum_{\lambda_1=s_2}^{s_1} \dots
  \sum_{\lambda_{r-1}=s_r}^{s_{r-1}} \sum_{\lambda_r=-s_r}^{s_r}
  \chi^{\so(2r)}_{\vec\lambda}(\vec x)
  \label{odd_branching}
\end{equation}

\begin{proof}
  Having in mind the character for the $\so(2r+1)$ irrep $\vec s$:
  \begin{equation}
    \chi_{\vec s}^{\so(2r+1)}(\vec x) = \frac{1}{\Delta^{(r)}(\, \vec
      \xi \, )} \sum_{\sigma \in \cS_r} \varepsilon(\sigma)
    \prod_{i=1}^r \zeta_{\sigma(i)}^{-1}(\tfrac{1}2)\,
    \zeta_{\sigma(i)}(s_i+r-i+\tfrac{1}2)\, ,
  \end{equation}
  with $\varepsilon(\sigma)$ the signature of the permutation
  $\sigma$, let us rewrite the sum of $\so(2r)$ of the irreps
  appearing in the branching rule of $\vec s$:
  \begin{eqnarray}
    \sum_{\lambda_1=s_2}^{s_1} \dots\sum_{\lambda_{r-1}=s_r}^{s_{r-1}}
    \sum_{\lambda_r=-s_r}^{s_r} \chi_{\vec\lambda}^{\so(2r)}(\vec x) &
    = & \frac{1}{2\,\Delta^{(r)}(\,\vec\xi\,)} \sum_{\sigma \in \cS_r}
    \varepsilon(\sigma) \prod_{i=1}^r \left(
    \sum_{\lambda_r=-s_r}^{s_r} \dots \sum_{\lambda_1=s_2}^{s_1}
    \xi_{\sigma(i)}(\lambda_i+r-i) \right) \nonumber \\ =
    \frac{\prod_{k=1}^r
      \zeta_k^{-1}(\tfrac{1}2)}{2\,\Delta^{(r)}(\,\vec\xi\,)}
    \ \sum_{\sigma \in \cS_r} & \varepsilon(\sigma) &
    \prod_{i=1}^{r-1} \Big( \zeta_{\sigma(i)}(s_i+r-i+\tfrac{1}2) -
    \zeta_{\sigma(i)}(s_{i+1}+r-i-\tfrac{1}2) \Big) \\ && \qquad
    \qquad \qquad \times \Big( \zeta_{\sigma(r)}(s_r+\tfrac{1}2) -
    \zeta_{\sigma(r)}(-s_r-\tfrac{1}2)\Big)\,, \nonumber
  \end{eqnarray}
  where we used Lemma \ref{LemmaF} to obtain the final line. Notice
  that $\zeta(-x) = -\zeta(x)$, as follows from the definition,
  therefore only the determinant involving the variables $\xi$ in
  \eqref{char_so_even} survives in the expression of the $\so(2r)$
  characters in the sum to begin with, and the last factor above
  becomes $2\,\zeta_{\sigma(r)}(s_r+\tfrac{1}2)$. Finally, only the
  term $\prod_{i=1}^r \zeta_{\sigma(r)}(s_i+r-i+\tfrac{1}2)$ in the
  product of the previous expression remains. Indeed, the second terms
  of the difference inside this product, namely
  $\zeta_{\sigma(i)}(s_{i+1}+r-i-\tfrac{1}2)$ gives rise to terms of
  the form $\zeta_{\sigma(i)}(s_{i+1}+r-i-\tfrac{1}2)
  \zeta_{\sigma(i+1)}(s_{i+1}+r-i-\tfrac{1}2)$ when expanding the
  product. These terms will automatically be cancelled when summing
  over all permutations, as there will always be two permutations
  $\sigma$ and $\sigma'$ such that for a fixed $i \in \{ 1,2,\cdots, r
  \}$, $\sigma(i)=\sigma'(i+1)\, , \, \sigma(i+1)=\sigma'(i)$ and
  $\sigma(j)=\sigma'(j)\, , \, \forall j \neq i$, and by definition
  the signature of $\sigma$ and $\sigma'$ differ by a minus sign. As a
  consequence, the above equation yields:
  \begin{eqnarray}
    \sum_{\lambda_1=s_2}^{s_1} \dots \sum_{\lambda_r=-s_r}^{s_r}
    \chi_{\vec\lambda}^{\so(2r)}(\vec x) & = &
    \frac{1}{\Delta^{(r)}(\,\vec\xi\,)\, \prod_{k=1}^r
      \zeta_k(\tfrac{1}2)} \ \sum_{\sigma \in \cS_r}
    \varepsilon(\sigma) \prod_{i=1}^r
    \zeta_{\sigma(i)}(s_i+r-i+\tfrac{1}2) = \chi^{\so(2r+1)}_{\vec
      s}(\vec x)\,, \quad
  \end{eqnarray}
  which proves \eqref{odd_branching}.
\end{proof}

\begin{example}
  Consider the simple, low rank, case of $\so(5) \downarrow \so(4)$:
  \begin{itemize}
  \item On the one hand,
    \begin{equation}
      \chi^{\so(5)}_{(s,t)}(x_1, x_2) = \frac{1}{\Delta^{(2)}(\vec
        \xi) \zeta_1(\tfrac{1}2) \zeta_2(\tfrac{1}2)} \left(
      \zeta_1(s+\tfrac{3}2) \zeta_2(t+\tfrac{1}2) -
      \zeta_1(t+\tfrac{1}2) \zeta_2(s+\tfrac{3}2) \right)
    \end{equation}
  \item On the other hand,
    \begin{eqnarray}
      \sum_{\sigma=t}^s \sum_{\tau=-t}^t \chi^{\so(4)}_{(\sigma,
        \tau)}(x_1, x_2) & = & \frac{1}{2\,\Delta^{(2)}(\vec \xi)}
      \sum_{\sigma=t}^s \sum_{\tau=-t}^t \big(\xi_1(\sigma+1)
      \xi_2(\tau) - \xi_2(\sigma+1) \xi_1(\tau)\big) \\ & = &
      \frac{1}{2\,\Delta^{(2)}(\vec \xi) \zeta_1(\tfrac{1}2)
        \zeta_2(\tfrac{1}2)} \Big[ \left( \zeta_1(s+\tfrac{3}2) -
        \zeta_1(t+\tfrac{1}2) \right) \left( \zeta_2(t+\tfrac{1}2)
        -\zeta_2(-t-\tfrac{1}2) \right) \nonumber \\ && \qquad \qquad
        - \left( \zeta_1(t+\tfrac{1}2) -\zeta_1(-t-\tfrac{1}2)
        \right)\left( \zeta_2(s+\tfrac{3}2) - \zeta_2(t+\tfrac{1}2)
        \right)\Big]
    \end{eqnarray}
  \end{itemize}
  The terms $\zeta_1(t+\tfrac{1}2) \zeta_2(t+\tfrac{1}2)$ cancel, and
  using $\zeta(-x) = -\zeta(x)$, we are left with:
  \begin{equation}
    \chi^{\so(5)}_{(s,t)}(x_1, x_2) = \sum_{\sigma=t}^s
    \sum_{\tau=-t}^t \chi^{\so(4)}_{(\sigma, \tau)}(x_1, x_2)
  \end{equation}
\end{example}

\subsection{$\so(2r) \downarrow \so(2r-1)$}
In even dimensions, the branching rule at the character level reads:
\begin{equation}
  \chi^{\so(2r)}_{\vec s_+}(\vec x) + \chi^{\so(2r)}_{\vec s_-}(\vec
  x) = \sum_{k=1}^r \cA_{k, \vec s}^{(r)}(\vec x) \,
  \sum_{\lambda_1=s_2}^{s_1} \dots \sum_{\lambda_{r-1}=s_r}^{s_{r-1}}
  \chi^{\so(2r-1)}_{\vec\lambda} (\hat{\vec x}_k)\, ,
  \label{even_branching}
\end{equation}
with $\vec s_\pm = (s_1, \dots, \pm s_r)$, $\hat{\vec x}_k := (x_1,
\dots, x_{k-1}, x_{k+1}, \dots, x_r)$ and
\begin{equation}
  \cA_{k, \vec s}^{(r)}(x_1, \dots, x_r) := \xi_k(s_r)
  \frac{\Delta^{(r)}(\,\vec\xi\,)
    \big|_{\xi_k=2}}{\Delta^{(r)}(\,\vec\xi\,)}
\end{equation}
Notice that when $s_r = 0$, $\xi_k(s_r) = 2$ and the above
identity reduces to a statement involving only one character.

\begin{proof}
   We can rewrite the sum of the characters for $\vec s_+$ and $\vec
   s_-$, by explicitly expanding the only remaining determinant, the
   one involving the variables $\xi$. Indeed, either the last
   component $s_r$ of $s_\pm$ vanishes, hence a whole column in
   $\det(\zeta_j(s_r+r-i))$ vanishes and thereby the whole determinant
   vanishes; or the two determinants involving the variables $\zeta$
   will cancel each other as the last column of one will be
   $\zeta_i(s_r)$ and $\zeta_i(-s_r) = -\zeta_i(s_r)$ for the other
   one. The resulting sum of characters reads:
  \begin{eqnarray}
    \chi_{\vec s_+}^{\so(2r)}(\vec x) +
    \chi_{\vec s_-}^{\so(2r)}(\vec x) & = & \frac{1}{\Delta^{(r)}(\,
      \vec \xi\, )}
    \begin{vmatrix}
      \xi_1(s_1+r-1) & \dots & \xi_1(s_{r-1}+1) & \xi_1(s_r) \\ \vdots
      & \ddots & \vdots & \vdots \\ \xi_r(s_1+r-1) & \dots &
      \xi_r(s_{r-1}+1) & \xi_r(s_r)
    \end{vmatrix}
    \\ & = & \frac{1}{\Delta^{(r)}(\, \vec \xi \, )}\, \sum_{k=1}^r
    (-)^{k+r} \xi_k(s_r)\, \left( \sum_{\sigma \in \cS_{r-1}}
    \varepsilon(\sigma) \prod_{i \in \{ 1,2,\cdots, r\}\,,\, i \neq k}
    \xi_{\sigma(i)}(s_i+r-i) \right), \quad \nonumber
  \end{eqnarray}
  One the other hand, the sum of the $\so(2r-1)$ characters
  corresponding to the irreps appearing in the branching rule of
  $\vec s_\pm$ reads:
  \begin{eqnarray}
    \sum_{\lambda_{r-1} = 0}^{s_{r-1}} \sum_{\lambda_{r-2} =
      s_{r-1}}^{s_{r-2}} \dots \sum_{\lambda_1 = s_2}^{s_1}
    \chi_{\vec\lambda}^{\so(2r-1)}(\vec x) & = & \nonumber
    \\ && \hspace{-120pt} \frac{1}{\Delta^{(r-1)}(\, \vec \xi \, )}
    \sum_{\sigma \in \cS_{r-1}} \varepsilon(\sigma) \prod_{i=1}^{r-1}
    \zeta_{\sigma(i)}^{-1}(\tfrac{1}2) \sum_{\lambda_{r-1} =
      0}^{s_{r-1}} \sum_{\lambda_{r-2} = s_{r-1}}^{s_{r-2}} \dots
    \sum_{\lambda_1 = s_2}^{s_1}
    \zeta_{\sigma(i)}(\lambda_i+r-\tfrac{1}2) \\ && \hspace{-120pt} =
    \frac{1}{\Delta^{(r-1)}(\, \vec \xi \, )} \sum_{\sigma \in
      \cS_{r-1}} \varepsilon(\sigma) \prod_{i=1}^{r-1}
    \zeta_{\sigma(i)}^{-2}(\tfrac{1}2) \Big( \xi_{\sigma(i)}(s_i+r-i)
    - \xi_{\sigma(i)}(s_{i+1}+r-[i+1]) \Big) \nonumber
  \end{eqnarray}
  At this point, one can notice the following identities:
  \begin{equation}
    \zeta^2(\tfrac{1}2) = (x^{1/2}-x^{-1/2})^2 = x + x^{-1} - 2 =
    \xi(1) - 2\, ,
  \end{equation}
  and
  \begin{eqnarray}
    \Delta^{(r)}(\vec \xi)\rvert_{\xi_k=2} & = & \prod_{1\leqslant i <
      j \leqslant r\, ,\, i,j \neq k} (\xi_i - \xi_j)
    \prod_{n=1}^{k-1} (\xi_n-2) \prod_{m=k+1}^r (2-\xi_m) \nonumber
    \\ & = & (-)^{r+k} \prod_{1\leqslant i \leqslant r\, ,\, i \neq k}
    (\xi_i-2) \, \Delta^{(r-1)}(\hat{\vec \xi}_k) = (-)^{k+r}
    \prod_{1\leqslant i \leqslant r\, ,\, i \neq k}
    \zeta_i^2(\tfrac{1}2) \, \Delta^{(r-1)}(\hat{\vec \xi}_k)\,.
    \label{trick}
  \end{eqnarray}
  The above sum of $\so(2r-1)$ can therefore be rewritten as:
  \begin{eqnarray}
    \sum_{\lambda_{r-1} = 0}^{s_{r-1}} \sum_{\lambda_{r-2} =
      s_{r-1}}^{s_{r-2}} & \dots & \sum_{\lambda_1 = s_2}^{s_1}
    \chi_{\vec \lambda}^{\so(2r-1)}(\vec x) = \\ &&
    \frac{1}{\Delta^{(r)}(\vec \xi)|_{\xi_r=2}} \sum_{\sigma \in
      \cS_{r-1}} \varepsilon(\sigma) \prod_{i=1}^{r-1} \left(
    \xi_{\sigma(i)}(s_i+r-i) - \xi_{\sigma(i)}(s_{i+1}+r-[i+1])
    \right)\,. \nonumber
  \end{eqnarray}
  Now taking the following linear combination, and using
  \eqref{trick}:
  \begin{eqnarray}
    \sum_{k=1}^r \frac{\Delta^{(r)}(\vec
      \xi)\rvert_{\xi_k=2}}{\Delta^{(r)}(\vec \xi)} \, \xi_k(s_r)
    \sum_{\lambda_{r-1} = 0}^{s_{r-1}} \dots \sum_{\lambda_1 =
      s_2}^{s_1} \chi_{\vec \lambda}^{\so(2r-1)}(\hat{\vec x}_{k}) & =
    & \\ && \hspace{-200pt} \frac{1}{\Delta^{(r)}(\vec \xi)}\,
    \sum_{k=1}^r (-)^{k+r} \xi_k(s_r) \sum_{\sigma \in \cS_{r-1}}
    \varepsilon(\sigma) \prod_{i\in \{ 1,2,\cdots, r \}\,,\, i \neq k}
    \left( \xi_{\sigma(i)}(s_i+r-i) - \xi_{\sigma(i)}(s_{i+1}+r-[i+1])
    \right)\,, \nonumber
  \end{eqnarray}
  one can easily recognise the sum $\chi_{\vec s_+}^{\so(2r)}(\vec x)
  + \chi_{\vec s_-}^{\so(2r)}(\vec x)$ by isolating the contribution
  $\prod_{i\in \{ 1,\cdots, r \}\,,\, i \neq k}
  \xi_{\sigma(i)}(s_i+r-i)$ in the expansion of the final product. It
  turns out that all the other terms in this expansion cancel one
  another for the same reason as in the previous case $\so(2r+1)
  \downarrow \so(2r)$: the remaining terms are of the form
  $\xi_{\sigma(i)}(k_j)\xi_{\sigma(l)}(k_j)$ and one can check that
  there will always be two permutations $\sigma$ and $\sigma'$ only
  exchanging $i$ and $l$ and whose signature differs by a minus sign.
\end{proof}

\begin{example}
  Consider the simple, low rank, case of $\so(4) \downarrow \so(3)$:
  \begin{itemize}
  \item On the one hand:
    \begin{eqnarray}
      \chi_{(s,t)}^{\so(4)}(x_1,x_2) + \chi_{(s,-t)}^{\so(4)}(x_1,x_2)
      & = & \frac{1}{\Delta^{(2)}(\,\vec\xi\,)} \big( \xi_1(s+1)\,
      \xi_2(t) - \xi_1(t)\, \xi_2(s+1) \big) \\ & = &
      \frac{1}{\Delta^{(2)}(\,\vec\xi\,)} \big( \left[ \xi_1(s+1) -
        \xi_1(t) \right]\, \xi_2(t) - \xi_1(t)\, \left[ \xi_2(s+1) -
        \xi_2(t) \right] \big) \nonumber
    \end{eqnarray}
  \item On the other hand:
    \begin{equation}
      \chi_{(\sigma)}^{\so(3)}(x) =
      \frac{\zeta(\sigma+\tfrac{1}2)}{\zeta(\tfrac{1}2)} \Rightarrow
      \sum_{\sigma=t}^s \chi_{(\sigma)}^{\so(3)}(x) =
      \zeta^{-2}(\tfrac{1}2) \big( \xi(s+1) - \xi(t) \big)\, .
    \end{equation}
  \end{itemize}
  Putting this altogether, we end up with:
  \begin{eqnarray}
    \chi_{(s,t)}^{\so(4)}(x_1,x_2) + \chi_{(s,-t)}^{\so(4)}(x_1,x_2) =
    \sum_{\sigma=t}^s
    \tfrac{\Delta^{(2)}(\,\vec\xi\,)_{\xi_1=2}}{\Delta^{(2)}(\,\vec\xi\,)}
    \xi_2(t)\, \chi_{(\sigma)}^{\so(3)}(x_2) +
    \tfrac{\Delta^{(2)}(\,\vec\xi\,)_{\xi_2=2}}{\Delta^{(2)}(\,\vec\xi\,)}
    \xi_1(t)\, \chi_{(\sigma)}^{\so(3)}(x_1)
  \end{eqnarray}
\end{example}

\section{Non-unitary mixed-symmetry massless fields}\label{app:nonunitary}

In this appendix we spell out the characters corresponding to the
non-unitary massless mixed-symmetry fields in both de Sitter and
anti-de Sitter, and comment on their flat limit as well.

\subsection{Anti-de Sitter case}
As shown by Metsaev \cite{Metsaev:1995re, Metsaev:1997nj}, massless
mixed-symmetry fields in anti-de Sitter are unitary only when their
first block is ``activated'' by gauge transformations, i.e., its gauge
parameter takes values in the $\so(d)$ Young diagram obtained by
removing one box in the last row of the first block of the Young
diagram of the gauge field. However, if one ignores the unitarity of
the representation and is only interested in its irreducibility, all
intermediary block of the gauge field's Young diagram can be
activated.

The conformal weight of a gauge field with symmetry $\Y =
(\ell_1^{h_1}, \dots, \ell_B^{h_B})$ (a diagram with $B$ blocks of
respective lengths $\ell_I$) whose $I$th block is activated is
$\Delta_I := \ell_I + d - p_I - 1$, where $p_I := \sum_{J=1}^I h_J$ is
the cumulated height of the $I$ first blocks. This $\so(2) \oplus
\so(d)$ highest-weight can be found in the BGG sequences for $\so(2,d)$
detailed in \hyperref[app:char]{Appendix \ref{app:char}} at level
$(\lambda)_{d-p_I}$ for $d=2r+1$ and at level $(\lambda)_{r-p_I}$ for
$d=2r$. Using the general formulae derived in
\hyperref[app:char]{Appendix \ref{app:char}}, we can therefore write
down the corresponding character:
\begin{equation}\label{nonuniteq1}
  \chi^\AdS_{[\Delta_I, \Y]}(q, \vec x) = q^{\Delta_I}
  \chi^{\so(d)}_{\Y} (\vec x)\, \Pd d (\vec x) + \sum_{k=1}^{p_I}
  (-1)^{p_I+k+1} q^{\ell_k+d-k}
  \chi^{\so(d)}_{\check{Y}^{(k)}_I}(\vec x)\, \Pd d (\vec x)\, .
\end{equation}
where $\check{\Y}^{(k)}_I$ is obtained from $\Y$ by (i) adding an
additional row to the $I$th block (of the same length, i.e., $\ell_I$)
and (ii) deleting the $k$th row in this new diagram together with
removing one box in each of the rows under the one just removed and
until the $p_I$th (i.e., the end of the $I$th block). More explicitly:
\begin{equation}
  \check{\Y}^{(k)}_I = (s_1, \dots, s_{k-1}, \pos{s_{k+1} - 1}{k\th},
  \dots, s_{p_I}-1,\pos{s_{p_I}-1}{p_I\th}, s_{p_I+1},\dots, s_r)\, .
\end{equation}
Again, when taking the flat limit of these characters one obtains
first a sequence of $\so(d)$ irreps to branch onto $\so(d-1)$,
corresponding to the Young diagrams of the massless fields, its gauge
parameter and its reducibility. This precise combination of
$\so(d)$-irreps in \eqref{nonuniteq1} is such that, when branched onto
$\so(d-1)$, it produces all possible Young diagrams resulting from the
branching rule of the gauge field Young diagram where the block
activated by the gauge symmetry is left untouched. The proof is
identical to the analysis performed in
\hyperref[proof_exc_odd]{Subsection \ref{proof_exc_odd}} when deriving
the flat limit of exceptional series UIRs, and therefore we will not
reproduce it here. The spectrum of massless fields produced by taking
the flat limit of a single non-unitary mixed-symmetry field with Young
diagram $\Y=(\ell_1^{h_1}, \dots, \ell_B^{h_B})$ in AdS$_{d+1}$ whose
$I$th block is touched by gauge symmetry is therefore:
\begin{eqnarray}
  \Sp(\Y) & = & \Big\{ \Y' = (\ell_1^{h_1-1}, \ell_1 - n_1, \dots,
  \ell_{I-1}^{h_{I-1}}, \ell_{I-1}-n_{I-1}, \ell_I^{h_I},
  \ell_{I+1}^{h_{I+1}}, \ell_{I+1}-n_{I+1}, \dots,
  \ell_B^{h_B-1},\ell_B-n_B)\, , \nonumber \\ && \qquad \qquad \qquad
  \qquad 0 \leqslant n_i \leqslant s_i - s_{i+1}\, , i \in
  \{1,2,\cdots, B \}\, , \, i \neq I\Big\}\, .
\end{eqnarray}

\begin{example}
  Let us consider a gauge field with mixed-symmetry given by
  $\Yboxdim{4pt} \Ylinethick{0.8pt} \Y = \gyoung(;;;,;;,;;,;)$ whose
  second block is activated, thus with conformal weight $\Delta_{I=2}
  = d-2$ since $\ell_2=2$ and $p_2=3$. Its character reads:
  \begin{eqnarray}
    \chi^\AdS_{[d-2;\ \Yboxdim{3pt} \Ylinethick{0.8pt}
        \gyoung(;;;,;;,;;,;)\,]}(q, \vec x) & = & \Big( q^{d-2}
    \chi^{\so(d)}_{\Yboxdim{3pt} \Ylinethick{0.8pt}
      \gyoung(;;;,;;,;;,;)} (\vec x) - q^{d-1}
    \chi^{\so(d)}_{\Yboxdim{3pt} \Ylinethick{0.8pt}
      \gyoung(;;;,;;,;,;)} (\vec x) + q^{d}
    \chi^{\so(d)}_{\Yboxdim{3pt} \Ylinethick{0.8pt}
      \gyoung(;;;,;,;,;)}(\vec x) - q^{d+2}
    \chi^{\so(d)}_{\Yboxdim{3pt} \Ylinethick{0.8pt}
      \gyoung(;,;,;,;)}(\vec x) \Big) \Pd d (q, \vec x) \\ & \flimit
    & \Big( \chi^{\so(d)}_{\Yboxdim{3pt} \Ylinethick{0.8pt}
      \gyoung(;;;,;;,;;,;)} (\vec x) - \chi^{\so(d)}_{\Yboxdim{3pt}
      \Ylinethick{0.8pt} \gyoung(;;;,;;,;,;)} (\vec x) +
    \chi^{\so(d)}_{\Yboxdim{3pt} \Ylinethick{0.8pt}
      \gyoung(;;;,;,;,;)}(\vec x) - \chi^{\so(d)}_{\Yboxdim{3pt}
      \Ylinethick{0.8pt} \gyoung(;,;,;,;)}(\vec x) \Big) \Pf d (\vec
    x)
  \end{eqnarray}
  Branching all diagrams appearing in the previous formula (and
  sorting the result by lexicographic ordering):
  \begin{eqnarray}
    \Yboxdim{5pt} \Ylinethick{0.8pt} \gyoung(;;;,;;,;;,;) & \branching
    & \Yboxdim{5pt} \Ylinethick{0.8pt} \gyoung(;;;,;;,;;,;)\oplus
		\gyoung(;;;,;;,;;) \oplus \gyoung(;;;,;;,;,;) \oplus
    \gyoung(;;;,;;,;)\oplus
    \gyoung(;;,;;,;;,;) \oplus \gyoung(;;,;;,;;)\oplus
    \gyoung(;;,;;,;,;) \oplus
     \gyoung(;;,;;,;)   \\ \Yboxdim{5pt} \Ylinethick{0.8pt}
    \gyoung(;;;,;;,;,;) & \branching & \Yboxdim{5pt}
    \Ylinethick{0.8pt} \gyoung(;;;,;;,;,;) \oplus \gyoung(;;;,;;,;)
		\oplus \gyoung(;;;,;,;,;) \oplus
    \gyoung(;;;,;,;)
		 \oplus \gyoung(;;,;;,;,;)
     \oplus \gyoung(;;,;;,;) \oplus
    \gyoung(;;,;,;,;)  \oplus \gyoung(;;,;,;) \\ \Yboxdim{5pt}
    \Ylinethick{0.8pt} \gyoung(;;;,;,;,;) & \branching & \Yboxdim{5pt}
    \Ylinethick{0.8pt} \gyoung(;;;,;,;,;) \oplus \gyoung(;;;,;,;)
    \oplus \gyoung(;;,;,;,;) \oplus \gyoung(;;,;,;) \oplus
    \gyoung(;,;,;,;) \oplus \gyoung(;,;,;) \\ \Yboxdim{5pt}
    \Ylinethick{0.8pt} \gyoung(;,;,;,;) &\branching & \Yboxdim{5pt}
    \Ylinethick{0.8pt} \gyoung(;,;,;,;) \oplus \gyoung(;,;,;)
  \end{eqnarray}
  One is therefore left with:
  \begin{eqnarray}
    \chi^\AdS_{[d-1;\ \Yboxdim{3pt} \Ylinethick{0.8pt}
        \gyoung(;;;,;;,;;,;)\,]}(q, \vec x) & \flimit & \Big(
    \chi^{\so(d)}_{\Yboxdim{3pt} \Ylinethick{0.8pt}
      \gyoung(;;;,;;,;;,;)} (\vec x) + \chi^{\so(d)}_{\Yboxdim{3pt}
      \Ylinethick{0.8pt} \gyoung(;;;,;;,;;)}(\vec x) +
    \chi^{\so(d)}_{\Yboxdim{3pt} \Ylinethick{0.8pt} \gyoung(;;,;;,;;,;)}
    (\vec x) + \chi^{\so(d)}_{\Yboxdim{3pt} \Ylinethick{0.8pt}
      \gyoung(;;,;;,;;)}(\vec x) \Big) \Pf d (\vec x) \\ && =
    \chi^{\Poinc}_{[0;\ \Yboxdim{3pt} \Ylinethick{0.8pt}
        \gyoung(;;;,;;,;;,;)\,]} (\vec x) +
    \chi^{\Poinc}_{[0;\ \Yboxdim{3pt} \Ylinethick{0.8pt}
        \gyoung(;;;,;;,;;)\,]}(\vec x) +
    \chi^{\Poinc}_{[0;\ \Yboxdim{3pt} \Ylinethick{0.8pt}
        \gyoung(;;,;;,;;,;)\,]} (\vec x) +
    \chi^{\Poinc}_{[0;\ \Yboxdim{3pt} \Ylinethick{0.8pt}
        \gyoung(;;,;;,;;)\,]}(\vec x)\, ,
  \end{eqnarray}
  i.e., as expected, only appear massless fields with Young diagrams
  obtained from branching $\Y$ from $\so(d)$ onto $\so(d-1)$ with the
  exception of leaving the second block untouched.
\end{example}

\subsection{de Sitter case}
Irreducible representations of $\so(1,d+1)$ were (to our knowledge)
first spelled out in \cite{Hirai1962, Hirai1962b} then completed in
\cite{Schwarz1971,Gavrilik:1975ae, Klimyk:1976ac}. In these early
papers, one can find the classification of irreps, irrespectively of
their unitary character.

Irreps of the exceptional series are labeled by \cite{Hirai1962,
  Hirai1962b} the conformal weight $\Delta_c = d+n-p_I-1$, and a Young
diagram $\Y_{p_I} = (s_1,\dots,s_r)$, such that $s_{p_I+1} > n
\geqslant s_{p_I+2}\, , \, n \in \N$. This set of data should describe
a gauge field with symmetry $\Y_{n,p_I}: = (s_1, \dots, s_{p_I}, n,
s_{p_I+2}, \dots, s_B)$ whose $I$th block is activated (having in mind
that as in the previous subsection, $p_I$ is the cumulated height of
the first $I$ blocks of this diagram, whose total height is $p_B$)
whose gauge parameter has symmetry $\check{\Y}_{n,p_I}^{(p_I)} :=
(s_1, \dots, s_{p_I-1}, s_{p_I+1}-1, s_{p_I+2}, \dots,
s_{p_B})$. Those representations are unitary only for $n=0$, that is
when the activated block is the last one. More generically, the
characters of the exceptional series are:
\begin{itemize}
\item Even spacetime dimensions:
  \begin{eqnarray}
    \chi^\dS_{[d+n-p_I-1; \Y_{n,p_I}]}(q, \vec x) & = & (q^{p_I+1-n} -
    q^{d-p_I-1+n}) \chi^{\so(d)}_{\Y_{p_I}}(\vec x)\, \Pd d (q, \vec
    x) \nonumber \\ && \quad + \sum_{m=1}^{p_B-p_I} (-1)^m
    (q^{p_I+1+m-s_{p_I+1+m}} - q^{d-p_I-1-m+s_{p_I+1+m}})
    \chi^{\so(d)}_{\Yp{m}}(\vec x)\, \Pd d (q, \vec x) \nonumber \\ &&
    \quad \qquad + \sum_{m=1}^{r-p_B} (-)^{p_B+p_I+m} (q^{p_B + 1 + m}
    - q^{d-p_B - 1 - m})\chi^{\so(d)}_{(\Yp{p_B-p_I},\1^m)}(\vec x)\,
    \Pd d (q, \vec x) \nonumber \\ && \quad \qquad \qquad +
    \sum_{\ell=1}^{p_I+1} (-)^{\ell+p_I+1} q^{s_\ell + d - \ell}
    \chi^{\so(d)}_{\check{\Y}^{(\ell)}_{n,p_I}}(\vec x)\, \Pd d (q,
    \vec x)\, ,
  \end{eqnarray}
  where $\Yp{m}$ is the Young diagram obtained by adding one box in
  each of the $m$ row under the $(p_I+1)$th of $\Y_{n,p_I}$, i.e.,
  \begin{equation}
    \Yp{m} = (s_1,\dots,s_{p_I}, n+1, s_{p_I+2}+1, \dots, s_{p_I+m}+1,
    s_{p_I+m+2}, \dots, s_{p_B},0,\dots,0)
  \end{equation}
  and $\check{\Y}^{(\ell)}_{n,p_I}$ is the diagram obtained by
  removing the $\ell$th row together with one box in each of the rows
  after the $\ell$th one until the $(p_I+1)$th from
  $\Y_{n,p_I}$,
  \begin{equation}
    \check{\Y}^{(\ell)}_{n,p_I} = (s_1, \dots, s_{\ell-1},
    s_{\ell+1}-1, \dots, s_{p_I+1}-1, n, s_{p_I+2}, \dots, s_{p_B}, 0,
    \dots, 0)\, .
  \end{equation}
  Taking the flat limit ($q \rightarrow 1$) of the above expression,
  one is left with an alternating sum of $\so(d)$ characters of the same
  type as in the unitary case or the above detailed AdS$_{d+1}$ case:
  \begin{equation}
    \chi^\dS_{[d+n-p_I-1; \Y_{n,p_I}]}(q, \vec x) \flimit
    \sum_{\ell=1}^{p_I+1} (-)^{\ell+p_I+1}
    \chi^{\so(d)}_{\check{\Y}^{(\ell)}_{n,p_I}}(\vec x)
  \end{equation}
  The Young diagrams appearing in this sum correspond to a gauge field
  with symmetry $\Y_{n,p_I}$, its gauge parameter having symmetry
  $\check{\Y}_{n,p_I}^{(p_I)}$ and its reducibility parameters, and
  therefore by the same arguments used in
  \hyperref[proof_exc_odd]{Subsection \ref{proof_exc_odd}} one is left
  with the following spectrum of massless fields in flat space:
  \begin{eqnarray}
    &&\Sp(\Y_{n,p_I}) = \nonumber\\ && \Big\{ \Y' = (\ell_1^{h_1-1},
    \ell_1-n_1,\dots, \ell_{I-1}^{h_{I-1}-1}, \ell_{I-1}-n_{I-1},
    \ell_{I}^{h_{I}}, n - n_p, \ell_{I+1}^{h_{I+1}-1},
    \ell_{I+1}-n_{I+1}, \dots, \ell_B^{h_B-1}, \ell_B-n_B)\, ,
    \nonumber \\ && \qquad \qquad 0 \leqslant s_i-s_{i+1}\, , i \in\{
    1,\cdots, B \}\,,\, i \neq I\, , \, 0 \leqslant n_p \leqslant
    n-s_{I+1} \Big\}
  \end{eqnarray}

  \begin{remark}
    From our earlier analysis of the unitary irreps of the exceptional
    series, we learned that the character obtained from resolving the
    module of the shadow of what we called the gauge field's curvature
    (using the BGG sequences recalled in \hyperref[app:char]{Appendix
      \ref{app:char}}) matches the character obtained in
    \cite{Hirai1965}. We therefore applied the same technique for
    non-unitary representation, i.e., we computed the character
    corresponding to generalized Verma module with highest-weight
    $[\Delta_c\,; \vec s\,] = [p_I+1-n\,; \Y_{p_I}]$ (remember that
    the curvature is caracterized by the same Young diagram and
    conformal weight $d-\Delta_c$).
  \end{remark}
  
\item Odd spacetime dimensions:
  \begin{eqnarray}
    \chi^\dS_{[d+n-p_I-1; \Y_{n,p_I}]}(q, \vec x) & = & (q^{p_I+1-n} +
    q^{d-p_I-1+n}) \chi^{\so(d)}_{\Y_{n,p_I}}(\vec x)\, \Pd d (q, \vec
    x) \nonumber \\ && \quad + \sum_{m=1}^{p_B-p_I} (-1)^m
    (q^{p_I+1+m-s_{p_I+1+m}} + q^{d-p_I-1-m+s_{p_I+1+m}})
    \chi^{\so(d)}_{\Yp{m}}(\vec x)\, \Pd d (q, \vec x) \nonumber \\ &&
    \quad \quad + \sum_{m=1}^{r-p_B-1} (-)^{p_B+p_I+m} (q^{p_B+1+m} +
    q^{d-p_B-1-m})\chi^{\so(d)}_{(\Yp{p_B-p_I},\1^m)}(\vec x)\, \Pd d
    (q, \vec x) \nonumber \\ && \quad \quad \quad + (-)^{r+p_I}
    q^{d/2} \Big(\chi^{\so(d)}_{(\Yp{p_B-p_I}, \1^{p_B-r}_+)}(\vec x)
    + \chi^{\so(d)}_{(\Yp{p_B-p_I}, \1^{p_B-r}_-)}(\vec x)\Big) \Pd d
    (q, \vec x) \nonumber \\ && \quad \quad \quad\quad - 2
    \sum_{\ell=1}^{p_I+1} (-)^{\ell+p_I+1} q^{s_\ell + d - \ell}
    \chi^{\so(d)}_{\check{\Y}^{(\ell)}_{n,p_I}}(\vec x)\, \Pd d (q,
    \vec x)\, ,
  \end{eqnarray}
  For the same reason as in the case of UIRs in the exceptional series
  in even spacetime dimensions treated in
  \hyperref[proof_exc_odd]{Subsection \ref{proof_exc_odd}}, the flat
  limit of the character of their nonunitary counterpart does not
  appear to produce a result that can be interpreted as a sum of
  Poincar\'e characters for massless fields that could be part of a
  BMV-type mechanism.
\end{itemize}

\newpage
\bibliographystyle{utphys}
\bibliography{biblio}

\end{document}